\documentclass[sigplan, screen, nonacm]{acmart}
\usepackage{algorithm}
\usepackage{algorithmic}
\usepackage{amsmath}
\usepackage{bm}         
\usepackage{enumitem}   
\usepackage{listings} 
\usepackage{pifont}     
\usepackage{subcaption} 
\usepackage{xcolor}     
\newcommand{\Rop}{\mathop{\scalebox{1.5}{\raisebox{0pt}{$\mathcal{R}$}}}\limits}
\AtBeginDocument{%
  }

\copyrightyear{2026}
\acmYear{2026}
\setcopyright{cc}
\setcctype{by}
\acmConference[ASPLOS '26] {Proceedings of the 30th ACM International Conference on Architectural Support for Programming Languages and Operating Systems, Volume 2}{March 21--26, 2026}{Pittsburgh, PA, USA.}
\acmBooktitle{Proceedings of the 30th ACM International Conference on Architectural Support for Programming Languages and Operating Systems, Volume 2 (ASPLOS '26), March 21--26, 2026, Pittsburgh, PA, USA}
\acmISBN{979-8-4007-2359-9/2026/03}
\acmDOI{10.1145/3779212.3790209}
\acmSubmissionID{V2fp3794}

\begin{document}

\title{RedFuser: An Automatic Operator Fusion Framework for Cascaded Reductions on AI Accelerators}

\author{Xinsheng Tang}
\authornote{All three authors contributed equally to this research.}
\affiliation{%
  \institution{Alibaba Cloud Computing}
  \city{Shanghai}
  \country{China}
}

\author{Yangcheng Li}
\authornotemark[1]
\affiliation{%
  \institution{Alibaba Cloud Computing}
  \city{Shanghai}
  \country{China}
}

\author{Nan Wang}
\authornotemark[1]
\affiliation{%
  \institution{Alibaba Cloud Computing}
  \city{Shanghai}
  \country{China}
}

\author{Zhiyi Shu}
\affiliation{%
  \institution{Alibaba Cloud Computing}
  \city{Shanghai}
  \country{China}
}

\author{Xingyu Ling}
\affiliation{%
  \institution{Alibaba Cloud Computing}
  \city{Shanghai}
  \country{China}
}

\author{Junna Xing}
\affiliation{%
  \institution{Alibaba Cloud Computing}
  \city{Shanghai}
  \country{China}
}

\author{Peng Zhou}
\affiliation{%
  \institution{Alibaba Cloud Computing}
  \city{Sunnyvale}
  \country{USA}
}

\author{Qiang Liu}
\affiliation{%
  \institution{Alibaba Cloud Computing}
  \city{Shenzhen}
  \country{China}
}

\renewcommand{\shortauthors}{Xinsheng Tang et al.}

\begin{abstract}
Operator fusion, as a key performance optimization technique in 
the deployment of AI models,
significantly improves execution efficiency and has been widely
adopted in modern AI compilers.
However, for cascaded reduction operations involving multiple loops
with inter-loop data dependencies, such as the safe softmax
followed by GEMM within attention mechanisms, existing compilers lack
effective automated fusion and kernel generation
capabilities. Although some works have addressed specific instances
through hand-crafted fusion strategies, their solutions are
limited in generality and difficult to extend to other similar
structures. Given the prevalence of such computational patterns in
deep learning models, there remains significant untapped potential in
achieving general and automated fusion optimization.

In this paper, we present a formal theoretical methodology for analyzing
cascaded reductions which can fuse them into a single loop and
introduce an incremental computation form. 
Based on this methodology, we design 
\textbf{Red}uction \textbf{Fuser} (RedFuser), a framework
that automatically identifies supported cascaded reduction patterns and generates optimized fused kernels.
Experiments show that RedFuser successfully fuses diverse
workloads, achieving up to 2$\times$ to 5$\times$ speedup over state-of-the-art AI
compilers and matching the performance of highly optimized hand-written kernels.
The code is available at https://github.com/alibaba/redfuser

\end{abstract}



\maketitle

\section{Introduction}
In recent years, artificial intelligence (AI) models have
demonstrated remarkable advances in various technical domains
\cite{gpt4.0, multimodal, chameleonteam2025,
crossdomain, reed2022generalistagent, contentsynthesis, esser2024},
with state-of-the-art models \cite{model1, model2, model3, model4, kimik2}
consistently achieving benchmark performance that approaches or
exceeds human-level baselines. However, these technological
achievements have introduced significant scaling challenges. Modern
cut-ting-edge large language models (LLMs) now exceed trillion
parameters (e.g., Kimi K2 \cite{kimik2} has 1 trillion parameters),
greatly increasing computational cost. 
To meet the extreme demands for cost and speed, enhancing
the efficiency of model training and inference becomes critically important.

\begin{figure}[htbp]
    \centering
    \begin{subfigure}[t]{0.35\textwidth}
        \includegraphics[width=\linewidth]{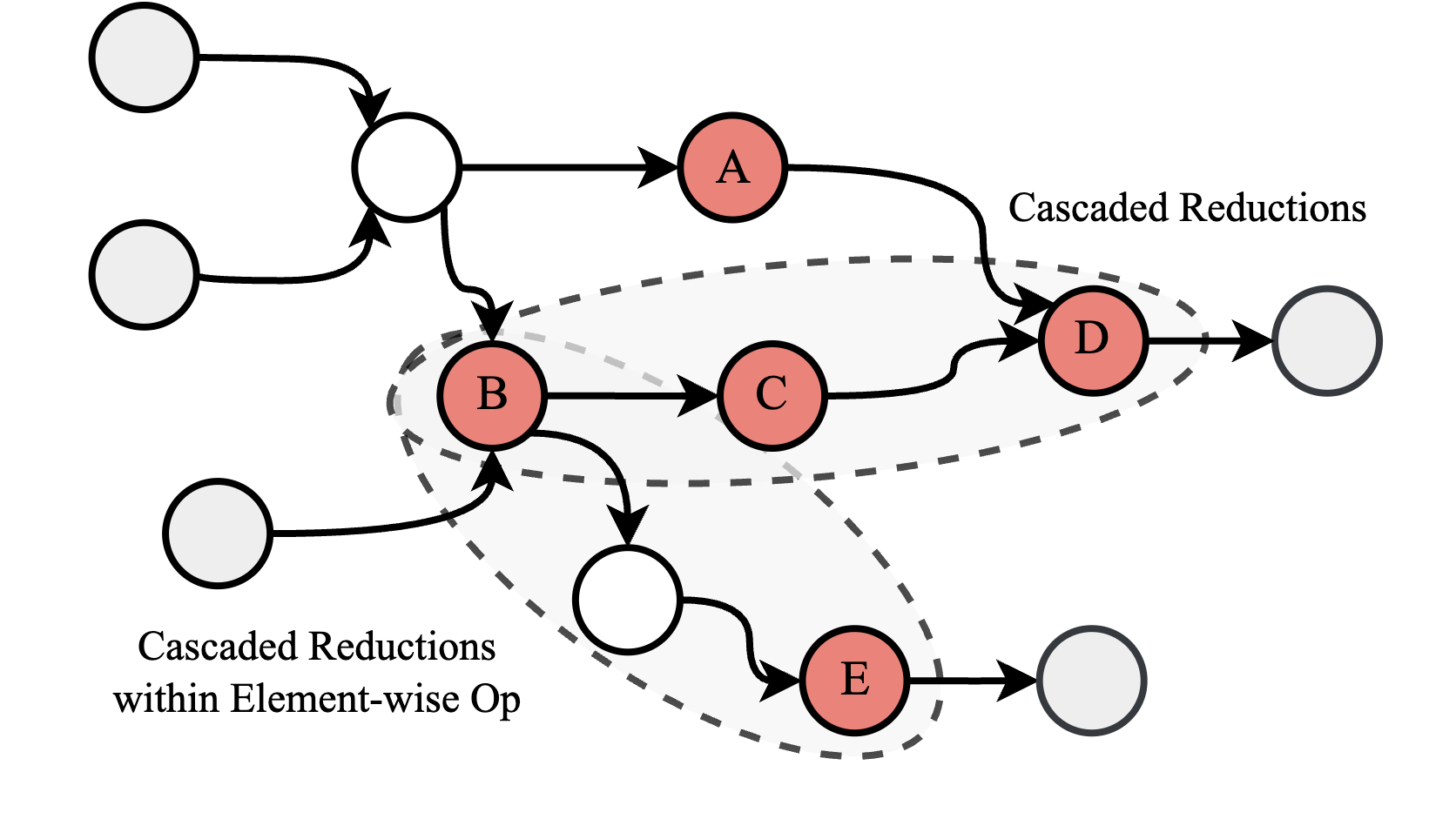}
        \caption{Cascaded reduction patterns}
        \label{fig:cascaded_reduction_patterns}
    \end{subfigure}
    \hfill
    \begin{subfigure}[t]{0.45\textwidth}
        \includegraphics[width=\linewidth]{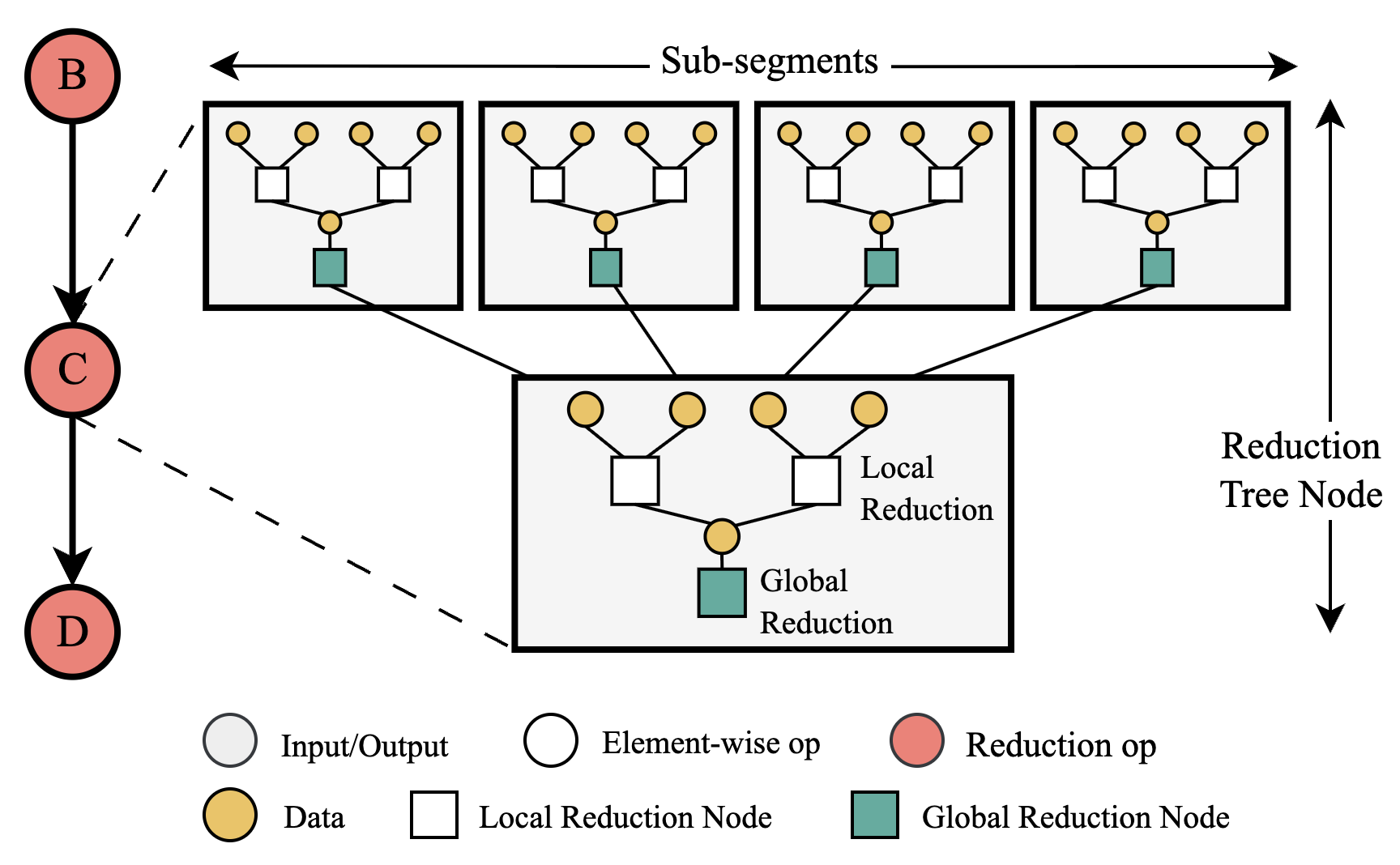}
        \caption{The chain of reduction trees}
        \label{fig:reduction_tree_chains}
    \end{subfigure}
    \caption{The reduction tree structure in cascaded reductions}
    \Description{The reduction tree structure in cascaded reductions}
    \label{fig:intro}
\end{figure}

Reductions are fundamental computational patterns in AI workloads that derive a single value from an array through loop-based mathematical operations, such as summation and maximum.
As shown in Figure~\ref{fig:intro}, a common parallel strategy for
reductions on AI accelerators is the
\textbf{reduction tree} structure. In this approach, the input sequence is partitioned into
sub-segments for local computation and caching, followed by global
reductions to aggregate results into a final output. Notably, the general matrix multiplication (GEMM) operation, 
which multiples an M $\times$ K matrix A with an K $\times$ N matrix B, 
producing an M $\times$ N matrix C, can be expressed as a triple
nested loop consisting of M $\times$ N independent reduction operations along the K-axis, 
making it fundamentally a reduction operation.

In deep learning, sequences of reduction operations with data
dependencies, referred to as \textbf{cascaded reductions}, are prevalent
computational patterns. Examples include attention mechanisms and
normalization layers. Cascaded reductions involving multiple loops naturally form chains of reduction trees. Unlike inherently parallelizable operations such as element-wise computations, cascaded reductions
exhibit a critical computational characteristic -- \textbf{data dependency}: each subsequent
reduction requires the output of its predecessor, specifically, the
root node of the prior reduction tree. This dependency introduces two key performance bottlenecks:

\begin{itemize}[leftmargin=*, labelindent=0pt]
  \item \textbf{Redundant Memory Access}:
  Each reduction tree incurs independent memory accesses for its input data.
  \item \textbf{Limited Parallelism}:
  Reduction trees cannot be executed in parallel, nor can memory accesses be overlapped with reduction computations to hide latency, significantly limiting overall performance. 
\end{itemize}
Moreover, each level of a reduction tree must cache the complete result from the previous level, increasing cache pressure and constraining the achievable reduction length per level.

To address these challenges, \textbf{operation fusion} \cite{kernelfusion, loopfusion, SRNPU,
tangram} is a widely adopted optimization technique that merges multiple consecutive operators into a single operator. 
However, existing AI compilers \cite{XLA2020, TVM2018, PyTorch2024, DNNFusion2021, AStitch2022, Bolt2022, Chimera2023, MCFuser2024, SOUFFLE2024}
exhibit critical limitations in optimizing cascaded reductions
through operator fusion. Specifically, these approaches lack
the capability to perform detailed analysis, leading to suboptimal
fusion strategies. While some fine-grained manual
optimization techniques \cite{dac2024softmax, FLAT, SC, FlashAttention2022, FlashDecoding2023} have addressed
specific instances, they require careful case-by-case derivation
and exhibit limited universality across different patterns.

To overcome these limitations, our work makes the
following key contributions:

\begin{itemize}[leftmargin=*, labelindent=0pt]
  \item \textbf{Cascaded Reductions Fusion Methodology}:
  We propose a formal methodology that fuses cascaded reductions into a single reduction, thereby eliminating redundant data loads and enabling parallel execution of dependent reductions.  To mitigate the constraints of cache capacity at each level on the corresponding reduction length,  we introduce an incremental computation scheme that  updates partial results progressively.
  The methodology is designed to be general and compatible with the compilation stacks of modern AI compilers.
  
  \item \textbf{Lightweight Operator Fusion Framework}:
 We implement our methodology in \textbf{RedFuser}, a two-stage lightweight operator fusion framework that systematically optimizes cascaded reductions. First, it employs a symbolic deduction engine to analyze the fusibility of cascaded reductions, and derive the fused reduction expressions and corresponding incremental computation forms. Second, it performs hardware-aware code generation by lowering the fused expression into a tile-based representation and generating high-performance kernels.
 \item \textbf{Comparative Performance Evaluation of Various \\Workloads}: We evaluate RedFuser on modern GPU architectures (NVIDIA A10 and H800) across four representative workloads: Multi-Head Attention (MHA), Multi-Latent Attetion (MLA), MoE routing, and FP8 Quant + GEMM. Results show that RedFuser achieves 2$\times$ to 5$\times$ speedup over state-of-the-art AI compilers (e.g., TVM\cite{TVM2018}, Dynamo\cite{PyTorch2024}), and its generated kernels match or exceed the performance of expert-optimized implementations, achieving state-of-the-art (SOTA) results.
\end{itemize}

\section{Background and Motivation}
\label{chap:Background}

\subsection{Parallel Reduction on GPUs}

Modern Graphics Processing Units (GPUs) have emerged as the cornerstone of accelerating compute-intensive applications, due to their massive parallelism and high memory bandwidth.
At the core of GPU computing lies a hierarchical programming model~\cite{CUDA} that organizes parallel execution into three levels: threads, thread blocks, and grids.
Each thread possesses its own private local memory and executes instructions independently. Threads are grouped into thread blocks, within which they can cooperate efficiently through a shared memory space, enabling fine-grained data reuse and synchronization. Direct synchronization or communication mechanisms between blocks are absent in the current programming model, unless via multiple kernel launches or features such as cooperative groups. All blocks collectively form a grid that executes a single kernel function.
The memory hierarchy includes per-thread local memory, block-scoped shared memory, and globally accessible global memory, providing a balanced trade-off between capacity and access latency.
Programming frameworks such as CUDA enable developers to implement parallel GPU algorithms through techniques such as tiling, data reuse, and explicit memory management.

Parallel reduction on GPUs is a well-studied problem.
The seminal work by Harris~\cite{Harris2007} established a systematic methodology for efficient GPU-based reduction.
The input data are first partitioned into multiple segments, with each thread block responsible for reducing one segment locally. Specifically, each thread loads its assigned elements from global memory into shared memory. Within the block, a sequence of reduction steps is performed using an interleaved addressing pattern: in each iteration, pairs of elements separated by a stride $s$ are combined (e.g., via addition or maximization), where $s$ starts at half the block size and is halved in each subsequent step. This process effectively constructs a binary tree-like reduction topology, progressively collapsing the data within the block. The final partial result is held by a designated thread (e.g., thread 0) and written back to global memory. After all partial results are generated by all thread blocks, another separate kernel is launched to reduce these intermediates into the final output.

\subsection{Cascaded Reductions in Deep Learning}

Cascaded reductions are pervasive in deep learning models, where their performance directly impacts end-to-end performance.

A canonical example is the safe softmax computation. It first computes the maximum value along a specified dimension (a max reduction), subtracts this maximum from the input, computes the sum of exponentials (a sum reduction), and finally normalizes the result. This process involves two consecutive reductions with strict data dependence, making it a prototypical instance of cascaded reductions.
This pattern naturally extends to more complex model components:
\begin{itemize}[leftmargin=*, labelindent=0pt]
    \item \textbf{Attention}.
    The entire forward pass of attention mechanism of Transformers~\cite{model1} can be abstracted as GEMM $+$ Softmax $+$ GEMM, forming complex cascaded reductions that become a critical performance bottleneck.
    \item \textbf{MoE routing}.
    In Mixture-of-Experts (MoE) models, the router often employs a softmax $+$ top-k pipeline: softmax normalizes the gating scores (involving max and sum reductions), followed by selecting the top-k experts.
    \item \textbf{FP8 Quant $+$ GEMM}.
    In FP8 per-token quantization, dynamic scaling factors are computed for each token, often based on the absolute maximum value (an abs-max reduction). The quantized FP8 tensors are then used in GEMM operations, yielding a Quant $+$ GEMM cascade.
\end{itemize}

\subsection{Limitations of Previous Works}

Existing deep learning compilers exhibit limitations in optimizing cascaded reductions.
On the one hand, systems such as DNNFusion\cite{DNNFusion2021} and AStitch\cite{AStitch2022} lack dedicated fusion rules for these patterns. On the other hand, while frameworks like Bolt\cite{Bolt2022}, Chimera\cite{Chimera2023}, MCFuser\cite{MCFuser2024}, and SOUFFLE\cite{SOUFFLE2024} may incidentally handle certain instances, such as the attention pattern, they do not formally model cascaded reductions as a structured computation pattern, nor do they perform in-depth analysis of their dependency and memory access characteristics. As a result, these compilers fail to generate loop-level fused kernels across multiple reduction stages, leading to suboptimal fusion strategies. In contrast, hand-optimized kernels such as FlashAttention\cite{FlashAttention2022} and FlashDecoding\cite{FlashDecoding2023} achieve remarkable performance by leveraging tiling and online softmax techniques, but their implementations rely on expert-level GPU programming and manual scheduling. Moreover, their optimization logic is tightly coupled to specific computational patterns, making them difficult to generalize to other cascaded reduction structures, such as the softmax $+$ top-k pipeline in MoE routing or the Quant $+$ GEMM chain in FP8 computation. Together, these limitations highlight a compelling opportunity: the need for a compiler framework that not only formally characterizes the structure of cascaded reductions, but also enables automated, semantics-preserving fusion transformations, thereby bridging the gap between high performance and broad applicability.

\section{Optimization Methods for Cascaded Reductions}
\label{chap:Methodology}

This chapter presents a
mathematical definition of the cascaded reductions and
formally describes its typical implementation on AI accelerators—the
chain of reduction trees. Subsequently, we derive a method
that fuses chain of reduction trees into a single reduction
tree. Building upon this, an incremental computation formulation is proposed.

\subsection{Formal Definition of Cascaded Reductions}
\label{chap:formaldefinition}

\begin{figure}[htbp]
  \centering
  \centerline{\includegraphics[width=0.95\linewidth]{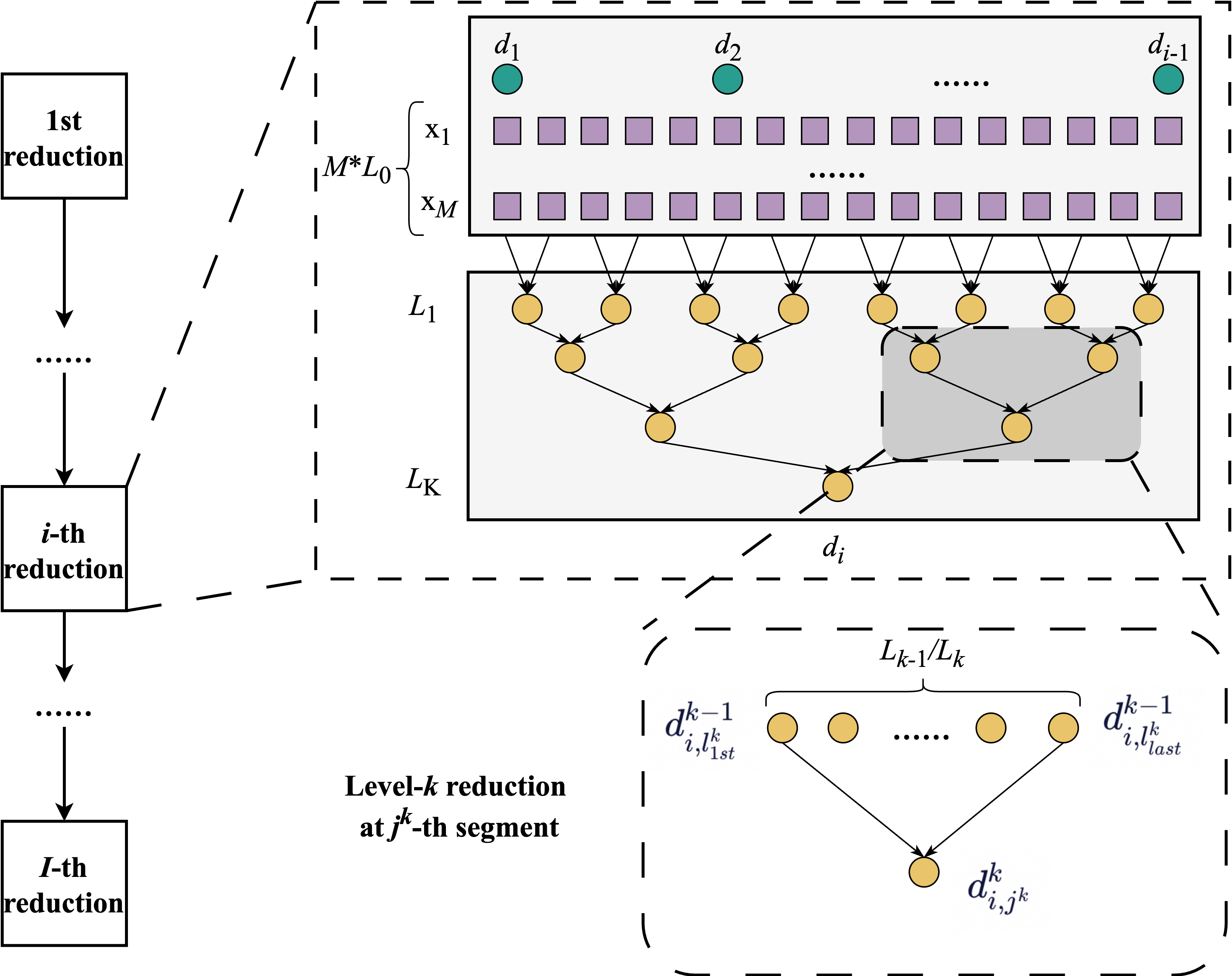}}
  \caption{The formal definition of cascaded reductions}
  \label{fig:formal_def}
  \Description{The formal definition of cascaded reductions.}
\end{figure}

As shown in Figure~\ref{fig:formal_def}, consider $I$ cascaded reduction operations defined over a set
$\mathbb{S}$, with input consisting of $M$ data vectors
$\{\bm{X}_1, \bm{X}_2,\allowbreak \ldots, \bm{X}_M\}$, each of
length $L_0$, denoted as $\bm{X} \in \mathbb{S}^{M \times L_0}$.
Let $d_i$ be the output of the $i$-th reduction operation, which
depends on the previous $i-1$ reduction results $\bm{D}_i =
\{d_1, d_2, \ldots, d_{i-1}\}$. All reduction results form a vector
$\bm{D} = \{d_1, d_2, \ldots, d_N\}$ in dependency order. The
$i$-th reduction expression can be formally defined as:
\begin{equation}
  d_i = \sideset{}{_i}\Rop_{l=1}^{L_0}
  \mathrm{F}_i(\bm{X}[l], \bm{D}_i),
  \label{eq:d_i}
\end{equation}
where:
\begin{itemize}[leftmargin=*, labelindent=0pt]
  \item $\sideset{}{_i}\Rop$ denotes the $i$-th reduction
    operation, whose underlying associative operator is $\oplus_i$;
  \item $\bm{X}[l] = \{\bm{X}_1[l], \bm{X}_2[l], \ldots,
    \bm{X}_M[l]\}$ represents the set of input elements at position $l$;
  \item $\mathrm{F}_i: \mathbb{S}^M \times \mathbb{S}^{i-1} \to
    \mathbb{S}$ is the mapping function for the $i$-th reduction,
    operating on the input and dependent results.
\end{itemize}

\subsubsection{Reduction Tree}
If the binary operator $\oplus_i$ satisfies
the following properties:
\begin{itemize}[leftmargin=*, labelindent=0pt]
  \item \textbf{Associativity}: $\forall s_1,s_2,s_3 \in \mathbb{S}$,
    $s_1 \oplus_i (s_2 \oplus_i s_3) = (s_1 \oplus_i s_2) \oplus_i s_3$;
  \item \textbf{Commutativity}: $\forall s_1,s_2 \in \mathbb{S}$,
    $s_1 \oplus_i s_2 = s_2 \oplus_i s_1$;
\end{itemize}
then the reduction operation can be split into multiple
independent segments computed in parallel, whose outputs are
later combined via the associative operator. 
This property allows the reduction process to be organized into 
a reduction tree structure, formally defined as follows.

Suppose the $i$-th reduction operation structured into $K$
levels, where the output length at level $k$ is denoted by
$L_k$ ($k = 1,2,\ldots,K$), satisfying $L_0 > L_1 > \cdots > L_K =
1$.

For the $j^1$-th segment at the first level ($k=1$), its
output is defined as:
\begin{equation}
  d_{i,j^1}^1 = \sideset{}{_i}\Rop_{l =
  l_{1st}^1}^{l_{last}^1} \mathrm{F}_i(\bm{X}[l], \bm{D}_i),
\end{equation}
where $l_{1st}^1 = (j^1 - 1)\frac{L_0}{L_1} + 1$ and $l_{last}^1 =
j^1\frac{L_0}{L_1}$ represent the index range of the input for this segment.

For level $k > 1$, each level $k$ partitions the $L_{k-1}$ outputs of the 
previous level into segments of length $L_{k-1}/L_k$, 
which serve as the inputs of the current level. Let $d_{i,j^k}^k$ denote the
output of the $j^k$-th segment at level $k$. Then:
\begin{equation}
  d_{i,j^k}^k = \sideset{}{_i}\Rop_{j^{k-1} =
  l_{1st}^k}^{l_{last}^k} d_{i,j^{k-1}}^{k-1}, \quad k > 1,
\end{equation}
where $l_{1st}^k = (j^k - 1)\frac{L_{k-1}}{L_k} + 1$ and $l_{last}^k
= j^k\frac{L_{k-1}}{L_k}$.



\subsubsection{Chain of Reduction Trees}
It is important to note that the $i$-th reduction
depends on the results $\bm{D}_i$ of the preceding
$i-1$ reductions, that is, it must wait for all roots of the
preceding reduction trees before it can begin. This strong sequential
dependency forces the reduction trees to execute in a chain-like
order, forming a \textbf{Chain of Reduction Trees}. 

\subsection{Fused Cascaded Reductions}

\begin{figure}[htbp]
  \centering
  \begin{subfigure}[b]{\linewidth}
    \centering
    \includegraphics[width=0.9\linewidth]{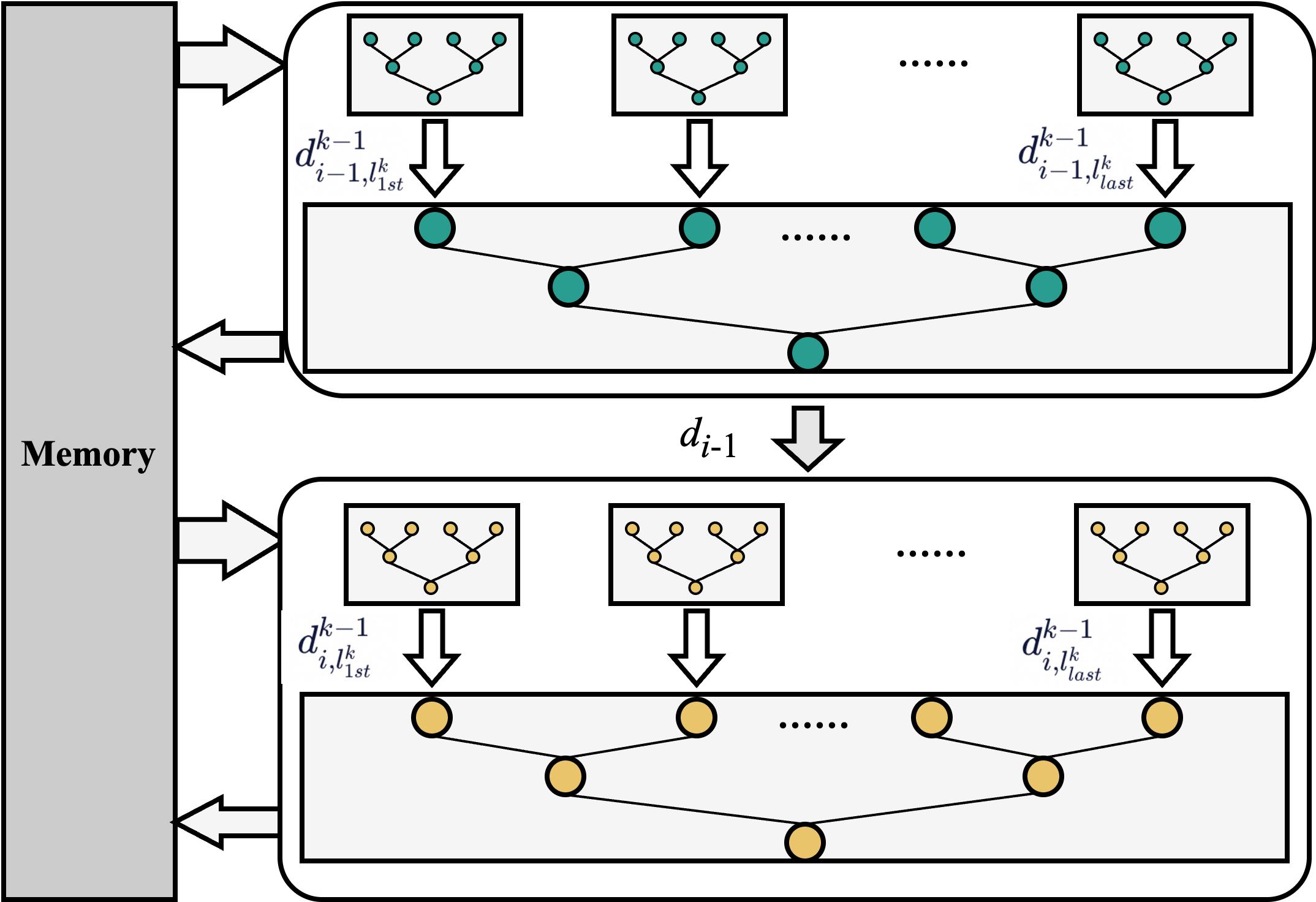}
    \caption{Cascaded reductions before fusion}
    \label{fig:before-fusion}
  \end{subfigure}
  \begin{subfigure}[b]{\linewidth}
    \centering
    \includegraphics[width=0.9\linewidth]{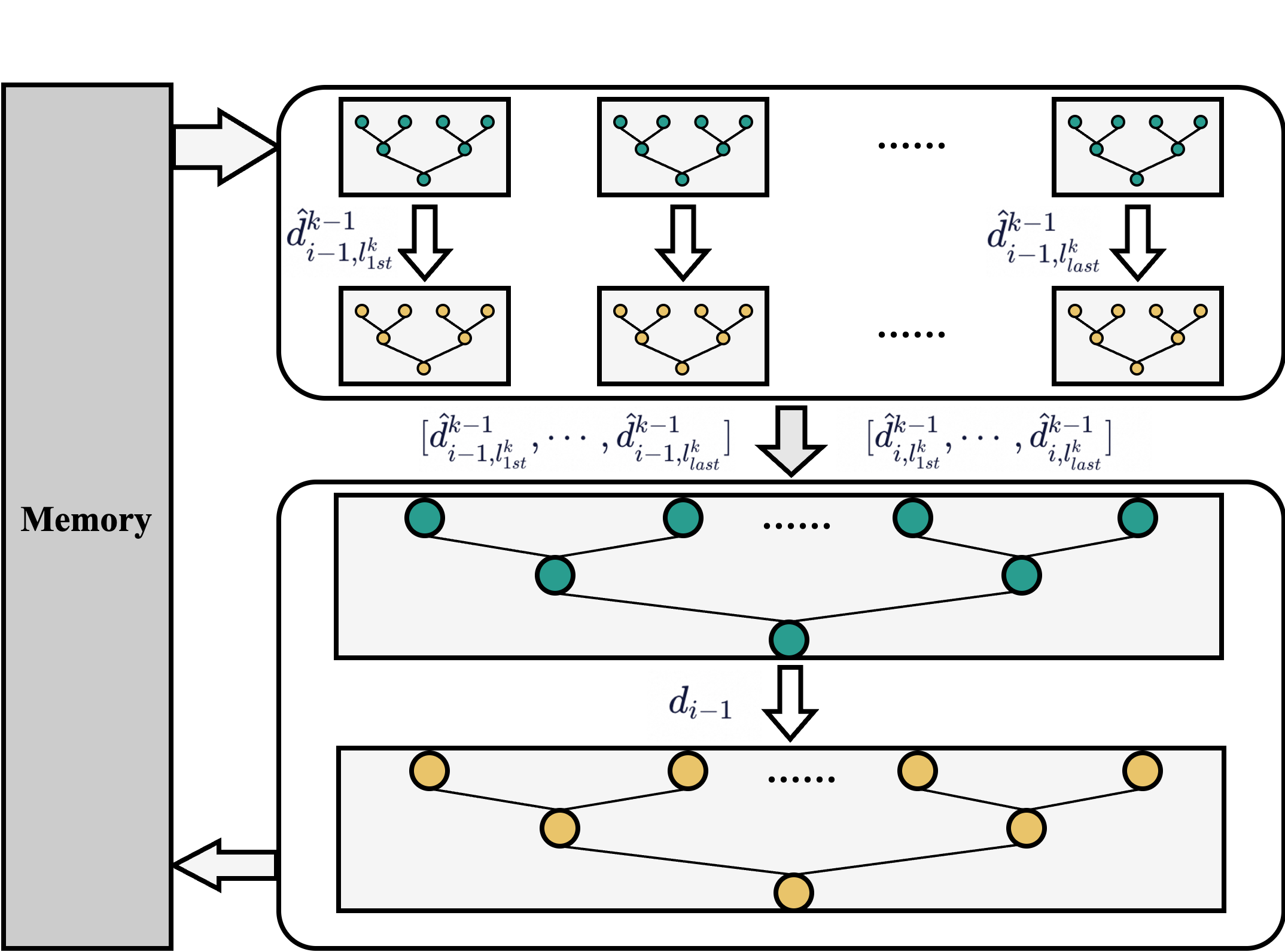}
    \caption{Cascaded reductions after fusion}
    \label{fig:after-fusion}
  \end{subfigure}
  \caption{Comparison of computation and memory access patterns in the cascaded reductions before and after fusion.
  (a)~Unfused cascaded reductions: each reduction re-loads the input data and accesses the results from prior reductions.
  (b)~Fused cascaded reductions: the input is loaded only once, and memory accesses to results of preceding reductions are eliminated.}
  \Description{Cascaded reductions before and after fusion.}
  \label{fig:unfuse_fuse}
\end{figure}

As shown in Figure~\ref{fig:unfuse_fuse}, this subsection derives a
fusion transformation to break
inter-reduction dependencies and enable cross-reduction expression fusion.

\subsubsection{Fusion Feasibility Conditions}
\label{chap:FusionConditions}

To support cross-reduction fusion, we require the $i$-th
reduction operation~\eqref{eq:d_i} to satisfy the following three conditions:

\begin{itemize}[leftmargin=*, labelindent=0pt]
  \item \textbf{Decomposability}: There exist functions
    $\mathrm{G}_i: \mathbb{S}^M \to \mathbb{S}$, $\mathrm{H}_i:
    \mathbb{S}^{i-1} \to \mathbb{S}$, and a binary operator $\otimes_i$ such that:
    \begin{equation}
      \mathrm{F}_i(\bm{X}[l], \bm{D}_i) =
      \mathrm{G}_i(\bm{X}[l]) \otimes_i \mathrm{H}_i(\bm{D}_i).
    \end{equation}

  \item \textbf{Algebraic Structure}: $(\mathbb{S}, \otimes_i)$ forms a
    commutative monoid, i.e., it satisfies:
    \begin{itemize}
      \item Associativity: $\forall s_1,s_2,s_3 \in \mathbb{S}$, $s_1
        \otimes_i (s_2 \otimes_i s_3) = (s_1 \otimes_i s_2) \otimes_i s_3$
      \item Commutativity: $\forall s_1,s_2 \in \mathbb{S}$, $s_1
        \otimes_i s_2 = s_2 \otimes_i s_1$
      \item Existence of identity: $\exists e \in \mathbb{S}$ such
        that $\forall s \in \mathbb{S}$, $e \otimes_i s = s \otimes_i e = s$
    \end{itemize}

  \item \textbf{Distributivity}: The reduction operator $\oplus_i$
    distributes over $\otimes_i$:
    \begin{equation}
      (s_1 \oplus_i s_2) \otimes_i s_3 = (s_1 \otimes_i s_3) \oplus_i
      (s_2 \otimes_i s_3).
      \label{eq:distributivity}
    \end{equation}
\end{itemize}

These conditions are widely satisfied in common reductions,
such as GEMM and safe softmax.

\subsubsection{Derivation of Fusion Expressions}
\label{chap:derivation}
As illustrated in Figure~\ref{fig:after-fusion}, 
let $\hat{d}_{i,j^k}^k$ denote the output of the $j^k$-th segment at
level $k$ of the $i$-th reduction. This output depends on the
level-$k$ outputs of the preceding $i-1$ reductions, denoted by
$\hat{\bm{D}}^k = \{\hat{d}_{1,j^k}^k, \dots,
\hat{d}_{i-1,j^k}^k\}$.

At the first level, the output of the $j^1$-th segment is:
\begin{equation}
  \hat{d}^1 = \sideset{}{_i}\Rop_{l = l_{1st}^1}^{l_{last}^1}
  \mathrm{G}_i(\bm{X}[l]) \otimes_i \mathrm{H}_i(\hat{\bm{D}}^1).
  \label{eq:fused_d1}
\end{equation}

Using Equation~\eqref{eq:distributivity},
$\mathrm{H}_i(\hat{\bm{D}}^1)$ can be factored out:
\begin{equation}
  \hat{d}^1 = \left( \sideset{}{_i}\Rop_{l =
  l_{1st}^1}^{l_{last}^1} \mathrm{G}_i(\bm{X}[l]) \right) \otimes_i
  \mathrm{H}_i(\hat{\bm{D}}^1).
\end{equation}

If $\mathrm{H}_i(\hat{\bm{D}}^1)$ is invertible under
$(\mathbb{S}, \otimes_i)$, with inverse denoted by
$\mathrm{H}_i(\hat{\bm{D}}^1)^{-1}$, then:
\begin{equation}
  \sideset{}{_i}\Rop_{l = l_{1st}^1}^{l_{last}^1}
  \mathrm{G}_i(\bm{X}[l]) = \hat{d}^1 \otimes_i
  \mathrm{H}_i(\hat{\bm{D}}^1)^{-1}.
  \label{eq:sum_G}
\end{equation}

Furthermore, the output of the $j^2$-th segment at level 2 is:
\begin{equation}
  \hat{d}^2 = \sideset{}{_i}\Rop_{j^1 =
  l_{1st}^2}^{l_{last}^2} \left( \sideset{}{_i}\Rop_{l =
    l_{1st}^1}^{l_{last}^1} \mathrm{G}_i(\bm{X}[l]) \otimes_i
  \mathrm{H}_i(\hat{\bm{D}}^2) \right).
  \label{eq:hatd_level2}
\end{equation}

Substituting Equation~\eqref{eq:sum_G} into Equation~\eqref{eq:hatd_level2}:
\begin{equation}
  \hat{d}^2 = \sideset{}{_i}\Rop_{j^1 =
  l_{1st}^2}^{l_{last}^2} \left( \hat{d}^1 \otimes_i
    \mathrm{H}_i(\hat{\bm{D}}^1)^{-1} \otimes_i
  \mathrm{H}_i(\hat{\bm{D}}^2) \right).
\end{equation}

Generalizing to level $k > 1$, the local result at segment $j^k$ can
be expressed as:
\begin{equation}
  \hat{d}^k = \sideset{}{_i}\Rop_{j^{k-1} =
    l_{1st}^k}^{l_{last}^k} \left( \hat{d}^{k-1} \otimes_i
  \mathrm{H}_i(\hat{\bm{D}}^{k-1})^{-1} \otimes_i
  \mathrm{H}_i(\hat{\bm{D}}^k) \right).
  \label{eq:fused_dk}
\end{equation}
Noted that $\mathrm{H}_i(\hat{\bm{D}}^{k-1})$ must be invertible under
$(\mathbb{S}, \otimes_i)$. For the non-invertible case, 
we present a correction to $\mathrm{H}_i(\hat{\bm{D}}^{k-1})$ in the 
Appendix~\ref{Appendix:Non-invertible}, ensuring that the 
Equation~\ref{eq:fused_dk} remains valid.


In the above expression, $\hat{d}^k$ depends on
the outputs at the same level
$\hat{\bm{D}}^k$ from the preceding $i-1$ reductions, 
rather than on the outputs $\bm{D}_i$ from the last level $K$. This implies:
\begin{itemize}[leftmargin=*, labelindent=0pt]
  \item Expressions from multiple reduction trees at the same level $k$
    can be merged into a single reduction tree;
  \item Input data $\bm{X}$ and dependent results
    $\hat{\bm{D}}^k$ can be cached on-chip and reused, avoiding
    redundant memory accesses.
\end{itemize}

\subsection{Incremental Computation Form}

Although the fused expression~\eqref{eq:fused_dk} eliminates redundant memory
accesses, it still requires the complete caching of the outputs
$\hat{d}_{i,j^{k-1}}^{k-1}$ from
the previous level $k-1$, as well as $\hat{\bm{D}}_{i,j^{k-1}}^{k-1}$
and $\hat{\bm{D}}_{i,j^{k-1}}^{k}$, before
computation can proceed.

\begin{figure}[htbp]
  \centering
  \begin{subfigure}[b]{\linewidth}
    \centering
    \includegraphics[width=0.9\linewidth]{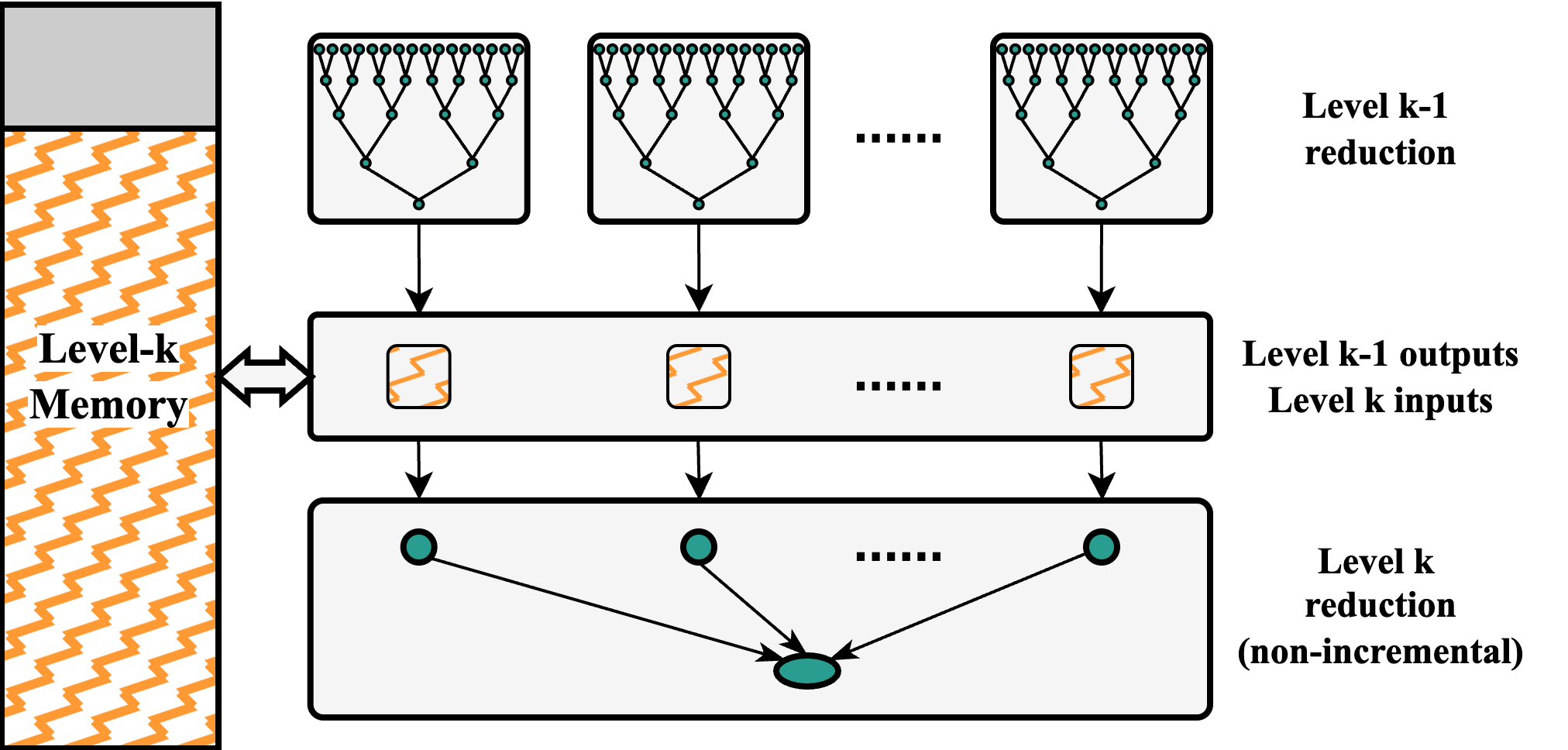}
    \caption{Non-incremental mode. Memory usage grows linearly with the input length.}
    \label{fig:non-incremental-detail}
  \end{subfigure}
  \begin{subfigure}[b]{\linewidth}
    \centering
    \includegraphics[width=0.9\linewidth]{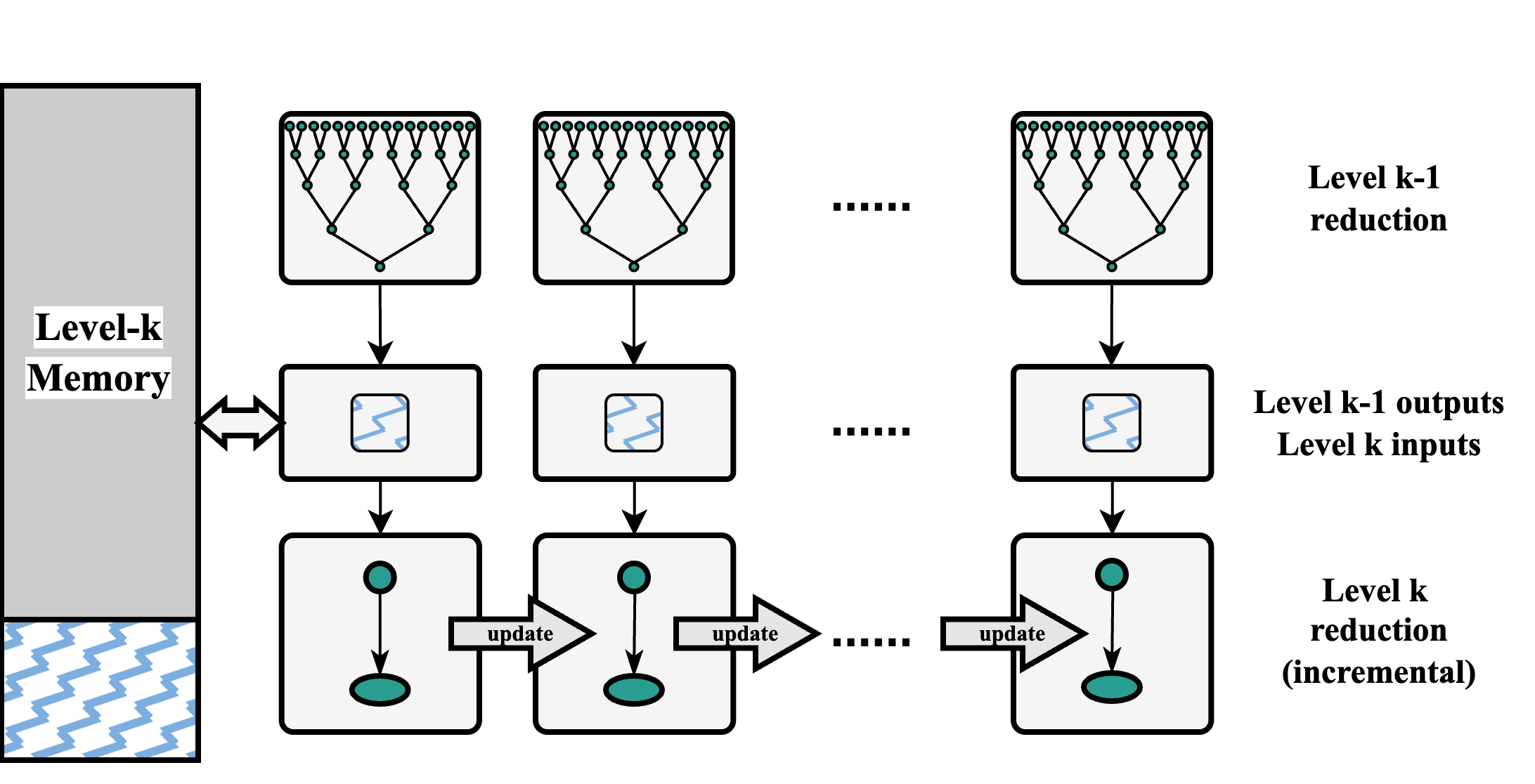}
    \caption{Incremental mode. A constant memory usage regardless of input length.}
    \label{fig:incremental-detail}
  \end{subfigure}

  \caption{Comparison between non-incremental and incremental computation. 
  (a)~Non-incremental computation: the $k$-th level reduction must wait until all inputs are available before execution and on-chip memory consumption grows with the input length. 
  (b)~Incremental computation: the result at level~$k$ can be updated immediately upon arrival of new input, maintaining constant on-chip memory footprint.}
  \Description{Illustration of memory access pattern in non-incremental and incremental modes.}
  \label{fig:offline-online-detail}
\end{figure}

To address this, we derive an
\textbf{incremental computation form}. As
shown in Figure~\ref{fig:offline-online-detail}, the core idea of
this method is to immediately update the current-level output 
upon completion of preceding level's computation.

\subsubsection{Derivation of Incremental Computation}

Consider the reduction process for the $j^k$-th segment at level $k$, $k >
1$. Suppose the first $L - 1$ inputs have been processed ($1 \le L
\leq L_{k-1}/L_k$), and the partial result is denoted by
$\hat{d}^k[L-1]$, given as:
\begin{equation}
  \hat{d}^k[L-1] = \sideset{}{_i}\Rop_{j^{k-1} =
    l_{1st}^k}^{l_{1st}^k + L - 1} \left( \hat{d}^{k-1} \otimes_i
  \mathrm{H}_i(\hat{\bm{D}}^{k-1})^{-1}  \otimes_i
  \mathrm{H}_i(\hat{\bm{D}}^k[L-1]) \right),
  \label{eq:incremental_dk_L-1}
\end{equation}
where $\hat{\bm{D}}^k[L-1]$ represents the set of partial
results from the preceding $i-1$ reductions at the same level and
segment, computed over $L-1$ input elements.

Same as Equation~\eqref{eq:sum_G}, we have:
\begin{equation}
  \sideset{}{_i}\Rop_{j^{k-1} = l_{1st}^k}^{l_{1st}^k + L -
  1} \hat{d}^{k-1} \otimes_i \mathrm{H}_i(\hat{\bm{D}}^{k-1})^{-1}
  = \hat{d}^k[L-1] \otimes_i \mathrm{H}_i(\hat{\bm{D}}^k[L-1])^{-1}.
  \label{eq:incremental_dk_H}
\end{equation}

When processing the $L$-th input, the partial result becomes:
\begin{equation}
  \hat{d}^k[L] = \sideset{}{_i}\Rop_{j^{k-1} =
    l_{1st}^k}^{l_{1st}^k + L - 1} \left( \hat{d}^{k-1} \otimes_i
  \mathrm{H}_i(\hat{\bm{D}}^{k-1})^{-1} \otimes_i
  \mathrm{H}_i(\hat{\bm{D}}^k[L]) \right).
\end{equation}

Splitting $\hat{d}^k[L]$ into the first $L-1$ terms and the $L$-th
term, and using Equation~\eqref{eq:incremental_dk_H}, we obtain:
\begin{equation}
  \begin{split}
  \hat{d}^k[L] = &
    \hat{d}^k[L-1] \otimes_i \mathrm{H}_i(\hat{\bm{D}}^k[L-1])^{-1} \otimes_i \mathrm{H}_i(\hat{\bm{D}}^k[L]) \oplus_i \\&
    \hat{d}^{k-1} \otimes_i \mathrm{H}_i(\hat{\bm{D}}^{k-1})^{-1} \otimes_i \mathrm{H}_i(\hat{\bm{D}}^k[L]).
  \end{split}
  \label{eq:incremental_dk_L}
\end{equation}

In particular, when $L = L_{k-1}/L_k$, we have $l_{1st}^k + L - 1 =
l_{last}^k$, and $\hat{d}^k[L]$ becomes the $\hat{d}^k$.

For the first-level reduction, where inputs are
$\bm{X}[l]$, a similar derivation applies. Suppose the first
$L-1$ elements have been processed; then the incremental update is:
\begin{equation}
  \begin{split}
  \hat{d}^1[L] = & \hat{d}^1[L-1] \otimes_i
    \mathrm{H}_i(\hat{\bm{D}}^1[L-1])^{-1} \otimes_i \mathrm{H}_i(\hat{\bm{D}}^1[L]) \\
    &\oplus_i \mathrm{G}_i(\bm{X}[L]) \otimes_i \mathrm{H}_i(\hat{\bm{D}}^1[L]).
  \end{split}
  \label{eq:incremental_d1_L}
\end{equation}

Under this incremental mechanism, the computation of $\hat{d}^k[L]$
depends only on the previous state $\hat{d}^k[L-1]$ and the current
segment output $\hat{d}^{k-1}$ (or $\bm{X}[L]$). This storage overhead is constant-sized. Thus, the
storage complexity is reduced from $O(L_{k-1})$ to $O(1)$.

\subsection{Case Study: FP8 PerToken Quant + GEMM}
\label{chap:case_study}
We present a concrete application of the proposed methodology using FP8 per-token quantization followed by GEMM, a common cascaded reduction pattern employed to reduce the computational and memory overhead in serving large language models.

For a given token $A$ to be quantized, the computation is expressed as:
\begin{equation}
\begin{split}
&m = d_1 = \max_{l=1}^{L_0} \vert A[l] \vert, \\
&c = d_2 = \sum_{l=1}^{L_0} \frac{\mathrm{MAX} * A[l]}{m}*W[l],
\end{split}  
\end{equation}
where $W[l]$ denotes the $l$-th row of the weight matrix $W$, and $\mathrm{MAX}$ is the maximum value in the FP8 format.

Under the fusion feasibility conditions (see
Section~\ref{chap:FusionConditions}), we define the functions
$\mathrm{G}_i(\cdot)$, $\mathrm{H}_i(\cdot)$, and binary operators $\otimes_i$ for each
reduction as follows:
\begin{equation}
  \begin{split}
    &\mathrm{G}_1(A[l]) = \vert A[l]\vert ,\quad \mathrm{H}_1(\cdot) = 0,\quad \otimes_1 = +, \\
    &\mathrm{G}_2(A[l], W[l]) = A[l] * W[l],\quad \mathrm{H}_2(m) = \frac{\mathrm{MAX}}{m},\quad \otimes_2 = *.
  \end{split}
  \label{eq:quantgemm_gh_def}
\end{equation}

\subsubsection{Fused Expression}
Substituting Equation~\eqref{eq:quantgemm_gh_def} into the general
fusion expression (Equation~\eqref{eq:fused_dk}), we obtain the fused computation
form for the level-$k$ reduction ($k > 1$) at the $j^k$-th segment:
\begin{equation}
\begin{split}
&\hat{m}^k = \max_{j^{k-1} =
l_{\mathrm{1st}}^{k}}^{l_{\mathrm{last}}^{k}} \hat{m}^{k-1},\\
&\hat{c}^k = \sum_{j^{k-1} =
l_{\mathrm{1st}}^{k}}^{l_{\mathrm{last}}^{k}}  \hat{c}^{k-1} * \frac{\hat{m}^{k-1}}{\hat{m}^k}.
\end{split}  
\end{equation}
For $k=1$, according to Equation~\eqref{eq:fused_d1}, we have:
\begin{equation}
\begin{split}
&\hat{m}^1 = \max_{l =
l_{\mathrm{1st}}^{1}}^{l_{\mathrm{last}}^{1}} \vert A[l] \vert, \\
&\hat{c}^1 = \sum_{l =
l_{\mathrm{1st}}^{1}}^{l_{\mathrm{last}}^{1}}  \frac{\mathrm{MAX} * A[l]}{\hat{m}^1}*W[l]. \\
\end{split}  
\end{equation}

\subsubsection{Incremental Computation Form}
Substituting Equation~\eqref{eq:quantgemm_gh_def} into the
Equation~\eqref{eq:incremental_dk_L}, we
obtain the incremental computation form for the level-$k$ reduction ($k > 1$)
on the $j^k$-th input segment:
\begin{equation}
\begin{split}
&\hat{m}^k[L] = \max(\hat{m}^{k}[L-1],\hat{m}^{k-1}), \\
&\hat{c}^k[L] = \hat{c}^k[L-1] * \frac{\hat{m}^{k}[L-1]}{\hat{m}^k[L]} + \hat{c}^{k-1} * \frac{\hat{m}^{k-1}}{\hat{m}^k[L]}.
\end{split}  
\end{equation}
Based on Equation~\eqref{eq:incremental_d1_L}, the incremental form at the first level can be derived as:
\begin{equation}
\begin{split}
&\hat{m}^1[L] = \max(\hat{m}^{1}[L-1],\vert a[L] \vert), \\
&\hat{c}^1[L] = \hat{c}^1[L-1] * \frac{\hat{m}^{1}[L-1]}{\hat{m}^1[L]} + \frac{\mathrm{MAX} * a[L]}{\hat{m}^{1}[L]}*w[L].
\end{split}
\end{equation}
More cases, including FlashAttention, are provided in the Appendix~\ref{Appendix:Fused_Cases} for reference.

\section{RedFuser}
\label{chap:RedFuser}

Building upon the theoretical methodology, this chapter presents RedFuser, an optimization framework designed to fuse cascaded
reductions. To minimize development effort and avoid redundant
infrastructure work, we build RedFuser on top of the
TVM\cite{TVM2018} compiler stack, leveraging its mature frontend and
runtime ecosystem. Specifically, we use TVM's frontend to lower
models from mainstream deep learning frameworks (e.g.,
PyTorch\cite{PyTorch2024}) into a unified Relax\cite{Relax2025}
computational graph, where we identify and partition the graph into
cascaded reduction subgraphs and other non-reduction components. The
cascaded reduction subgraphs are then processed by RedFuser.
We first traverse their abstract syntax trees (ASTs) to convert each
reduction operation into a formal mathematical expression. Next,
using our automatic fusion algorithm, we extract the key functional
components $\mathrm{G}_i(\cdot)$ and $\mathrm{H}_i(\cdot)$. 
Finally, guided by our fused reduction
programming model, we generate high-performance, memory-efficient
kernels tailored for modern GPU-based AI accelerators. 

\subsection{Mathematical Representation of Cascaded Reductions}

Given a Relax\cite{Relax2025} subgraph containing a cascaded reduction pattern, we lower it into TensorIR (TIR) using TVM's standard lowering pipeline. At the TIR level, we first apply a series of preprocessing transformations, including function inlining and loop reordering to normalize and optimize the computational structure. We then traverse the TIR's abstract syntax tree (AST) via a dedicated visitor to extract a formal mathematical expression that precisely characterizes the reduction chain. This expression captures both the computational semantics and data dependencies across consecutive reductions, serving as the input to our automatic fusion algorithm.

\subsection{Automatic Fusion Algorithm}
\label{chap:automatic_fusion_algorithm}
Building upon theoretical methodology in Chapter.~\ref{chap:Methodology}, this subsection proposes an
engineering-practical \textbf{Automatic Cascaded Reductions Fusion
(ACRF)} algorithm that automatically identifies fusion-capable
expressions and generates fused expression.

\subsubsection{Domain-Specific Decomposition Feasibility}

The key to fusion lies in decomposing the reduction function $\mathrm{F}_i(\mathbf{x}[l],
\mathbf{d}_i)$ into the form that satisfies the conditions specified in Chapter~\ref{chap:FusionConditions}.
In general
function spaces, this decomposition has an enormous search space and
no closed-form solution. However, for machine learning workloads, we
can introduce the following domain-specific assumptions to
significantly narrow the search space:

\begin{itemize}[leftmargin=*, labelindent=0pt]
  \item The set $\mathbb{S}$ is the set of real numbers $\mathbb{R}$;
  \item The underlying operators $\oplus_i$ for common reduction
    operations $\sideset{}{_i}\Rop$ are limited and primarily include:
    \begin{itemize}
      \item Summation ($\sum$, $\oplus_i = +$)
      \item Product ($\prod$, $\oplus_i = *$)
      \item Extrema ($\max$, $\min$)
    \end{itemize}
\end{itemize}

Table~\ref{tab:reduction_ops} lists common reduction operations in
machine learning along with their corresponding $\oplus_i$ and
compatible $\otimes_i$.

\begin{table}[htbp]
  \centering
  \caption{Common reduction operations in machine learning and their corresponding binary operators}
  \label{tab:reduction_ops}
  \begin{tabular}{lll}
    \toprule
    Reduction Operation $\sideset{}{_i}\Rop$ & $\oplus_i$ &
    $\otimes_i$ \\
    \midrule
    Max, ArgMax, TopK, etc. & $\max$ & $+$ \\
    Min, ArgMin, etc. & $\min$ & $+$ \\
    Sum, Inner Product, Matrix Multiply, etc. & $+$ &
    $\ast$ \\
    Prod\footnotemark[1] & $+$ & $\ast$ \\
    \bottomrule
  \end{tabular}
\end{table}
\footnotetext[1]{$\prod \mathrm{F}_i $ can be transformed into summation via 
$\prod \mathrm{F}_i = \mathrm{sgn}(\cdot) 2^{\sum \log_2 |\mathrm{F}_i|}$.}

Given a reduction operation $\sideset{}{_i}\Rop$, its
corresponding $\otimes_i$ can be directly determined from the Table~\ref{tab:reduction_ops}.
This simplifies the decomposition problem to: do there exist functions
$\mathrm{G}_i(\cdot)$ and $\mathrm{H}_i(\cdot)$ such that
$\mathrm{F}_i(\mathbf{x}[l], \mathbf{d}_i) =
\mathrm{G}_i(\mathbf{x}[l]) \otimes_i \mathrm{H}_i(\mathbf{d}_i)$?

\subsubsection{Automatic Function Decomposability}

To determine the feasibility of such a decomposition, we introduce
the \textbf{Fixed-Point Analysis Method}. Select a constant input
point $(\mathbf{x}[l]_0, \mathbf{d}_{i0})$, requiring that
$\mathrm{F}_i(\mathbf{x}[l]_0, \mathbf{d}_{i0})$ has an inverse under
$(\mathbb{R}, \otimes_i)$ (e.g., non-zero when $\otimes_i = *$).

If the following identity holds:
\begin{equation}
  \mathrm{F}_i(\mathbf{x}[l], \mathbf{d}_i) \otimes_i
  \mathrm{F}_i(\mathbf{x}[l]_0, \mathbf{d}_{i0}) =
  \mathrm{F}_i(\mathbf{x}[l], \mathbf{d}_{i0}) \otimes_i
  \mathrm{F}_i(\mathbf{x}[l]_0, \mathbf{d}_i)
  \label{eq:decomposability}
\end{equation}
then $\mathrm{F}_i$ is decomposable, with:
\begin{align}
  \mathrm{G}_i(\mathbf{x}[l]) &= \mathrm{F}_i(\mathbf{x}[l], \mathbf{d}_{i0}) \\
  \mathrm{H}_i(\mathbf{d}_i) &= \mathrm{F}_i(\mathbf{x}[l]_0,
  \mathbf{d}_i) \otimes_i \mathrm{F}_i(\mathbf{x}[l]_0, \mathbf{d}_{i0})^{-1}
\end{align}

\textbf{Proof}: Rearranging Equation~\eqref{eq:decomposability} yields:
\begin{equation}
  \mathrm{F}_i(\mathbf{x}[l], \mathbf{d}_i) =
  \mathrm{F}_i(\mathbf{x}[l], \mathbf{d}_{i0}) \otimes_i \left(
    \mathrm{F}_i(\mathbf{x}[l]_0, \mathbf{d}_i) \otimes_i
  \mathrm{F}_i(\mathbf{x}[l]_0, \mathbf{d}_{i0})^{-1} \right)
\end{equation}
The right-hand side consists of two terms: the first depends only on
$\mathbf{x}[l]$ (defining $\mathrm{G}_i$), and the second only on
$\mathbf{d}_i$ (defining $\mathrm{H}_i$). Hence, the decomposition is valid.

\subsubsection{Automatic Fusion Algorithm Workflow}

As shown in Algorithm~\ref{alg:acrf}, the ACRF algorithm proceeds as follows:

\begin{algorithm}[htbp]
  \caption{Automatic Cascaded Reductions Fusion (ACRF)}
  \label{alg:acrf}
  \begin{algorithmic}[1]
    \REQUIRE Function $\mathrm{F}_i(\mathbf{x}[l],
    \mathbf{d}_i)$, reduction operator $\sideset{}{_i}\Rop$
    \ENSURE Fused and incremental expressions, or failure flag
    \STATE Determine $\otimes_i$ by looking up
    Table~\ref{tab:reduction_ops} based on $\oplus_i$
    \STATE Select a fixed point $(\mathbf{x}[l]_0, \mathbf{d}_{i0})$
    such that $\mathrm{F}_i(\mathbf{x}[l]_0, \mathbf{d}_{i0}) \ne 0$
    (if $\otimes_i = *$)
    \IF{Equation~\eqref{eq:decomposability} does not hold}
    \RETURN \textsc{NotFusable}
    \ENDIF
    \STATE Compute $\mathrm{G}_i(\mathbf{x}[l]) \gets
    \mathrm{F}_i(\mathbf{x}[l], \mathbf{d}_{i0})$
    \STATE Compute $\mathrm{H}_i(\mathbf{d}_i) \gets
    \mathrm{F}_i(\mathbf{x}[l]_0, \mathbf{d}_i) \otimes_i
    \mathrm{F}_i(\mathbf{x}[l]_0, \mathbf{d}_{i0})^{-1}$
    \STATE Instantiate Equation~\eqref{eq:fused_d1}, \eqref{eq:fused_dk}, 
    \eqref{eq:incremental_d1_L} and \eqref{eq:incremental_dk_L} using computed $\mathrm{G}_i(\mathbf{x}[l])$ and $\mathrm{H}_i(\mathbf{d}_i)$
    \RETURN Instantiated Equation~\eqref{eq:fused_d1}, \eqref{eq:fused_dk}, 
    \eqref{eq:incremental_d1_L} and \eqref{eq:incremental_dk_L}
  \end{algorithmic}
\end{algorithm}

The algorithm can be implemented using symbolic computation tools
(e.g., SymPy\cite{SymPy2017}).

\subsection{Programming Model on GPU}
\label{chap:ProgrammingModel}

The classic programming model for reduction
on GPUs follows a four-level hierarchy:
\textbf{intra-thread}, \textbf{intra-warp}, \textbf{intra-block}, and
\textbf{inter-block}. In the context of hierarchical reduction tree,
each level's output length corresponds to a specific granularity of
execution resources: $L_1$ is the number of threads, $L_2$ the number
of warps, $L_3$ the number of CTAs (Cooperative Thread Arrays), and
$L_4 = 1$. We adopt this established programming model to implement 
fused reduction operations on GPU.

Notably, the non-incremental mode, constrained by on-chip resources,
struggles to support reduction of long
sequences within a single CTA; whereas the incremental mode, 
through streaming updates,
allows the system to partition inputs into segments of arbitrary length
within limited storage, thereby enabling control over
parallelism---another critical factor affecting operator performance.
Leveraging the key advantage of incremental computation, we propose two implementation
strategies that adapt to varying hardware resource constraints and
parallelization requirements.

  

\begin{itemize}[leftmargin=*, labelindent=0pt]
  \item \textbf{Single-Segment Strategy}:
    The Single-Segment strategy adopts the incremental computation mode, 
    so the entire reduction can be completed within a single CTA even when the input length exceeds the capacity of a single CTA. 
    In this way, it no longer suffers from hardware resource limitations and avoids the extra overhead of inter-block communication and synchronization.

  \item \textbf{Multi-Segment Strategy}:
    The Multi-Segment strategy builds on the Single-Segment strategy by partitioning the input into multiple segments 
    that are processed in parallel by different CTAs, and then using the Equation~\eqref{eq:fused_dk} to merge their partial results into the final output. 
    This strategy improves hardware utilization under varying workloads, particularly in low-concurrency scenarios.
\end{itemize}

While both strategies offer distinct advantages, their performance is heavily dependent on workload characteristics and hardware constraints. 
To achieve optimal performance across diverse scenarios, RedFuser simultaneously generates TIR representations 
for both strategies from the fused computation expressions, enabling subsequent code generation.

\subsection{Code Generation}
\label{chap:code_generation}

After converting fused expressions into TIR, RedFuser performs \textbf{Tensorization}, a transformation from scalar-level to tile-level IR that bridges high-level computation with efficient hardware execution. 
This process consists of three stages:

\begin{itemize}[leftmargin=*, labelindent=0pt]
    \item \textbf{Blockization}: Restructures scalar loop nests into independently schedulable block-level tiles.
    
    \item \textbf{Block-level Buffer Management}: For each tile, allocates and optimizes on-chip buffers by 
        (i) inserting explicit Load/Store statements for global I/O, 
        (ii) inferring buffer scopes (e.g., register or shared memory), and 
        (iii) compacting buffer size to the tile's minimal footprint.
    \item \textbf{Conversion to TileOp}: Maps tiles to standardized TileOps, enabling TileLang~\cite{TileLang2025} code generation (see Appendix~\ref{sec:tileop_specification}).
\end{itemize}

Following Tensorization, RedFuser applies \textbf{Parallelization} to the tile-level IR by binding each block-level tile to a distinct block index (e.g., \texttt{blockIdx.x}), producing a TileLang program. 
RedFuser leverages TileLang to apply three key optimizations:

\begin{itemize}[leftmargin=*, labelindent=0pt]
    \item \textbf{Thread-level Mapping}: Partitions block tiles across threads with explicit assignment of per-thread work and hardware resources.
    \item \textbf{Software Optimization}: Employs various techniques to hide latency. Key methods include analyzing TileOp dependencies to construct software pipelines, and splitting memory and compute tasks across warps.
    \item \textbf{Hardware-aware Implementations}: Selects architecture-optimized code paths. 
        For example, \texttt{copy} uses \texttt{cp.async} (Ampere) or TMA (Hopper) with automatic vectorization and bank conflict avoidance, while \texttt{gemm} maps to MMA (Ampere) or WGMMA (Hopper) for peak throughput.
\end{itemize}

With TileLang, RedFuser automatically generates efficient GPU kernels. 
To ensure portable performance across diverse platforms and workload characteristics, 
the framework incorporates \textbf{Auto-Tuning} mechanism that optimizes key parameters, 
including block tile size, threads per block, software pipeline depth,
and (in the Multi-Segment strategy) the number of segments. 
We construct an empirical search space for these parameters and 
perform runtime configuration selection, ensuring that the generated kernels are highly optimized.

Implementation details are provided in Appendix~\ref{sec:implementation_details}, using \textbf{FlashAttention} and \textbf{FlashDecoding} as illustrative examples.

\section{Performance Evaluation}

This chapter evaluates the performance of RedFuser and compares it against state-of-the-art deep learning compilers.

\subsection{Experimental Setup}

\textbf{Platform.}
We evaluate RedFuser on two modern GPU architectures: a 24GB NVIDIA A10 GPU and an 80GB NVIDIA H800 GPU. All experiments are conducted under Ubuntu 22.04 with CUDA 12.8.

\noindent
\textbf{Workloads.}
We evaluate RedFuser on four representative subgraphs. The selected workloads are: 
\begin{itemize}[leftmargin=*, labelindent=0pt]
    \item \textbf{Attention} module.
    We consider two variants, Multi-Head Attention (\textbf{MHA}) from BERT~\cite{BERT2019}, ViT~\cite{ViT2020} and LLaMA-65B~\cite{LLaMA2023}, and Multi-Latent Attention (\textbf{MLA}) from \\ DeepSeek-R1~\cite{DeepSeekR1-2025}.
    \item \textbf{MoE routing} module.
    The routing function consists of a GEMM for calculating scores for each expert, followed by a softmax $+$ topk to select experts. We use different parameter configurations in different models (Switch Transformer~\cite{Switch2022}, ERNIE~\cite{ERNIE2025}, DeepSeek-V2-Lite~\cite{DeepSeekV2-2024} and \\ Qwen3~\cite{Qwen3-2025}) for the experiment.
    \item \textbf{FP8 PerToken Quant $+$ GEMM} module.
    This module first computes dynamic scaling factors via reduction operations, then applies quantization and performs FP8 GEMM calculation. We select configurations from existing FP8 quantized MoE models (ERNIE~\cite{ERNIE2025}, DeepSeek-R1~\cite{DeepSeekR1-2025} and Qwen3~\cite{Qwen3-2025}) for testing.
\end{itemize}

The detailed configurations of all modules are shown in Appendix~\ref{Appendix:Workload-Configurations}.

We also conduct experiments on non-ML workloads, such as variance computation. Details on the experimental setup and results are provided in Appendix~\ref{Appendix:non-ML-Workloads}.

\noindent
\textbf{Baselines.}
We compare RedFuser against the following strong baselines.
\begin{itemize}[leftmargin=*, labelindent=0pt]
    \item \textbf{PyTorch (v2.7) Eager}.
    The native PyTorch implementation serves as the primary baseline.
    \item \textbf{PyTorch Dynamo}.
    We use \texttt{torch.compile} with the default backend (Inductor) to enable graph capture, operator fusion and code generation. The Inductor backend will generate Triton~\cite{Triton2019} code on GPUs.
    \item \textbf{TVM (v0.21)}.
    We compile all workloads using TVM's Relax~\cite{Relax2025} frontend and perform the default optimization pipeline. In particular, we do not enable TVM's CUTLASS~\cite{CUTLASS2023} or FlashInfer~\cite{Flashinfer2025} backends in order to compare TVM's own operator fusion capabilities.
    \item \textbf{Hand-Optimized Libraries}.
    We evaluate FlashAttention2~\cite{FlashAttention2-2023} and FlashMLA~\cite{FlashMLA2025}, which provide highly optimized CUDA kernels for MHA and MLA patterns.
\end{itemize}
 
\subsection{Subgraph Performance}
\label{chap:subgraph_perf}
\subsubsection{MHA and MLA}

Figure~\ref{fig:exp_mha} and ~\ref{fig:exp_mla} show the normalized performance of MHA and MLA workloads, with all measurements reported relative to PyTorch Eager.

For MHA, the fused kernel generated by RedFuser achieves performance on par with hand-optimized implementations, averaging 1.09$\times$ that of FlashAttention2, while outperforming it in several configurations (H1-H5). On the LLaMA-65B configuration, RedFuser delivers 2.8$\times$ and 2.6$\times$ speedup over PyTorch Dynamo and TVM.
For MLA, our generated kernel achieves 102\% of the performance of FlashMLA, while significantly outperforming PyTorch Dynamo and TVM by 2.4$\times$ and 8.7$\times$.
The performance gains stem from our fine-grained fusion of the full attention reduction chain.

\subsubsection{MoE routing and Quant GEMM}

\begin{figure*}[t]
    \centering
    \begin{subfigure}[t]{0.49\textwidth}
        \includegraphics[width=\linewidth]{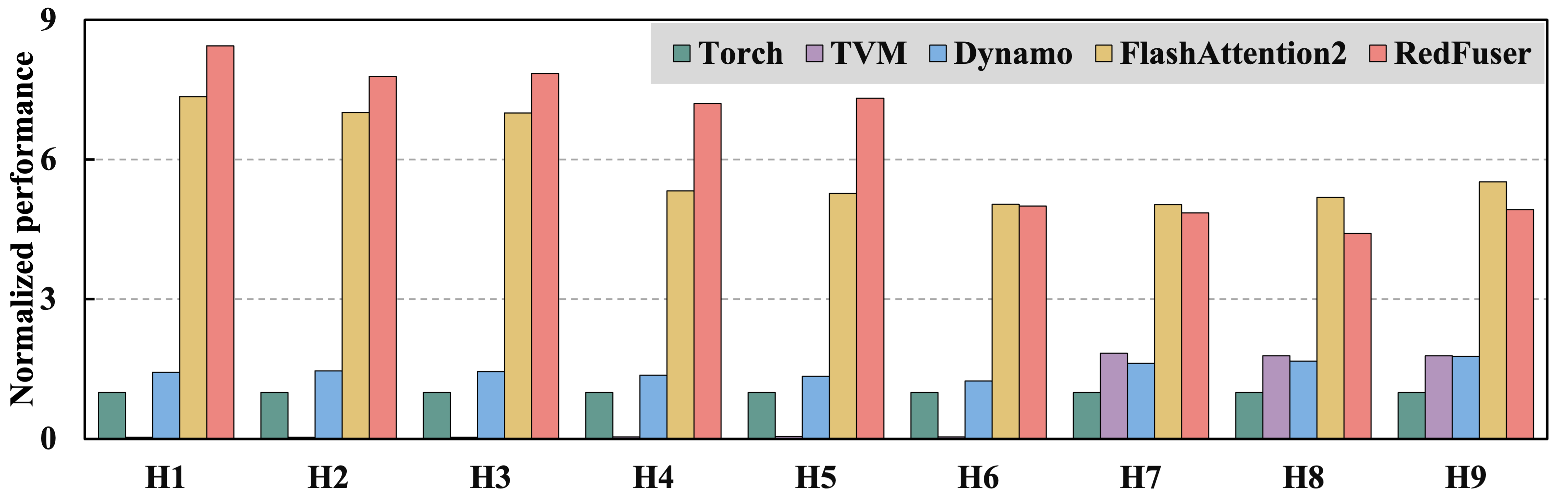}
        \caption{MHA on A10}
        \label{fig:exp_mha}
    \end{subfigure}
    \hfill
    \begin{subfigure}[t]{0.49\textwidth}
        \includegraphics[width=\linewidth]{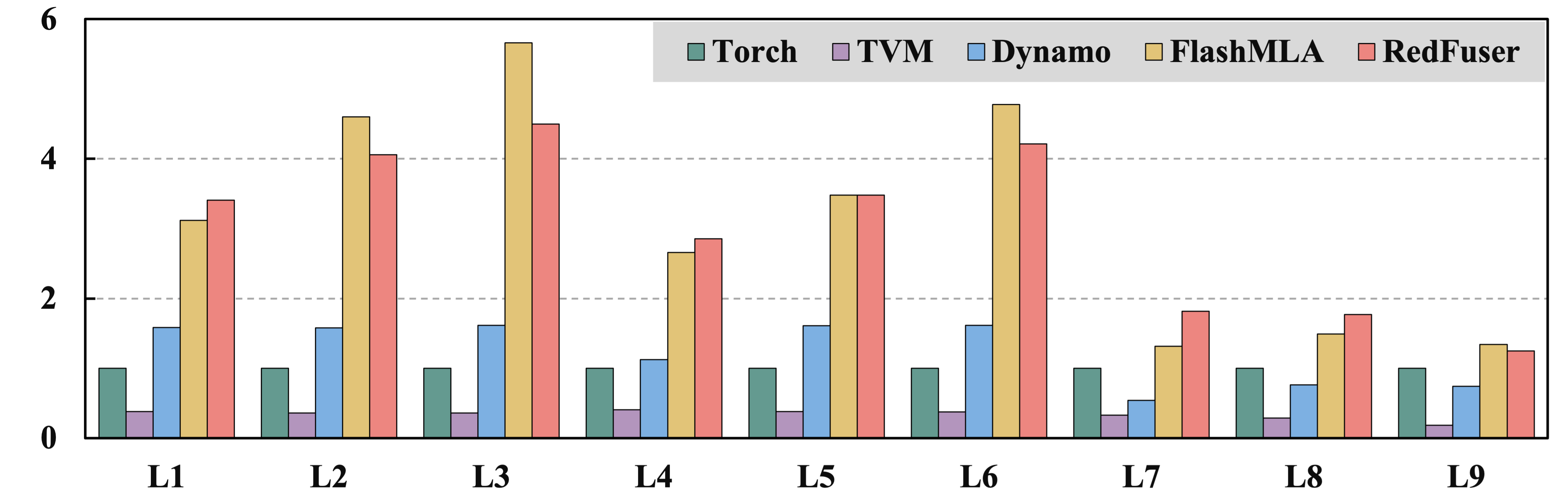}
        \caption{MLA on H800}
        \label{fig:exp_mla}
    \end{subfigure}

    \begin{subfigure}[t]{0.49\textwidth}
        \includegraphics[width=\linewidth]{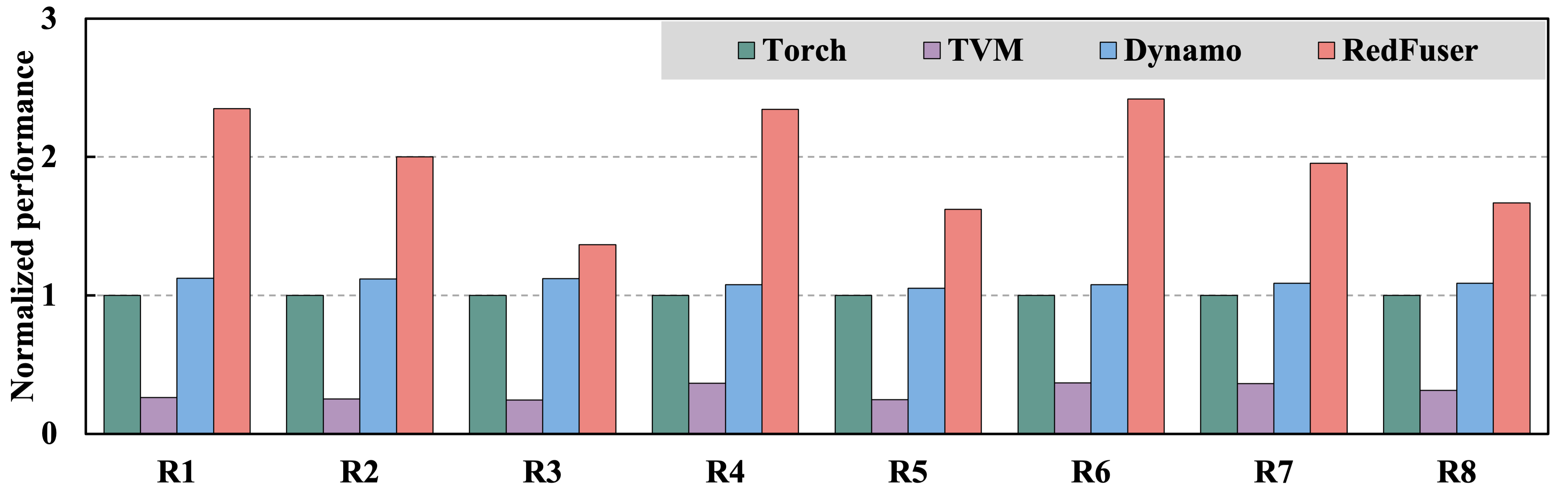}
        \caption{MoE routing on A10}
        \label{fig:exp_moe_routing}
    \end{subfigure}
    \hfill
    \begin{subfigure}[t]{0.49\textwidth}
        \includegraphics[width=\linewidth]{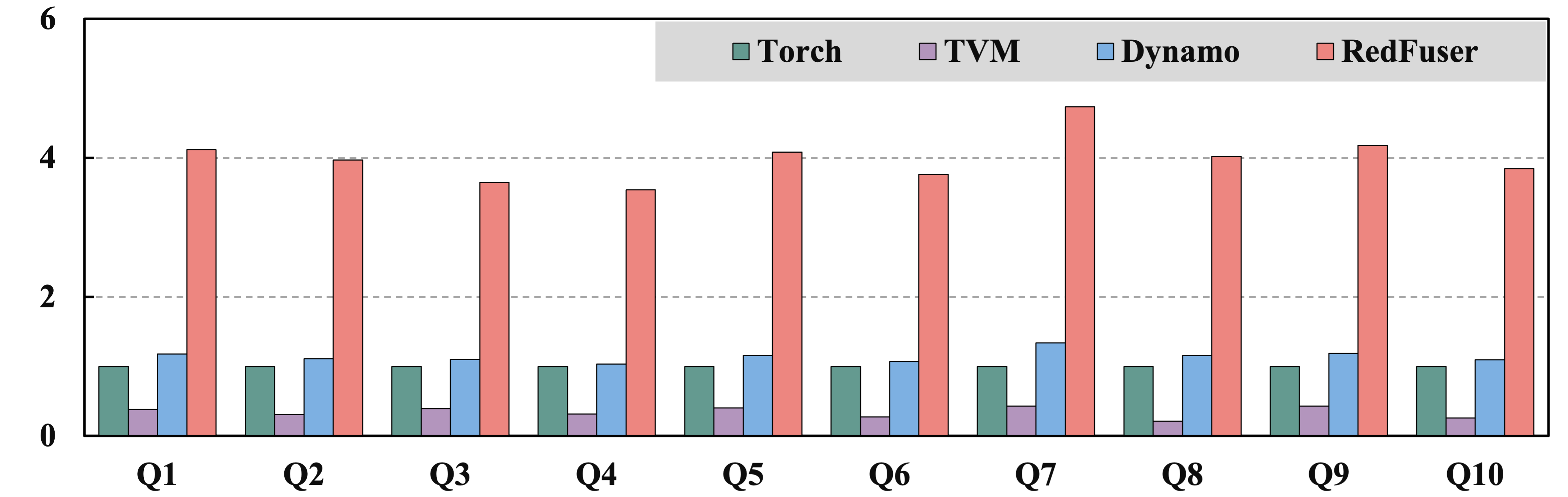}
        \caption{FP8 PerToken Quant $+$ GEMM on H800}
        \label{fig:exp_quant_gemm}
    \end{subfigure}
    \caption{The normalized performance of fusing four selected modules on GPUs.}
    \Description{The normalized performance of fusing four selected modules on GPUs.}
    \label{fig:exp_all}
\end{figure*}

Due to lack of widely adopted hand-optimized implementations, we compare RedFuser only against general-purpose compilers.
Figure~\ref{fig:exp_moe_routing} and ~\ref{fig:exp_quant_gemm} present the performance of MoE routing and Quant$+$GEMM, normalized to PyTorch Eager. 

RedFuser achieves significant speedups across both workloads. In MoE routing, it delivers 1.7$\times$ and 6.6$\times$ speedup over PyTorch Dynamo and TVM, respectively. 
For Quant $+$ GEMM, the improvements are even more pronounced, reaching 3.4$\times$ over PyTorch Dynamo and 12.1$\times$ over TVM.

These experiments confirm that RedFuser exhibits strong performance across various representative workloads, rivaling hand-optimized implementations.
We evaluate RedFuser on multiple hardware platforms to assess its portability. Detailed results are provided in Appendix~\ref{Appendix:Additional_Evaluations}.

\subsection{Latency of Different-Level Fusion}
\label{sec:evaluation_fusion_levels}

\begin{figure*}[t]
  \centering
  \begin{subfigure}[t]{0.49\textwidth}
      \includegraphics[width=\linewidth]{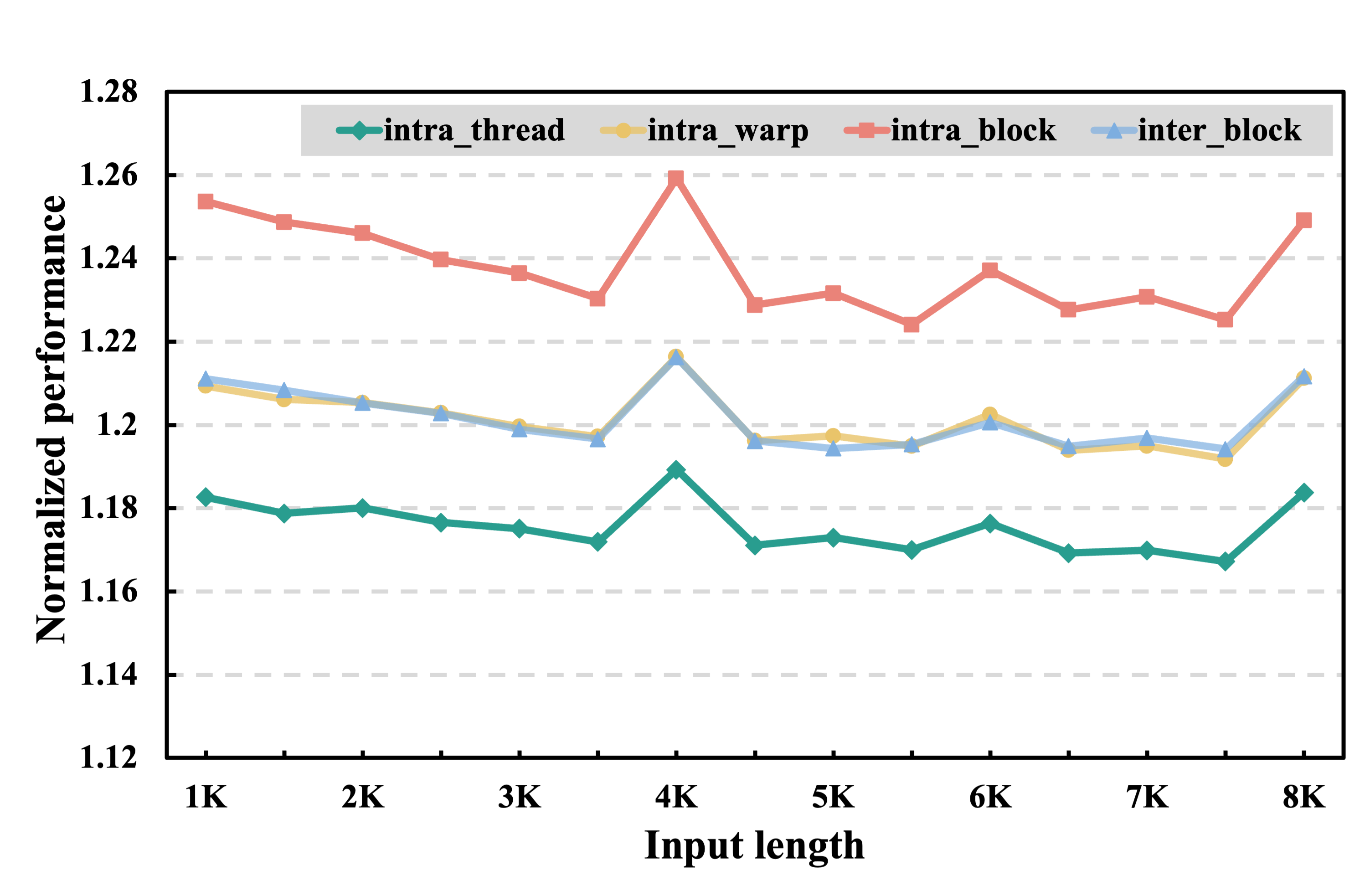}
      \caption{Normalized performance comparison of kernels fused at different levels.
      Intra-block fusion achieves the best performance.}
      \label{fig:fusion_different_level}
  \end{subfigure}
  \hfill
  \begin{subfigure}[t]{0.49\textwidth}
      \includegraphics[width=\linewidth]{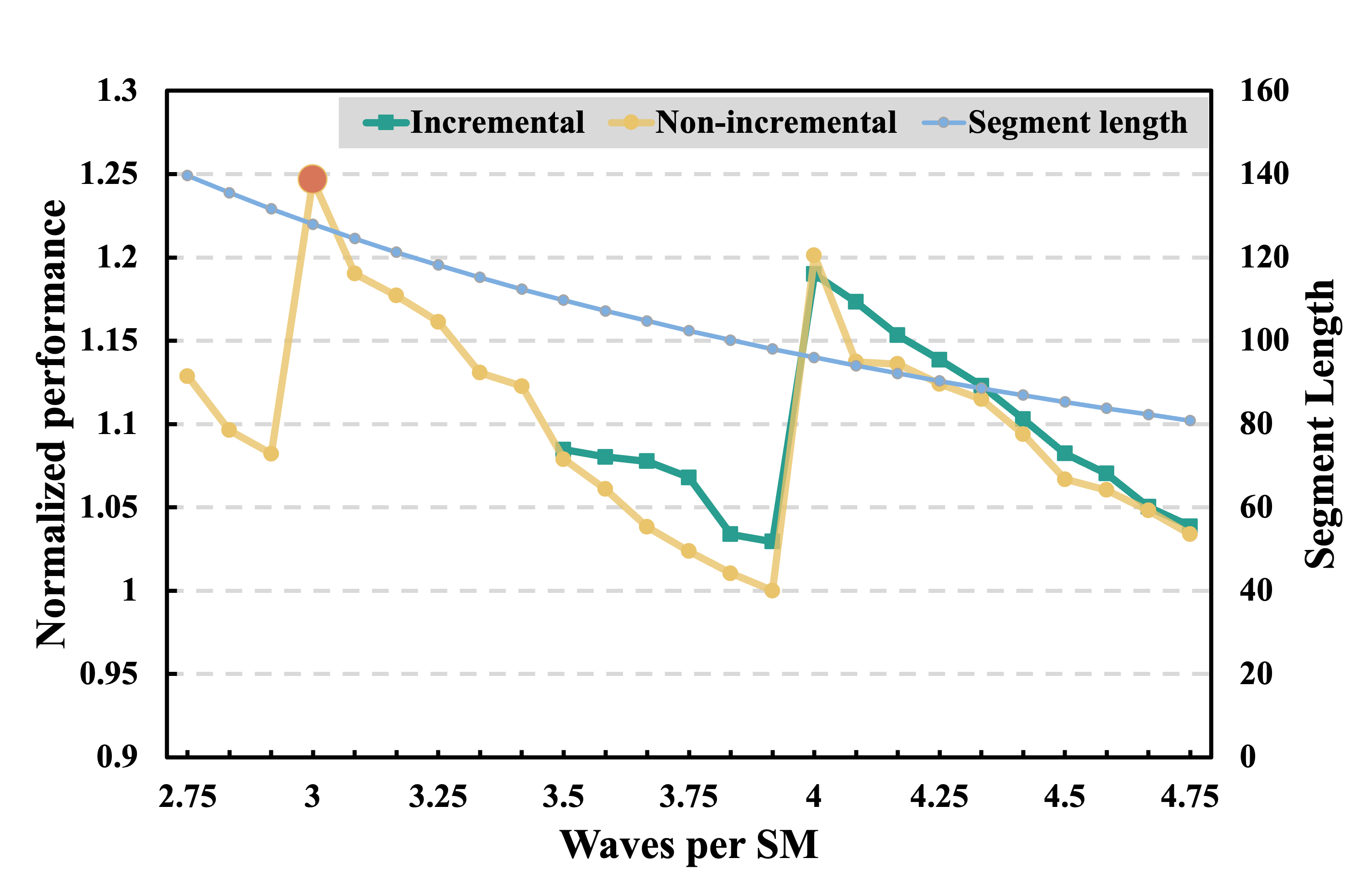}
      \caption{Normalized performance of incremental and non-incremental
          computation across different parallelism levels.}
      \label{fig:incremental_latency}
  \end{subfigure}
  \caption{Performance comparison results. (a)~fusion at different levels. 
  (b)~incremental and non-incremental computation.}
  \Description{Performance comparison results at different levels.}
\end{figure*}

We evaluate four fusion strategies for the safe softmax
operator -- \textit{intra-thread}, \textit{intra-warp},
\textit{intra-block}, and \textit{inter-block} -- over input sizes from
1K to 8K to assess the impact of fusion at different levels on 
the performance of the fused operator. 
All latency measurements are normalized with
respect to the unfused kernels. 
Figure~\ref{fig:fusion_different_level} reports the
normalized performance of different fusion strategies.

The results show that all fusion strategies reduce latency compared
to the unfused baseline.
And the latency ordering is: intra-block < inter-block
$\approx$ intra-warp < intra-thread.
The performance differences are analyzed as follows:

\begin{figure}[ht]
  \centering
  \begin{subfigure}[b]{\linewidth}
    \centering
    \includegraphics[width=0.95\linewidth]{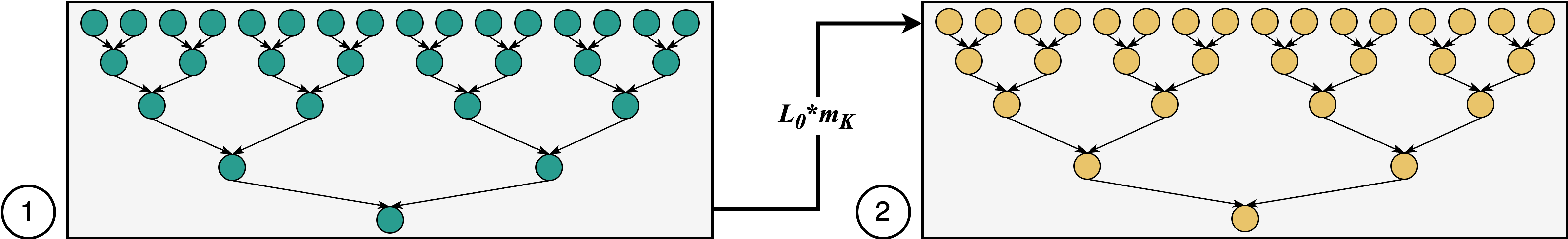}
    \caption{Unfused cascaded reduction trees}
    \label{fig:fusion_level1}
  \end{subfigure}
  \begin{subfigure}[b]{\linewidth}
    \centering
    \includegraphics[width=0.95\linewidth]{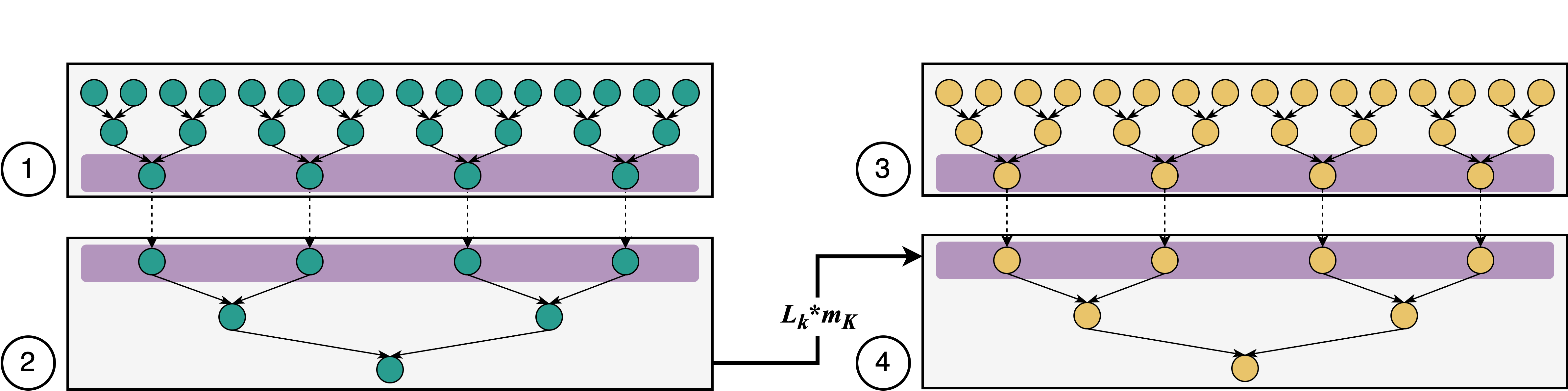}
    \caption{Fusion at level-$k$}
    \label{fig:fusion_levelk}
  \end{subfigure}
  \caption{Illustration of the impact of fusion at different levels on memory access patterns. 
  (a)~No fusion: $d^{K}$ needs to be loaded $L_0$ times to compute $\mathrm{F}_i(\cdot)$. 
  (b)~Fusion at level~$k$: reduces accesses to the final result $d^{K}$ to $L_k$ times. 
  }
    \Description{Illustration of fusion at different levels.}
  \label{fig:fusion_levels}
\end{figure}
As shown in Figure~\ref{fig:fusion_levels}, fusing cascaded
reductions eliminates redundant memory accesses, yielding performance
gains across all levels. However, as Equation~\eqref{eq:fused_dk}
shows, fusion at level $k$ requires correcting intermediate outputs
$\hat{d}^k$ of length $L_k$, introducing computation overhead linear
in $L_k$. Fortunately, in the fused reduction tree at level $k$ ($k <
K$), subtrees \ding{173} and \ding{174} are data-independent,
enabling instruction-level parallelism and overlap between memory
access and computation. As $k$ increases, the deeper subtree
\ding{174} provides longer computation paths, enhancing latency hiding.

\begin{itemize}[leftmargin=*, labelindent=0pt]
  \item \textbf{Intra-thread Fusion:}
    Requires $L_1$ corrections -- the highest overhead -- and the
    sub-reduction tree \ding{174} spans only \textit{level 1},
    offering insufficient computation to hide memory latency. Thus,
    it delivers the smallest performance gain.

  \item \textbf{Intra-warp Fusion:}
    Spans \textit{level-1} and \textit{level-2}, increasing
    computational intensity and better hiding global memory latency.
    With lower correction cost than \textit{intra-thread}, it
    achieves better performance.

  \item \textbf{Intra-block Fusion:}
    Introduces $L_3$ additional corrections but enables the best
    overlap between accessing $d^K$ and computing subtree \ding{174}.
    Experimental results show that the latency-hiding benefit outweighs 
    the overhead, resulting in the lowest overall latency.

  \item \textbf{Inter-block Fusion:}
    No correction is needed since $\hat{d}^K = d^K$. However,
    strict data dependencies prevent memory-computation overlap,
    limiting performance despite zero correction cost.
\end{itemize}





\subsection{Latency of Incremental vs.\ Non-Incremental Mode}
\label{sec:incremental_vs_nonincremental}

This section evaluates the performance of incremental computation
versus non-incremental computation under varying degrees of
parallelism. We use the attention computation pattern from the
BERT-base model as our representative workload. With a fixed problem
size, we systematically control the level of GPU parallelism by
adjusting the sequence length of kv processed per CTA, 
thereby modulating the number of CTAs.

In our experiments, we normalize all latency measurements with
respect to the maximum observed execution latency across all
configurations, which serves as the performance baseline. The
normalized performance results are shown in
Figure~\ref{fig:incremental_latency}. Based on the experimental
findings, we draw the following conclusions:

\begin{itemize}[leftmargin=*, labelindent=0pt]
  \item \textbf{On-chip cache capacity limits the applicability of
    non-incremental computation}:
    Non-incremental computation requires caching the complete
    previous reduction results in on-chip shared memory or
    registers, making its feasibility constrained by the SM's on-chip
    storage resources. In our experiments, the non-incremental
    approach only succeeds when the input sequence length is $\leq
    112$, corresponding to \texttt{Waves per SM} $\geq 3.5$. In
    contrast, incremental computation avoids the need to retain full
    results on-chip, thereby eliminating strict
    constraints on segment length and enabling more flexible
    parallelism configurations.

  \item \textbf{Non-incremental computation exhibits superior
    performance at matched parallelism levels}:
    As shown in Equation~\eqref{eq:incremental_dk_L},
    incremental computation introduces
    additional computational overhead due to the need to correct
    the result in each iteration. Consequently, under
    equivalent parallelism and hardware resource allocations,
    non-incremental computation outperforms its incremental
    counterpart, achieving lower latency and higher overall performance.

  \item \textbf{Parallelism level has a nonlinear impact on performance}:
    Operator performance exhibits local optima when \texttt{Waves per
    SM} takes on integer values, indicating peak SM resource
    utilization at these configurations. Notably, performance peaks
    at \texttt{Waves per SM} = 3, achieving up to a 1.25$\times$
    speedup over the baseline configuration. However, this
    optimal configuration corresponds to a segment length that
    exceeds the on-chip memory capacity required by the
    non-incremental approach, and thus can only be realized using
    incremental computation. This highlights the necessity of
    incremental computation in specific high-performance configurations.
\end{itemize}

In summary, although incremental computation incurs additional
computational overhead due to correction operations, it significantly
reduces on-chip resource requirements and greatly expands the
feasible range of parallelism configurations. This characteristic
enlarges the optimization search space, enabling
high-performance configurations even in low-parallelism and
long-sequence scenarios—such as the decode phase of LLM
inference, and thereby providing greater flexibility and potential for
performance tuning in practical deployments.


\section{Related Work}

\textbf{Parallel Reduction.}
The design of efficient parallel reduction algorithms on GPUs has been extensively studied.
Harris\cite{Harris2007} proposed a series of optimized algorithms for parallel reduction as early as 2007, based on a segmented reduction strategy that leverages shared memory and sequential addressing to maximize memory bandwidth utilization.
Recent work\cite{TC-Reduction2020} has explored the use of GPU Tensor Cores for accelerating parallel reduction.
RedFuser builds upon well-established parallel reduction algorithms and extends them to cascaded reductions, under specific structural constraints and execution conditions.

\noindent
\textbf{Operator Fusion in DL Compilers.}
Many popular deep learning compilers have invested significant effort in supporting operator fusion, and some have already explored fusing reduction operators with other computational operators.
Modern compilers such as TVM\cite{TVM2018}, XLA\cite{XLA2020}, and PyTorch Dynamo\cite{PyTorch2024} support fusion of elementwise operators with a single reduction, based on either schedule templates (TVM), rule-based heuristics (XLA) or pattern matching (Dynamo).
DNNFusion\cite{DNNFusion2021} and AStitch\cite{AStitch2022} classify operators into different types and apply fusion algorithms among them.
Bolt\cite{Bolt2022} enables operator fusion by leveraging CUTLASS\cite{CUTLASS2023} templates to fuse sequences of GEMM and convolution operators.
Chimera\cite{Chimera2023} and MCFuser\cite{MCFuser2024} both focus on fusing memory-bound compute-intensive operators, such as chains of batched GEMM and convolution. 
SOUFFLE\cite{SOUFFLE2024} adopts a top-down approach, aggressively fuses reduction operators with adjacent computation intensive operators.
RedFuser formally defines the structural properties of cascade reductions, and presents a novel algorithm that enables cross-reduction expression fusion.

\noindent
\textbf{Hand-optimized Fusion.}
FlashAttention\cite{FlashAttention2022} is a hand-optimized implementation of the attention operator. Its key insight is to apply tiling to decompose the $QK^T$ and $PV$ GEMMs into smaller, manageable blocks, enabling intermediate values, such as attention scores and normalization factors to reside entirely in on-chip memory (e.g., shared memory) and thereby avoiding expensive round trips to global memory. To ensure numerical correctness across tiles, FlashAttention employs an online softmax algorithm: it dynamically maintains running values of the maximum and sum of exponentials over processed tiles, and upon processing each new tile, rescales the accumulated results to incorporate the updated maximum before merging. This incremental update strategy allows FlashAttention to compute the exact softmax normalization in a memory-efficient manner.
FlashDecoding\cite{FlashDecoding2023} extends the tiling paradigm of FlashAttention to optimize long-context inference. Specifically, it partitions the KV cache into multiple chunks along the sequence axis, and applies FlashAttention independently within each chunk to compute local attention outputs and partial normalization statistics. These intermediate results are preserved and later aggregated by an additional reduction kernel, which performs a global max and sum reduction across all chunks to produce the final, numerically correct output.
RedFuser derives an incremental computation form based on the fused reduction expressions. Under the guidance of our programming model, FlashAttention and FlashDecoding emerge as special cases of our framework. RedFuser not only subsumes these hand-optimized designs, but also unifies them within a single, automated compilation framework, enabling a transition from ad-hoc optimization to general-purpose fusion.
\section{Conclusion}
In this paper, we present a general fusion methodology for cascaded
reduction operations, ubiquitous in AI models, which
derives fused expressions and their corresponding
incremental computation forms directly from the mathematical
representations of reductions. Our approach eliminates redundant
memory accesses and enables reductions of arbitrary length at any
level, overcoming cache capacity limitations. Based on this
methodology, we design an operator fusion framework named RedFuser that
automatically analyzes cascaded reductions and generates
high-performance, GPU-optimized fused kernels. Extensive experiments
demonstrate that our generated kernels outperform state-of-the-art AI
compilers and achieve performance on par with hand-written operators.
While our methodology begins to bridge the gap between hand-crafted, highly specialized optimizations and general-purpose compiler automation, several challenges and opportunities remain. First, while fusion can reduce memory traffic, it introduces additional computational overhead and increases pressure on hardware resources such as register usage, making it suboptimal for certain workloads. We plan to develop a systematic cost model to guide fusion decisions. Second, our current approach applies only to operations satisfying specific conditions; extending it to more relaxed constraints is an important direction for future work. Finally, RedFuser has been primarily implemented and evaluated on GPU architectures. We aim to generalize the framework to support a broader range of AI accelerators, further enhancing its applicability in diverse computing environments.

\newpage

\bibliographystyle{ACM-Reference-Format}
\balance
\bibliography{References}

@article{Harris2007,
  title={{Optimizing parallel reduction in CUDA}},
  author={Harris, Mark and others},
  journal={Nvidia developer technology},
  volume={2},
  number={4},
  pages={70},
  year={2007},
  publisher={Nvidia Corporation Santa Clara, CA, USA}
}

@article{TC-Reduction2020,
  title={{GPU tensor cores for fast arithmetic reductions}},
  author={Navarro, Crist{\'o}bal A and Carrasco, Roberto and Barrientos, Ricardo J and Riquelme, Javier A and Vega, Raimundo},
  journal={IEEE Transactions on Parallel and Distributed Systems},
  volume={32},
  number={1},
  pages={72--84},
  year={2020},
  publisher={IEEE}
}

@inproceedings{TVM2018,
  title={{TVM: An automated End-to-End optimizing compiler for deep learning}},
  author={Chen, Tianqi and Moreau, Thierry and Jiang, Ziheng and Zheng, Lianmin and Yan, Eddie and Shen, Haichen and Cowan, Meghan and Wang, Leyuan and Hu, Yuwei and Ceze, Luis and others},
  booktitle={13th USENIX Symposium on Operating Systems Design and Implementation (OSDI 18)},
  pages={578--594},
  year={2018}
}

@inproceedings{Relax2025,
  title={{Relax: composable abstractions for end-to-end dynamic machine learning}},
  author={Lai, Ruihang and Shao, Junru and Feng, Siyuan and Lyubomirsky, Steven and Hou, Bohan and Lin, Wuwei and Ye, Zihao and Jin, Hongyi and Jin, Yuchen and Liu, Jiawei and others},
  booktitle={Proceedings of the 30th ACM International Conference on Architectural Support for Programming Languages and Operating Systems, Volume 2},
  pages={998--1013},
  year={2025}
}

@article{TileLang2025,
  title={{TileLang: A Composable Tiled Programming Model for AI Systems}},
  author={Wang, Lei and Cheng, Yu and Shi, Yining and Tang, Zhengju and Mo, Zhiwen and Xie, Wenhao and Ma, Lingxiao and Xia, Yuqing and Xue, Jilong and Yang, Fan and others},
  journal={arXiv preprint arXiv:2504.17577},
  year={2025}
}

@misc{XLA2020,
  title={{XLA : Compiling Machine Learning for Peak Performance}},
  author={Amit Sabne},
  year={2020}
}

@inproceedings{PyTorch2024,
  title={{Pytorch 2: Faster machine learning through dynamic python bytecode transformation and graph compilation}},
  author={Ansel, Jason and Yang, Edward and He, Horace and Gimelshein, Natalia and Jain, Animesh and Voznesensky, Michael and Bao, Bin and Bell, Peter and Berard, David and Burovski, Evgeni and others},
  booktitle={Proceedings of the 29th ACM International Conference on Architectural Support for Programming Languages and Operating Systems, Volume 2},
  pages={929--947},
  year={2024}
}

@inproceedings{DNNFusion2021,
  title={{Dnnfusion: accelerating deep neural networks execution with advanced operator fusion}},
  author={Niu, Wei and Guan, Jiexiong and Wang, Yanzhi and Agrawal, Gagan and Ren, Bin},
  booktitle={Proceedings of the 42nd ACM SIGPLAN International Conference on Programming Language Design and Implementation},
  pages={883--898},
  year={2021}
}

@inproceedings{AStitch2022,
  title={{AStitch: enabling a new multi-dimensional optimization space for memory-intensive ML training and inference on modern SIMT architectures}},
  author={Zheng, Zhen and Yang, Xuanda and Zhao, Pengzhan and Long, Guoping and Zhu, Kai and Zhu, Feiwen and Zhao, Wenyi and Liu, Xiaoyong and Yang, Jun and Zhai, Jidong and others},
  booktitle={Proceedings of the 27th ACM International Conference on Architectural Support for Programming Languages and Operating Systems},
  pages={359--373},
  year={2022}
}

@article{Bolt2022,
  title={{Bolt: Bridging the gap between auto-tuners and hardware-native performance}},
  author={Xing, Jiarong and Wang, Leyuan and Zhang, Shang and Chen, Jack and Chen, Ang and Zhu, Yibo},
  journal={Proceedings of Machine Learning and Systems},
  volume={4},
  pages={204--216},
  year={2022}
}

@software{CUTLASS2023,
author = {Thakkar, Vijay and Ramani, Pradeep and Cecka, Cris and Shivam, Aniket and Lu, Honghao and Yan, Ethan and Kosaian, Jack and Hoemmen, Mark and Wu, Haicheng and Kerr, Andrew and Nicely, Matt and Merrill, Duane and Blasig, Dustyn and Qiao, Fengqi and Majcher, Piotr and Springer, Paul and Hohnerbach, Markus and Wang, Jin and Gupta, Manish},
license = {BSD-3-Clause},
month = jan,
title = {{CUTLASS}},
url = {https://github.com/NVIDIA/cutlass},
version = {3.0.0},
year = {2023}
}

@inproceedings{Chimera2023,
  title={{Chimera: An analytical optimizing framework for effective compute-intensive operators fusion}},
  author={Zheng, Size and Chen, Siyuan and Song, Peidi and Chen, Renze and Li, Xiuhong and Yan, Shengen and Lin, Dahua and Leng, Jingwen and Liang, Yun},
  booktitle={2023 IEEE International Symposium on High-Performance Computer Architecture (HPCA)},
  pages={1113--1126},
  year={2023},
  organization={IEEE}
}

@inproceedings{MCFuser2024,
  title={{MCFuser: High-performance and rapid fusion of memory-bound compute-intensive operators}},
  author={Zhang, Zheng and Yang, Donglin and Zhou, Xiaobo and Cheng, Dazhao},
  booktitle={SC24: International Conference for High Performance Computing, Networking, Storage and Analysis},
  pages={1--15},
  year={2024},
  organization={IEEE}
}

@inproceedings{SOUFFLE2024,
  title={{Optimizing deep learning inference via global analysis and tensor expressions}},
  author={Xia, Chunwei and Zhao, Jiacheng and Sun, Qianqi and Wang, Zheng and Wen, Yuan and Yu, Teng and Feng, Xiaobing and Cui, Huimin},
  booktitle={Proceedings of the 29th ACM International Conference on Architectural Support for Programming Languages and Operating Systems, Volume 1},
  pages={286--301},
  year={2024}
}

@article{FlashAttention2022,
  title={{Flashattention: Fast and memory-efficient exact attention with io-awareness}},
  author={Dao, Tri and Fu, Dan and Ermon, Stefano and Rudra, Atri and R{\'e}, Christopher},
  journal={Advances in neural information processing systems},
  volume={35},
  pages={16344--16359},
  year={2022}
}

@article{FlashAttention2-2023,
  title={{Flashattention-2: Faster attention with better parallelism and work partitioning}},
  author={Dao, Tri},
  journal={arXiv preprint arXiv:2307.08691},
  year={2023}
}

@misc{FlashDecoding2023,
  author={Tri Dao and Daniel Haziza and Francisco Massa and Grigory Sizov},
  url={https://pytorch.org/blog/flash-decoding/},
  title={{Flash-Decoding for long-context inference}},
  year={2023}
}

@misc{gpt4.0,
      title={{GPT-4 Technical Report}}, 
      author={OpenAI and Josh Achiam and Steven Adler and Sandhini Agarwal and Lama Ahmad and Ilge Akkaya and Florencia Leoni Aleman and Diogo Almeida and Janko Altenschmidt and Sam Altman and Shyamal Anadkat and Red Avila and Igor Babuschkin and Suchir Balaji and Valerie Balcom and Paul Baltescu and Haiming Bao and Mohammad Bavarian and Jeff Belgum and Irwan Bello and Jake Berdine and Gabriel Bernadett-Shapiro and Christopher Berner and Lenny Bogdonoff and Oleg Boiko and Madelaine Boyd and Anna-Luisa Brakman and Greg Brockman and Tim Brooks and Miles Brundage and Kevin Button and Trevor Cai and Rosie Campbell and Andrew Cann and Brittany Carey and Chelsea Carlson and Rory Carmichael and Brooke Chan and Che Chang and Fotis Chantzis and Derek Chen and Sully Chen and Ruby Chen and Jason Chen and Mark Chen and Ben Chess and Chester Cho and Casey Chu and Hyung Won Chung and Dave Cummings and Jeremiah Currier and Yunxing Dai and Cory Decareaux and Thomas Degry and Noah Deutsch and Damien Deville and Arka Dhar and David Dohan and Steve Dowling and Sheila Dunning and Adrien Ecoffet and Atty Eleti and Tyna Eloundou and David Farhi and Liam Fedus and Niko Felix and Simón Posada Fishman and Juston Forte and Isabella Fulford and Leo Gao and Elie Georges and Christian Gibson and Vik Goel and Tarun Gogineni and Gabriel Goh and Rapha Gontijo-Lopes and Jonathan Gordon and Morgan Grafstein and Scott Gray and Ryan Greene and Joshua Gross and Shixiang Shane Gu and Yufei Guo and Chris Hallacy and Jesse Han and Jeff Harris and Yuchen He and Mike Heaton and Johannes Heidecke and Chris Hesse and Alan Hickey and Wade Hickey and Peter Hoeschele and Brandon Houghton and Kenny Hsu and Shengli Hu and Xin Hu and Joost Huizinga and Shantanu Jain and Shawn Jain and Joanne Jang and Angela Jiang and Roger Jiang and Haozhun Jin and Denny Jin and Shino Jomoto and Billie Jonn and Heewoo Jun and Tomer Kaftan and Łukasz Kaiser and Ali Kamali and Ingmar Kanitscheider and Nitish Shirish Keskar and Tabarak Khan and Logan Kilpatrick and Jong Wook Kim and Christina Kim and Yongjik Kim and Jan Hendrik Kirchner and Jamie Kiros and Matt Knight and Daniel Kokotajlo and Łukasz Kondraciuk and Andrew Kondrich and Aris Konstantinidis and Kyle Kosic and Gretchen Krueger and Vishal Kuo and Michael Lampe and Ikai Lan and Teddy Lee and Jan Leike and Jade Leung and Daniel Levy and Chak Ming Li and Rachel Lim and Molly Lin and Stephanie Lin and Mateusz Litwin and Theresa Lopez and Ryan Lowe and Patricia Lue and Anna Makanju and Kim Malfacini and Sam Manning and Todor Markov and Yaniv Markovski and Bianca Martin and Katie Mayer and Andrew Mayne and Bob McGrew and Scott Mayer McKinney and Christine McLeavey and Paul McMillan and Jake McNeil and David Medina and Aalok Mehta and Jacob Menick and Luke Metz and Andrey Mishchenko and Pamela Mishkin and Vinnie Monaco and Evan Morikawa and Daniel Mossing and Tong Mu and Mira Murati and Oleg Murk and David Mély and Ashvin Nair and Reiichiro Nakano and Rajeev Nayak and Arvind Neelakantan and Richard Ngo and Hyeonwoo Noh and Long Ouyang and Cullen O'Keefe and Jakub Pachocki and Alex Paino and Joe Palermo and Ashley Pantuliano and Giambattista Parascandolo and Joel Parish and Emy Parparita and Alex Passos and Mikhail Pavlov and Andrew Peng and Adam Perelman and Filipe de Avila Belbute Peres and Michael Petrov and Henrique Ponde de Oliveira Pinto and Michael and Pokorny and Michelle Pokrass and Vitchyr H. Pong and Tolly Powell and Alethea Power and Boris Power and Elizabeth Proehl and Raul Puri and Alec Radford and Jack Rae and Aditya Ramesh and Cameron Raymond and Francis Real and Kendra Rimbach and Carl Ross and Bob Rotsted and Henri Roussez and Nick Ryder and Mario Saltarelli and Ted Sanders and Shibani Santurkar and Girish Sastry and Heather Schmidt and David Schnurr and John Schulman and Daniel Selsam and Kyla Sheppard and Toki Sherbakov and Jessica Shieh and Sarah Shoker and Pranav Shyam and Szymon Sidor and Eric Sigler and Maddie Simens and Jordan Sitkin and Katarina Slama and Ian Sohl and Benjamin Sokolowsky and Yang Song and Natalie Staudacher and Felipe Petroski Such and Natalie Summers and Ilya Sutskever and Jie Tang and Nikolas Tezak and Madeleine B. Thompson and Phil Tillet and Amin Tootoonchian and Elizabeth Tseng and Preston Tuggle and Nick Turley and Jerry Tworek and Juan Felipe Cerón Uribe and Andrea Vallone and Arun Vijayvergiya and Chelsea Voss and Carroll Wainwright and Justin Jay Wang and Alvin Wang and Ben Wang and Jonathan Ward and Jason Wei and CJ Weinmann and Akila Welihinda and Peter Welinder and Jiayi Weng and Lilian Weng and Matt Wiethoff and Dave Willner and Clemens Winter and Samuel Wolrich and Hannah Wong and Lauren Workman and Sherwin Wu and Jeff Wu and Michael Wu and Kai Xiao and Tao Xu and Sarah Yoo and Kevin Yu and Qiming Yuan and Wojciech Zaremba and Rowan Zellers and Chong Zhang and Marvin Zhang and Shengjia Zhao and Tianhao Zheng and Juntang Zhuang and William Zhuk and Barret Zoph},
      year={2024},
      eprint={2303.08774},
      archivePrefix={arXiv},
      primaryClass={cs.CL},
      url={https://arxiv.org/abs/2303.08774}, 
}

@misc{chameleonteam2025,
      title={{Chameleon: Mixed-Modal Early-Fusion Foundation Models}}, 
      author={Chameleon Team},
      year={2025},
      eprint={2405.09818},
      archivePrefix={arXiv},
      primaryClass={cs.CL},
      url={https://arxiv.org/abs/2405.09818}, 
}

@misc{reed2022generalistagent,
      title={{A Generalist Agent}}, 
      author={Scott Reed and Konrad Zolna and Emilio Parisotto and Sergio Gomez Colmenarejo and Alexander Novikov and Gabriel Barth-Maron and Mai Gimenez and Yury Sulsky and Jackie Kay and Jost Tobias Springenberg and Tom Eccles and Jake Bruce and Ali Razavi and Ashley Edwards and Nicolas Heess and Yutian Chen and Raia Hadsell and Oriol Vinyals and Mahyar Bordbar and Nando de Freitas},
      year={2022},
      eprint={2205.06175},
      archivePrefix={arXiv},
      primaryClass={cs.AI},
      url={https://arxiv.org/abs/2205.06175}, 
}

@misc{esser2024,
      title={{Scaling Rectified Flow Transformers for High-Resolution Image Synthesis}}, 
      author={Patrick Esser and Sumith Kulal and Andreas Blattmann and Rahim Entezari and Jonas Müller and Harry Saini and Yam Levi and Dominik Lorenz and Axel Sauer and Frederic Boesel and Dustin Podell and Tim Dockhorn and Zion English and Kyle Lacey and Alex Goodwin and Yannik Marek and Robin Rombach},
      year={2024},
      eprint={2403.03206},
      archivePrefix={arXiv},
      primaryClass={cs.CV},
      url={https://arxiv.org/abs/2403.03206}, 
}

@ARTICLE{SRNPU,
  author={Lee, Juhyoung and Lee, Jinsu and Yoo, Hoi-Jun},
  journal={IEEE Journal on Emerging and Selected Topics in Circuits and Systems}, 
  title={{SRNPU: An Energy-Efficient CNN-Based Super-Resolution Processor With Tile-Based Selective Super-Resolution in Mobile Devices}}, 
  year={2020},
  volume={10},
  number={3},
  pages={320-334},
  doi={10.1109/JETCAS.2020.3014454}}

@inproceedings{kernelfusion,
author = {Qiao, Bo and Reiche, Oliver and Hannig, Frank and Teich, J\"{u}rgen},
title = {Automatic Kernel Fusion for Image Processing DSLs},
year = {2018},
isbn = {9781450357807},
publisher = {Association for Computing Machinery},
address = {New York, NY, USA},
url = {https://doi.org/10.1145/3207719.3207723},
doi = {10.1145/3207719.3207723},
booktitle = {Proceedings of the 21st International Workshop on Software and Compilers for Embedded Systems},
pages = {76–85},
numpages = {10},
location = {Sankt Goar, Germany},
series = {SCOPES '18}
}

@inproceedings{loopfusion,
author = {Qiao, Bo and Reiche, Oliver and Hannig, Frank and Teich, J\"{u}rgen},
title = {From loop fusion to kernel fusion: a domain-specific approach to locality optimization},
year = {2019},
isbn = {9781728114361},
publisher = {IEEE Press},
booktitle = {Proceedings of the 2019 IEEE/ACM International Symposium on Code Generation and Optimization},
pages = {242–253},
numpages = {12},
location = {Washington, DC, USA},
series = {CGO 2019}
}

@inproceedings{dac2024softmax,
author = {Xu, Lei and Mo, Zhiwen and Wang, Qin and Jiang, Jianfei and Jing, Naifeng},
title = {Enabling Multiple Tensor-wise Operator Fusion for Transformer Models on Spatial Accelerators},
year = {2024},
isbn = {9798400706011},
publisher = {Association for Computing Machinery},
address = {New York, NY, USA},
url = {https://doi.org/10.1145/3649329.3657317},
doi = {10.1145/3649329.3657317},
booktitle = {Proceedings of the 61st ACM/IEEE Design Automation Conference},
articleno = {232},
numpages = {6},
location = {San Francisco, CA, USA},
series = {DAC '24}
}

@inproceedings{SC,
author = {Wahib, Mohamed and Maruyama, Naoya},
title = {Scalable kernel fusion for memory-bound GPU applications},
year = {2014},
isbn = {9781479955008},
publisher = {IEEE Press},
url = {https://doi.org/10.1109/SC.2014.21},
doi = {10.1109/SC.2014.21},
booktitle = {Proceedings of the International Conference for High Performance Computing, Networking, Storage and Analysis},
pages = {191–202},
numpages = {12},
location = {New Orleans, Louisana},
series = {SC '14}
}

@misc{FLAT,
      title={{FLAT: An Optimized Dataflow for Mitigating Attention Bottlenecks}}, 
      author={Sheng-Chun Kao and Suvinay Subramanian and Gaurav Agrawal and Amir Yazdanbakhsh and Tushar Krishna},
      year={2022},
      eprint={2107.06419},
      archivePrefix={arXiv},
      primaryClass={cs.LG},
      url={https://arxiv.org/abs/2107.06419}, 
}

@article{model1,
  title={{Attention is all you need}},
  author={Vaswani, Ashish and Shazeer, Noam and Parmar, Niki and Uszkoreit, Jakob and Jones, Llion and Gomez, Aidan N and Kaiser, {\L}ukasz and Polosukhin, Illia},
  journal={Advances in neural information processing systems},
  volume={30},
  year={2017}
}

@misc{model2,
      title={{Cross-lingual Language Model Pretraining}}, 
      author={Guillaume Lample and Alexis Conneau},
      year={2019},
      eprint={1901.07291},
      archivePrefix={arXiv},
      primaryClass={cs.CL},
      url={https://arxiv.org/abs/1901.07291}, 
}

@misc{model3,
      title={{Exploring the Limits of Transfer Learning with a Unified Text-to-Text Transformer}}, 
      author={Colin Raffel and Noam Shazeer and Adam Roberts and Katherine Lee and Sharan Narang and Michael Matena and Yanqi Zhou and Wei Li and Peter J. Liu},
      year={2023},
      eprint={1910.10683},
      archivePrefix={arXiv},
      primaryClass={cs.LG},
      url={https://arxiv.org/abs/1910.10683}, 
}

@misc{model4,
      title={{FlauBERT: Unsupervised Language Model Pre-training for French}}, 
      author={Hang Le and Loïc Vial and Jibril Frej and Vincent Segonne and Maximin Coavoux and Benjamin Lecouteux and Alexandre Allauzen and Benoît Crabbé and Laurent Besacier and Didier Schwab},
      year={2020},
      eprint={1912.05372},
      archivePrefix={arXiv},
      primaryClass={cs.CL},
      url={https://arxiv.org/abs/1912.05372}, 
}

@inproceedings{multimodal,
author = {Jiang, Jiachen and Vosoughi, Soroush},
title = {Not Judging a User by Their Cover: Understanding Harm in Multi-Modal Processing within Social Media Research},
year = {2020},
isbn = {9781450381482},
publisher = {Association for Computing Machinery},
address = {New York, NY, USA},
url = {https://doi.org/10.1145/3422841.3423534},
doi = {10.1145/3422841.3423534},
booktitle = {Proceedings of the 2nd International Workshop on Fairness, Accountability, Transparency and Ethics in Multimedia},
pages = {6–12},
numpages = {7},
location = {Seattle, WA, USA},
series = {FATE/MM '20}
}

@inproceedings{crossdomain,
author = {Zheng, Yaowei and Zhang, Richong and Wang, Suyuchen and Mensah, Samuel and Mao, Yongyi},
title = {Anchored Model Transfer and Soft Instance Transfer for Cross-Task Cross-Domain Learning: A Study Through Aspect-Level Sentiment Classification},
year = {2020},
isbn = {9781450370233},
publisher = {Association for Computing Machinery},
address = {New York, NY, USA},
url = {https://doi.org/10.1145/3366423.3380034},
doi = {10.1145/3366423.3380034},
booktitle = {Proceedings of The Web Conference 2020},
pages = {2754–2760},
numpages = {7},
location = {Taipei, Taiwan},
series = {WWW '20}
}

@inproceedings{contentsynthesis,
author = {Paneva-Marinova, Desislava and Zlatkov, Lubomir and Pavlova, Lilia},
title = {Improved User Experience in Digital Library through Advanced Content Synthesizing},
year = {2019},
isbn = {9781450371889},
publisher = {Association for Computing Machinery},
address = {New York, NY, USA},
url = {https://doi.org/10.1145/3357419.3357432},
doi = {10.1145/3357419.3357432},
booktitle = {Proceedings of the 9th International Conference on Information Communication and Management},
pages = {170–174},
numpages = {5},
location = {Prague, Czech Republic},
series = {ICICM '19}
}

@inproceedings{tangram,
author = {Gao, Mingyu and Yang, Xuan and Pu, Jing and Horowitz, Mark and Kozyrakis, Christos},
title = {TANGRAM: Optimized Coarse-Grained Dataflow for Scalable NN Accelerators},
year = {2019},
isbn = {9781450362405},
publisher = {Association for Computing Machinery},
address = {New York, NY, USA},
url = {https://doi.org/10.1145/3297858.3304014},
doi = {10.1145/3297858.3304014},
booktitle = {Proceedings of the Twenty-Fourth International Conference on Architectural Support for Programming Languages and Operating Systems},
pages = {807–820},
numpages = {14},
location = {Providence, RI, USA},
series = {ASPLOS '19}
}

@article{SymPy2017,
  title = {SymPy: symbolic computing in Python},
  author = {Meurer, Aaron and Smith, Christopher P. and Paprocki, Mateusz and \v{C}ert\'{i}k, Ond\v{r}ej and Kirpichev, Sergey B. and Rocklin, Matthew and Kumar, AMiT and Ivanov, Sergiu and Moore, Jason K. and Singh, Sartaj and Rathnayake, Thilina and Vig, Sean and Granger, Brian E. and Muller, Richard P. and Bonazzi, Francesco and Gupta, Harsh and Vats, Shivam and Johansson, Fredrik and Pedregosa, Fabian and Curry, Matthew J. and Terrel, Andy R. and Rou\v{c}ka, \v{S}t\v{e}p\'{a}n and Saboo, Ashutosh and Fernando, Isuru and Kulal, Sumith and Cimrman, Robert and Scopatz, Anthony},
  year = 2017,
  month = jan,
  volume = 3,
  pages = {e103},
  journal = {PeerJ Computer Science},
  issn = {2376-5992},
  url = {https://doi.org/10.7717/peerj-cs.103},
  doi = {10.7717/peerj-cs.103}
}

@misc{CUDA,
  author       = {NVIDIA},
  title        = {CUDA C++ Programming Guide},
  url = {https://docs.nvidia.com/cuda/cuda-c-programming-guide},
  year         = {2025}
}

@misc{kimik2,
      title={{Kimi K2: Open Agentic Intelligence}}, 
      author={Kimi Team and Yifan Bai and Yiping Bao and Guanduo Chen and Jiahao Chen and Ningxin Chen and Ruijue Chen and Yanru Chen and Yuankun Chen and Yutian Chen and Zhuofu Chen and Jialei Cui and Hao Ding and Mengnan Dong and Angang Du and Chenzhuang Du and Dikang Du and Yulun Du and Yu Fan and Yichen Feng and Kelin Fu and Bofei Gao and Hongcheng Gao and Peizhong Gao and Tong Gao and Xinran Gu and Longyu Guan and Haiqing Guo and Jianhang Guo and Hao Hu and Xiaoru Hao and Tianhong He and Weiran He and Wenyang He and Chao Hong and Yangyang Hu and Zhenxing Hu and Weixiao Huang and Zhiqi Huang and Zihao Huang and Tao Jiang and Zhejun Jiang and Xinyi Jin and Yongsheng Kang and Guokun Lai and Cheng Li and Fang Li and Haoyang Li and Ming Li and Wentao Li and Yanhao Li and Yiwei Li and Zhaowei Li and Zheming Li and Hongzhan Lin and Xiaohan Lin and Zongyu Lin and Chengyin Liu and Chenyu Liu and Hongzhang Liu and Jingyuan Liu and Junqi Liu and Liang Liu and Shaowei Liu and T. Y. Liu and Tianwei Liu and Weizhou Liu and Yangyang Liu and Yibo Liu and Yiping Liu and Yue Liu and Zhengying Liu and Enzhe Lu and Lijun Lu and Shengling Ma and Xinyu Ma and Yingwei Ma and Shaoguang Mao and Jie Mei and Xin Men and Yibo Miao and Siyuan Pan and Yebo Peng and Ruoyu Qin and Bowen Qu and Zeyu Shang and Lidong Shi and Shengyuan Shi and Feifan Song and Jianlin Su and Zhengyuan Su and Xinjie Sun and Flood Sung and Heyi Tang and Jiawen Tao and Qifeng Teng and Chensi Wang and Dinglu Wang and Feng Wang and Haiming Wang and Jianzhou Wang and Jiaxing Wang and Jinhong Wang and Shengjie Wang and Shuyi Wang and Yao Wang and Yejie Wang and Yiqin Wang and Yuxin Wang and Yuzhi Wang and Zhaoji Wang and Zhengtao Wang and Zhexu Wang and Chu Wei and Qianqian Wei and Wenhao Wu and Xingzhe Wu and Yuxin Wu and Chenjun Xiao and Xiaotong Xie and Weimin Xiong and Boyu Xu and Jing Xu and Jinjing Xu and L. H. Xu and Lin Xu and Suting Xu and Weixin Xu and Xinran Xu and Yangchuan Xu and Ziyao Xu and Junjie Yan and Yuzi Yan and Xiaofei Yang and Ying Yang and Zhen Yang and Zhilin Yang and Zonghan Yang and Haotian Yao and Xingcheng Yao and Wenjie Ye and Zhuorui Ye and Bohong Yin and Longhui Yu and Enming Yuan and Hongbang Yuan and Mengjie Yuan and Haobing Zhan and Dehao Zhang and Hao Zhang and Wanlu Zhang and Xiaobin Zhang and Yangkun Zhang and Yizhi Zhang and Yongting Zhang and Yu Zhang and Yutao Zhang and Yutong Zhang and Zheng Zhang and Haotian Zhao and Yikai Zhao and Huabin Zheng and Shaojie Zheng and Jianren Zhou and Xinyu Zhou and Zaida Zhou and Zhen Zhu and Weiyu Zhuang and Xinxing Zu},
      year={2025},
      eprint={2507.20534},
      archivePrefix={arXiv},
      primaryClass={cs.LG},
      url={https://arxiv.org/abs/2507.20534}, 
}

@inproceedings{BERT2019,
  title={{Bert: Pre-training of deep bidirectional transformers for language understanding}},
  author={Devlin, Jacob and Chang, Ming-Wei and Lee, Kenton and Toutanova, Kristina},
  booktitle={Proceedings of the 2019 conference of the North American chapter of the association for computational linguistics: human language technologies, volume 1 (long and short papers)},
  pages={4171--4186},
  year={2019}
}

@article{ViT2020,
  title={{An image is worth 16x16 words: Transformers for image recognition at scale}},
  author={Dosovitskiy, Alexey and Beyer, Lucas and Kolesnikov, Alexander and Weissenborn, Dirk and Zhai, Xiaohua and Unterthiner, Thomas and Dehghani, Mostafa and Minderer, Matthias and Heigold, Georg and Gelly, Sylvain and others},
  journal={arXiv preprint arXiv:2010. 11929},
  year={2020}
}

@article{LLaMA2023,
  title={{Llama: Open and efficient foundation language models}},
  author={Touvron, Hugo and Lavril, Thibaut and Izacard, Gautier and Martinet, Xavier and Lachaux, Marie-Anne and Lacroix, Timoth{\'e}e and Rozi{\`e}re, Baptiste and Goyal, Naman and Hambro, Eric and Azhar, Faisal and others},
  journal={arXiv preprint arXiv:2302.13971},
  year={2023}
}

@article{DeepSeekR1-2025,
  title={{Deepseek-r1: Incentivizing reasoning capability in llms via reinforcement learning}},
  author={Guo, Daya and Yang, Dejian and Zhang, Haowei and Song, Junxiao and Zhang, Ruoyu and Xu, Runxin and Zhu, Qihao and Ma, Shirong and Wang, Peiyi and Bi, Xiao and others},
  journal={arXiv preprint arXiv:2501.12948},
  year={2025}
}

@article{Switch2022,
  title={{Switch transformers: Scaling to trillion parameter models with simple and efficient sparsity}},
  author={Fedus, William and Zoph, Barret and Shazeer, Noam},
  journal={Journal of Machine Learning Research},
  volume={23},
  number={120},
  pages={1--39},
  year={2022}
}

@misc{ERNIE2025,
      title={{ERNIE 4.5 Technical Report}},
      author={Baidu-ERNIE-Team},
      year={2025},
      primaryClass={cs.CL},
      howpublished={\url{https://ernie.baidu.com/blog/publication/ERNIE_Technical_Report.pdf}}
}

@article{DeepSeekV2-2024,
  title={{Deepseek-v2: A strong, economical, and efficient mixture-of-experts language model}},
  author={Liu, Aixin and Feng, Bei and Wang, Bin and Wang, Bingxuan and Liu, Bo and Zhao, Chenggang and Dengr, Chengqi and Ruan, Chong and Dai, Damai and Guo, Daya and others},
  journal={arXiv preprint arXiv:2405.04434},
  year={2024}
}

@article{Qwen3-2025,
  title={{Qwen3 technical report}},
  author={Yang, An and Li, Anfeng and Yang, Baosong and Zhang, Beichen and Hui, Binyuan and Zheng, Bo and Yu, Bowen and Gao, Chang and Huang, Chengen and Lv, Chenxu and others},
  journal={arXiv preprint arXiv:2505. 09388},
  year={2025}
}

@inproceedings{Triton2019,
  title={{Triton: an intermediate language and compiler for tiled neural network computations}},
  author={Tillet, Philippe and Kung, Hsiang-Tsung and Cox, David},
  booktitle={Proceedings of the 3rd ACM SIGPLAN International Workshop on Machine Learning and Programming Languages},
  pages={10--19},
  year={2019}
}

@article{Flashinfer2025,
  title={{Flashinfer: Efficient and customizable attention engine for llm inference serving}},
  author={Ye, Zihao and Chen, Lequn and Lai, Ruihang and Lin, Wuwei and Zhang, Yineng and Wang, Stephanie and Chen, Tianqi and Kasikci, Baris and Grover, Vinod and Krishnamurthy, Arvind and others},
  journal={arXiv preprint arXiv:2501.01005},
  year={2025}
}

@misc{FlashMLA2025,
      title={{FlashMLA: Efficient MLA decoding kernels}},
      author={Jiashi Li, Shengyu Liu},
      year={2025},
      publisher = {GitHub},
      howpublished = {\url{https://github.com/deepseek-ai/FlashMLA}},
}

\clearpage
\appendix

\section{Appendix}

\subsection{Non-invertible Condition}
\label{Appendix:Non-invertible}

In the derivation presented in Chapter~\ref{chap:derivation}, we
assumed that the function $\mathrm{H}_i(\hat{\mathbf{d}}^{k-1})$ is
invertible within $(\mathbb{S}, \otimes_i)$,
to support the existence of the inverse term
$\mathrm{H}_i(\hat{\mathbf{d}}^{k-1})^{-1}$ in the fused expression.
However, in certain practical scenarios (e.g., when $\otimes_i =
*$ and $\mathrm{H}_i(\cdot) = 0$), this inverse may not exist,
rendering the expression undefined.

To ensure semantic correctness of the fusion transformation in all
cases, we propose a \textbf{reversibility repair mechanism}: extend
$\mathrm{H}_i(\hat{\mathbf{d}}^{k-1})$ into a corrected function
$\mathrm{H}_i'(\hat{\mathbf{d}}^{k-1})$, defined as:
\begin{equation}
  \mathrm{H}_i'(\hat{\mathbf{d}}^{k-1}) =
  \begin{cases}
    \mathrm{H}_i(\hat{\mathbf{d}}^{k-1}), & \text{if it is invertible
    in } (\mathbb{S}, \otimes_i), \\
    e, & \text{otherwise},
  \end{cases}
  \label{eq:hi_prime}
\end{equation}
where $e$ is the identity element of $(\mathbb{S}, \otimes_i)$ (e.g.
$0$ for $(\mathbb{R}, +)$, $1$ for $(\mathbb{R}, \ast)$). Since $e$ is always
invertible ($e^{-1} = e$), this extension guarantees that
$\mathrm{H}_i'(\hat{\mathbf{d}}^{k-1})$ is invertible for all inputs.

Substituting $\mathrm{H}_i'$ into the Equation~\ref{eq:fused_dk}:
\begin{equation}
  \hat{d}^k = \sideset{}{_i^\sharp}\sum_{j^{k-1}=l_{1st}^k}^{l_{last}^k}
  \hat{d}^{k-1} \otimes_i \mathrm{H}_i'(\hat{\mathbf{d}}^{k-1})^{-1}
  \otimes_i \mathrm{H}_i'(\hat{\mathbf{d}}^k).
  \label{eq:fused_corrected}
\end{equation}
This corrected expression is computable in all cases and is
equivalent to the original expression whenever
$\mathrm{H}_i(\hat{\mathbf{d}}^{k-1})$ is invertible. Thus, it
ensures robustness and correctness of the fused reduction computation
across all input configurations.

\subsection{Cases of Fused and Incremental Computation Expressions}
\label{Appendix:Fused_Cases}

\subsubsection{Attention Mechanism}
\label{Appendix:Fused_Cases:Attention}
For a given token corresponding to a query vector, let $P =
\frac{QK^T}{\sqrt{d_k}}$. The output of the attention computation can
be expressed as:
\begin{equation}
  \begin{split}
    &m = \max_{l=1}^{L_0} P[l], \\
    &t = \sum_{l=1}^{L_0} \exp(P[l] - m), \\
    &O = \sum_{l=1}^{L_0} \frac{\exp(P[l] - m)}{t} * V[l],
  \end{split}
  \label{eq:attention_reductions}
\end{equation}
where $V[l]$ denotes the $l$-th row of the value matrix $V$.
    
The fused computation form for the first level reduction at the $j^1$-th segment is
\begin{equation}
  \begin{split}
    &\hat{m}^1 = \max_{l =
    l_{\mathrm{1st}}^{1}}^{l_{\mathrm{last}}^{1}} P[l], \\
    &\hat{t}^1 = \sum_{l =
    l_{\mathrm{1st}}^{1}}^{l_{\mathrm{last}}^{1}} 
    \exp\left(P[l] - \hat{m}^1\right), \\
    &\hat{O}^1 = \sum_{l =
    l_{\mathrm{1st}}^{1}}^{l_{\mathrm{last}}^{1}} 
    \frac{\exp\left(P[l] - \hat{m}^1\right)}{\hat{t}^1}.
  \end{split}
  \label{eq:fused_attention_levle1}
\end{equation}

For level-$k$:
\begin{equation}
  \begin{split}
    &\hat{m}^k = \max_{j^{k-1} =
    l_{\mathrm{1st}}^{k}}^{l_{\mathrm{last}}^{k}} \hat{m}^{k-1}, \\
    &\hat{t}^k = \sum_{j^{k-1} =
    l_{\mathrm{1st}}^{k}}^{l_{\mathrm{last}}^{k}} \hat{t}^{k-1} *
    \exp\left(\hat{m}^{k-1} - \hat{m}^k\right), \\
    &\hat{O}^k = \sum_{j^{k-1} =
    l_{\mathrm{1st}}^{k}}^{l_{\mathrm{last}}^{k}} \hat{O}^{k-1} *
    \exp\left(\hat{m}^{k-1} - \hat{m}^k\right) *
    \frac{\hat{t}^{k-1}}{\hat{t}^k}.
  \end{split}
  \label{eq:fused_attention_levelk}
\end{equation}

The incremental computation form for the first level on the $j^1$-th input segment is:
\begin{equation}
  \begin{split}
    &\hat{m}^1[L] = \max\left(\hat{m}^1[L-1],\, P[L]\right), \\
    &
    \begin{split}
    \hat{t}^1[L] = &\hat{t}^1[L-1] * \exp\left(\hat{m}^{1}[L-1] -
    \hat{m}^{1}[L]\right) \\
    & + \exp\left(P[L] - \hat{m}^{1}[L]\right), 
    \end{split}
    \\
    &
    \begin{split}
    \hat{O}^1[L] = &\hat{O}^1[L-1] * \exp\left(\hat{m}^{1}[L-1] -
    \hat{m}^{1}[L]\right) \\
    & * \frac{\hat{t}^1[L-1]}{\hat{t}^1[L]} + \frac{\exp\left(P[L] -
    \hat{m}^{1}[L]\right)}{\hat{t}^1[L]}.
    \end{split}
  \end{split}
  \label{eq:incremental_attention_level1}
\end{equation}

For level-$k$:
\begin{equation}
  \begin{split}
    &\hat{m}^k[L] = \max\left(\hat{m}^k[L-1],\, \hat{m}^{k-1}\right), \\
    &
    \begin{split}
    \hat{t}^k[L] = &\hat{t}^k[L-1] * \exp\left(\hat{m}^{k}[L-1] -
    \hat{m}^{k}[L]\right) \\
    & + \hat{t}^{k-1} * \exp\left(\hat{m}^{k-1} - \hat{m}^{k}[L]\right), 
    \end{split}
    \\
    &
    \begin{split}
    \hat{O}^k[L] = &\hat{O}^k[L-1] * \exp\left(\hat{m}^{k}[L-1] -
    \hat{m}^{k}[L]\right) * \frac{\hat{t}^k[L-1]}{\hat{t}^k[L]} \\
    & + \hat{O}^{k-1} * \exp\left(\hat{m}^{k-1} -
    \hat{m}^{k}[L]\right) * \frac{\hat{t}^{k-1}}{\hat{t}^k[L]}.
    \end{split}
  \end{split}
  \label{eq:incremental_attention_levelk}
\end{equation}
    
Notably, Equation~\eqref{eq:incremental_attention_levelk} is identical to the tiled attention computation formula proposed in FlashAttention, which corresponds to fusion at level $k=3$.
\subsubsection{MoE routing}
Since softmax preserves the relative ordering of input values, we partition the expert scores into segments of length $K'$ for input to the $\mathrm{TopK}$ operation. The full computation is:
\begin{equation}
\begin{split}
&m = \max_{l=1}^{L_0} x[l], \\
&t = \sum_{l=1}^{L_0} \exp(x[l] - m), \\
&\vec{s} = \mathrm{TopK}_{l=1}^{L_0/K'} \vec{x}[l].
\end{split}  
\end{equation}
where $\vec{s}$ and $\vec{x}$ are vectors of length $K'$.
The fused expression at level $k$ is defined as follows. For $k=1$:
\begin{equation}
\begin{split}
&\hat{m}^1 = \max_{l =
l_{\mathrm{1st}}^{1}}^{l_{\mathrm{last}}^{1}} x[l], \\
&\hat{t}^1 = \sum_{l =
l_{\mathrm{1st}}^{1}}^{l_{\mathrm{last}}^{1}} 
\exp\left(x[l] - \hat{m}^1\right), \\
&\hat{\vec{s}}^1 = \mathrm{TopK}_{l =
l_{\mathrm{1st}}^{1}}^{l_{\mathrm{last}}^{1}} \vec{x}[l].
\end{split}  
\end{equation}
For $k>1$:
\begin{equation}
\begin{split}
&\hat{m}^k = \max_{j^{k-1} =
l_{\mathrm{1st}}^{k}}^{l_{\mathrm{last}}^{k}} \hat{m}^{k-1}, k > 1\\
&\hat{t}^k = \sum_{j^{k-1} =
l_{\mathrm{1st}}^{k}}^{l_{\mathrm{last}}^{k}} \hat{t}^{k-1} *
\exp\left(\hat{m}^{k-1} - \hat{m}^k\right), \\
&\hat{\vec{s}}^k = \mathrm{TopK}_{j^{k-1} =
l_{\mathrm{1st}}^{k}}^{l_{\mathrm{last}}^{k}} \hat{\vec{s}}^{k-1}.
\end{split}  
\end{equation}
The incremental form at level $k=1$ is:
\begin{equation}
\begin{split}
&\hat{m}^1[L] = \max\left(\hat{m}^1[L-1],\, x[l]\right), \\
&
\begin{split}
\hat{t}^1[L] = &\hat{t}^1[L-1] * \exp\left(\hat{m}^{1}[L-1] -
\hat{m}^{1}[L]\right) \\
& + \exp\left(x[L] - \hat{m}^{1}[L]\right), 
\end{split}
\\
&
\begin{split}
\hat{\vec{s}}^1[L] = \mathrm{TopK} (\hat{\vec{s}}^{1}[L-1], \vec{x}[L]).
\end{split}
\end{split}
\end{equation}
For $k>1$:
\begin{equation}
\begin{split}
&\hat{m}^k[L] = \max\left(\hat{m}^k[L-1],\, \hat{m}^{k-1}\right), \\
&
\begin{split}
\hat{t}^k[L] = &\hat{t}^k[L-1] * \exp\left(\hat{m}^{k}[L-1] -
\hat{m}^{k}[L]\right) \\
& + \hat{t}^{k-1} * \exp\left(\hat{m}^{k-1} - \hat{m}^{k}[L]\right), 
\end{split}
\\
&
\begin{split}
\hat{\vec{s}}^k[L] = \mathrm{TopK} (\hat{\vec{s}}^{k}[L-1], \hat{\vec{s}}^{k-1}).
\end{split}
\end{split}
\end{equation}
The final output can be obtained by normalizing the selected top-$K'$ values: $\vec{s} / t$
\subsubsection{Sum + Sum}
This is a pattern used within the internal model, and its computational formulation is as follows:
\begin{equation}
\begin{split}
&m = \sum_{l=1}^{L_0} x_{1}[l]^2, \\
&s = \sum_{l=1}^{L_0} \frac{x_{1}[l]*x_{2}[l]}{\sqrt{\max{(m-10)}}}.
\end{split}
\end{equation}
The fused expression at level $k$ is defined as follows. For $k=1$:
\begin{equation}
\begin{split}
&\hat{m}^1 = \sum_{l =
l_{\mathrm{1st}}^{1}}^{l_{\mathrm{last}}^{1}} x_{1}[l]^2, \\
&\hat{s}^1 = \sum_{l =
l_{\mathrm{1st}}^{1}}^{l_{\mathrm{last}}^{1}} \frac{x_{1}[l]*x_{2}[l]}{\sqrt{\max{(\hat{m}^1-10)}}}. \\
\end{split}  
\end{equation}
For $k>1$:
\begin{equation}
\begin{split}
&\hat{m}^k = \sum_{j^{k-1} =
l_{\mathrm{1st}}^{k}}^{l_{\mathrm{last}}^{k}} \hat{m}^{k-1},\\
&\hat{s}^k = \sum_{j^{k-1} =
l_{\mathrm{1st}}^{k}}^{l_{\mathrm{last}}^{k}} \hat{s}^{k-1}*\sqrt{\frac{\max{(\hat{m}^{k-1}-10)}}{\max{(\hat{m}^{k}-10)}}}.
\end{split}  
\end{equation}
The incremental form at level $k=1$ is:
\begin{equation}
\begin{split}
&\hat{m}^1[L] = \hat{m}^1[L-1] + x_{1}[L]^2, \\
&
\begin{split}
\hat{s}^1[L] = & \hat{s}^1[L-1] * \sqrt{\frac{\max{(\hat{m}^{1}[L-1]-10)}}{\max{(\hat{m}^{1}[L]-10)}}} \\
&+ \frac{x_{1}[L]*x_{2}[L]}{\sqrt{\max{(\hat{m}^{1}[L]-10)}}}. \\
\end{split}  
\end{split}  
\end{equation}
For $k>1$:
\begin{equation}
\begin{split}
&\hat{m}^k[L] = \hat{m}^k[L-1] + \hat{m}^{k-1}, \\
&
\begin{split}
\hat{s}^k[L] = &\hat{s}^{k}[L-1] * \sqrt{\frac{\max{(\hat{m}^{k}[L-1]-10)}}{\max{(\hat{m}^{k}[L]-10)}}} \\
& + \hat{s}^{k-1} * \sqrt{\frac{\max{(\hat{m}^{k-1}-10)}}{\max{(\hat{m}^{k}[L]-10)}}}.
\end{split}
\end{split}
\end{equation}

\subsection{TileOp Specification}
\label{sec:tileop_specification}

We formalize the TileOps used in RedFuser in Figure~\ref{fig:tileop_specification}, with semantics as follows:

\begin{itemize}[leftmargin=*, labelindent=0pt]
    \item \texttt{copy(src, dst)}: Copies data from tile \texttt{src} to \texttt{dst}.
    
    \item \texttt{gemm(A, B, C)}: Computes \texttt{C += A * B}.
    
    \item \texttt{reduce(src, dst, axis = k, op)}: Reduces tile \texttt{src} along dimension \texttt{k} using binary operator \texttt{op}, storing the result in \texttt{dst}.
    
    \item \texttt{parallel(buf[idx+], f(args*), iters+, ranges+)}: Evaluates \texttt{buf[idx+] = f(args*)} in parallel over the iteration space defined by \texttt{iters+} and \texttt{ranges+}.
    
    \item \texttt{fill(tile, c)}: Fills \texttt{tile} with constant \texttt{c}.
\end{itemize}

\subsection{Implementation of FlashAttention and FlashDecoding}
\label{sec:implementation_details}

This section details the implementation of FlashAttention and FlashDecoding in RedFuser.

Beginning with the unfused TIR representation in Figure~\ref{fig:unfused_attention_TIR}, RedFuser analyzes the AST to both identify four distinct reduction operations and determine their respective axes. Crucially, it detects that reductions 2-4 share a common axis, forming a cascaded reduction pattern. RedFuser then lifts this structure into equivalent mathematical expressions to facilitate fusion.

Using the ACRF algorithm (Section~\ref{chap:automatic_fusion_algorithm}), RedFuser first obtains the fused expressions. These expressions are then realized into two scalar-level IR structures by applying the \textbf{Single-Segment} and \textbf{Multi-Segment} strategies (Section~\ref{chap:ProgrammingModel}), following a three-step reduction template: (1) store prior results, (2) correct current results, and (3) execute the reduction. 
Notably, RedFuser further eliminates redundant steps based on dataflow dependencies; for instance, step (2) is omitted for reductions with no dependencies (such as reduction 2), and step (1) is skipped for reductions whose outputs are not reused (such as reduction 4). The resulting scalar-level IR structures are shown in Figures~\ref{fig:flashattention_scalar} and~\ref{fig:flashdecoding_scalar}.

The scalar-level IR is subsequently elevated to tile-level IR via \textbf{Tensorization} and \textbf{Parallelization} (Section~\ref{chap:code_generation}), and further optimized by TileLang into high-performance GPU kernels. We configure a tile size of (\texttt{blk\_qs}, \texttt{blk\_kvs}) = (128, 128) for both FlashAttention and FlashDecoding, and set \texttt{num\_split} = 2 for FlashDecoding. The final tile-level IR structures are illustrated in Figures~\ref{fig:flashattention_tile} and~\ref{fig:flashdecoding_tile}.

\subsection{Workload Configurations}
\label{Appendix:Workload-Configurations}

We present in detail the configurations of all selected subgraphs.

\subsubsection{MHA and MLA}

For MHA, the input tensors are shaped as $[bs, hn, q, hd]$ for query and $[bs, hn, kv, hd]$ for key and value, where $bs$, $hn$, $q$, $kv$ and $hd$ denote for \textit{batch size}, \textit{head num}, \textit{sequence length q}, \textit{sequence length kv} and \textit{head dim}.
In MLA, we set the query length to always be 1 to simulate the decoding stage of LLM. The hidden dimensions of query and key are extended by the RoPE embedding dimension ($ped$).
Detailed configurations are shown in Table~\ref{tab:mha_configurations} and Table~\ref{tab:mla_configurations}.

\subsubsection{MoE routing and Quant GEMM}

In MoE routing, the expert scores are computed via a GEMM between an input tensor of shape $[s, hd]$ and a routing weight matrix of shape $[hd, en]$, where \textit{s}, \textit{hd} and \textit{en} denote for \textit{sequence length}, \textit{head dim} and \textit{expert num}.
For Quant$+$GEMM, we consider a dynamic quantization workflow where a matrix of shape $[M, K]$ is quantized to FP8 using a reduction-based scaling factor, followed by a GEMM with a weight matrix of shape $[K, N]$.
Table~\ref{tab:moe_routing_configurations} and Table~\ref{tab:quant_gemm_configurations} provide detailed configurations.

\begin{table*}[tp]
    \centering
    \caption{The configurations of four workloads}
    \label{tab:workload-configurations}
\begin{subtable}[t]{0.45\textwidth}
    \centering
    \begin{tabular}{lllllll}
        \toprule
        Name & bs & hn & q & kv & hd & model \\
        \midrule
        H1 & 32 & 8 &  512 & 512 & 64 & BERT-Small \\
        H2 & 32 & 12 & 512 & 512 & 64 & BERT-Base \\
        H3 & 32 & 16 & 512 & 512 & 64 & BERT-Large \\
        H4 & 32 & 12 & 256 & 256 & 64 & ViT-Base \\
        H5 & 32 & 16 & 256 & 256 & 64 & ViT-Large \\
        H6 & 32 & 16 & 256 & 256 & 80 & ViT-Huge \\
        H7 & 32 & 64 & 1 & 1024 & 128 & LLaMA-65B \\
        H8 & 32 & 64 & 1 & 2048 & 128 & LLaMA-65B \\
        H9 & 32 & 64 & 1 & 4096 & 128 & LLaMA-65B \\
        \bottomrule
    \end{tabular}
    \caption{The configurations of MHA}
    \label{tab:mha_configurations}
\end{subtable}
    \hfill
\begin{subtable}[t]{0.45\textwidth}
    \centering
    \begin{tabular}{llllll}
        \toprule
        Name & bs & hn & kv & hd & ped \\
        \midrule
        L1 & 32 & 128 & 1024 & 512 & 64 \\
        L2 & 32 & 128 & 2048 & 512 & 64 \\
        L3 & 32 & 128 & 4096 & 512 & 64 \\
        L4 & 16 & 128 & 1024 & 512 & 64 \\
        L5 & 16 & 128 & 2048 & 512 & 64 \\
        L6 & 16 & 128 & 4096 & 512 & 64 \\
        L7 & 1 & 128 & 1024 & 512 & 64 \\
        L8 & 1 & 128 & 2048 & 512 & 64 \\
        L9 & 1 & 128 & 4096 & 512 & 64 \\
        \bottomrule
    \end{tabular}
    \caption{The configurations of MLA}
    \label{tab:mla_configurations}
\end{subtable}

\begin{subtable}[b]{0.45\textwidth}
    \centering
    \begin{tabular}{llllll}
        \toprule
        &&&&& \\
        Name & s & hd & en & topk & model \\
        &&&&& \\
        \midrule
        R1 & 2048 & 768 & 128 & 1 & switch-base-128 \\
        R2 & 2048 & 1024 & 128 & 1 & switch-large-128 \\
        R3 & 2048 & 4096 & 128 & 1 & switch-xxl-128 \\
        R4 & 2048 & 2560 & 64 & 6 & ERNIE-21B-A3B \\
        R5 & 2048 & 8192 & 64 & 8 & ERNIE-300B-A47B \\
        R6 & 2048 & 2048 & 64 & 6 & DeepSeek-V2-Lite \\
        R7 & 2048 & 2048 & 128 & 8 & Qwen3-30B-A3B \\
        R8 & 2048 & 4096 & 128 & 8 & Qwen3-235B-A30B \\
        \bottomrule
    \end{tabular}
    \caption{The configurations of MoE routing}
    \label{tab:moe_routing_configurations}
\end{subtable}
    \hfill
\begin{subtable}[b]{0.45\textwidth}
    \centering
    \begin{tabular}{lllll}
        \toprule
        Name & M & N & K & model \\
        \midrule
        Q1 & 4096 & 1536 & 2560 & ERNIE-21B-A3B \\
        Q2 & 4096 & 2560 & 1536 & ERNIE-21B-A3B \\
        Q3 & 4096 & 3584 & 8192 & ERNIE-300B-A47B \\
        Q4 & 4096 & 8192 & 3584 & ERNIE-300B-A47B \\
        Q5 & 4096 & 7168 & 2048 & DeepSeek-R1 \\
        Q6 & 4096 & 2048 & 7168 & DeepSeek-R1 \\
        Q7 & 4096 & 2048 & 768 & Qwen3-30B-A3B \\
        Q8 & 4096 & 768 & 2048 & Qwen3-30B-A3B \\
        Q9 & 4096 & 4096 & 1536 & Qwen3-235B-A30B \\
        Q10 & 4096 & 1536 & 4096 & Qwen3-235B-A30B \\
        \bottomrule
    \end{tabular}
    \caption{The configurations of Quant $+$ GEMM}
    \label{tab:quant_gemm_configurations}
\end{subtable}
\end{table*}

\subsection{Evaluation on non-ML Workloads}
\label{Appendix:non-ML-Workloads}

For non-ML workloads, we evaluate two representative examples: (1) variance computation, and (2) moment of inertia computation in 3D space relative to the center of mass.
The formula for variance is as follows:
\begin{equation}
\begin{split}
& m=\frac{1}{L_0}\sum_{l=1}^{L_0}x[l], \\
& var=\frac{1}{L_0}\sum_{l=1}^{L_0}(x[l]-m)^2.
\end{split}
\end{equation}

The moment of inertia about the center of mass is computed as:
\begin{equation}
\begin{split}
& M=\sum_{l=1}^{L_0}m[l], \\
& \vec{c}=\frac{1}{M}\sum_{l=1}^{L_0}m[l]*\vec{x[l]}, \\
& I=\sum_{l=1}^{L_0}m[l] *\lVert \vec{x[l]} - \vec{c} \rVert^{2}.
\end{split}
\end{equation}

As both formulas involve sequences of inter-dependent reductions, they exhibit cascaded reduction structures. Hence, they are amenable to fusion optimization within the RedFuser framework.

The configurations for non-ML workloads are summarized in Table~\ref{tab:non-ml-workload-configurations}.
For variance computation, the input $x$ is a 2D tensor of shape $[bs, l]$ where $bs$ denotes the \textit{batch size} and $l$ denotes the \textit{number of data points} in each batch.
For moment of inertia computation, the mass input $m$ is a 2D tensor of shape $[bs, n]$, and the position input $x$ is a 3D tensor of shape $[bs, n, dim]$, where $bs$ is the \textit{batch size}, $n$ is the \textit{number of particles}, and $dim$ (set to 3) represents the \textit{spatial dimensions of 3D coordinates}.

\begin{figure*}[b]
    \centering
    \begin{subfigure}[t]{0.49\textwidth}
        \includegraphics[width=\linewidth]{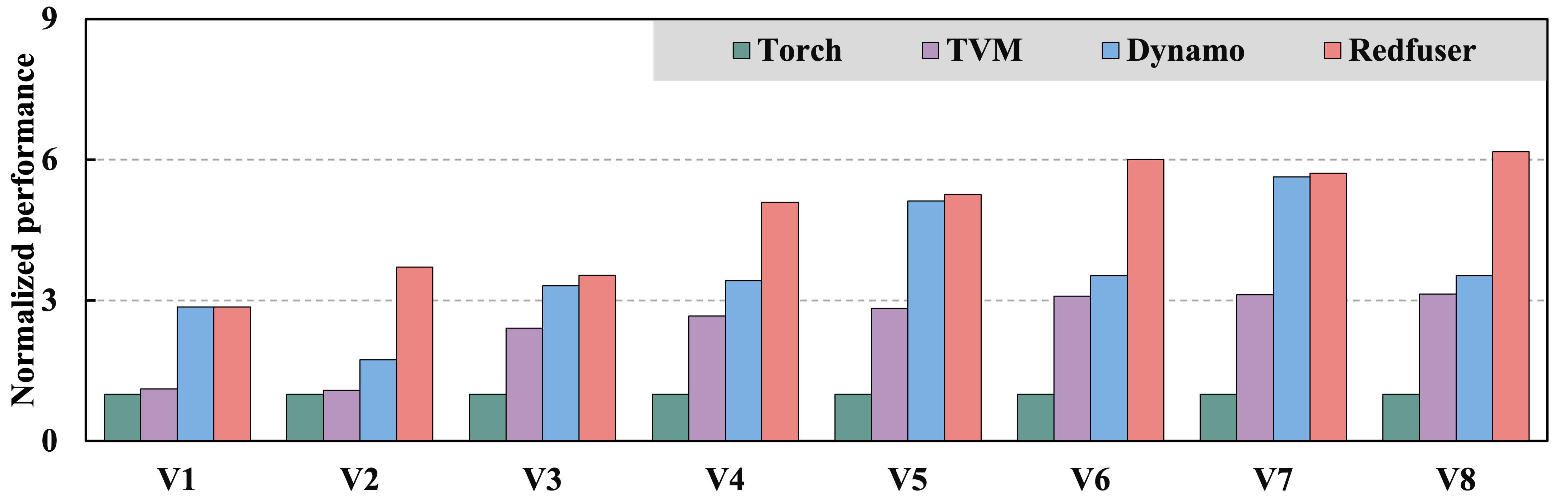}
        \caption{Variance on A10}
        \label{fig:exp_var_A10}
    \end{subfigure}
    \hfill
    \begin{subfigure}[t]{0.49\textwidth}
        \includegraphics[width=\linewidth]{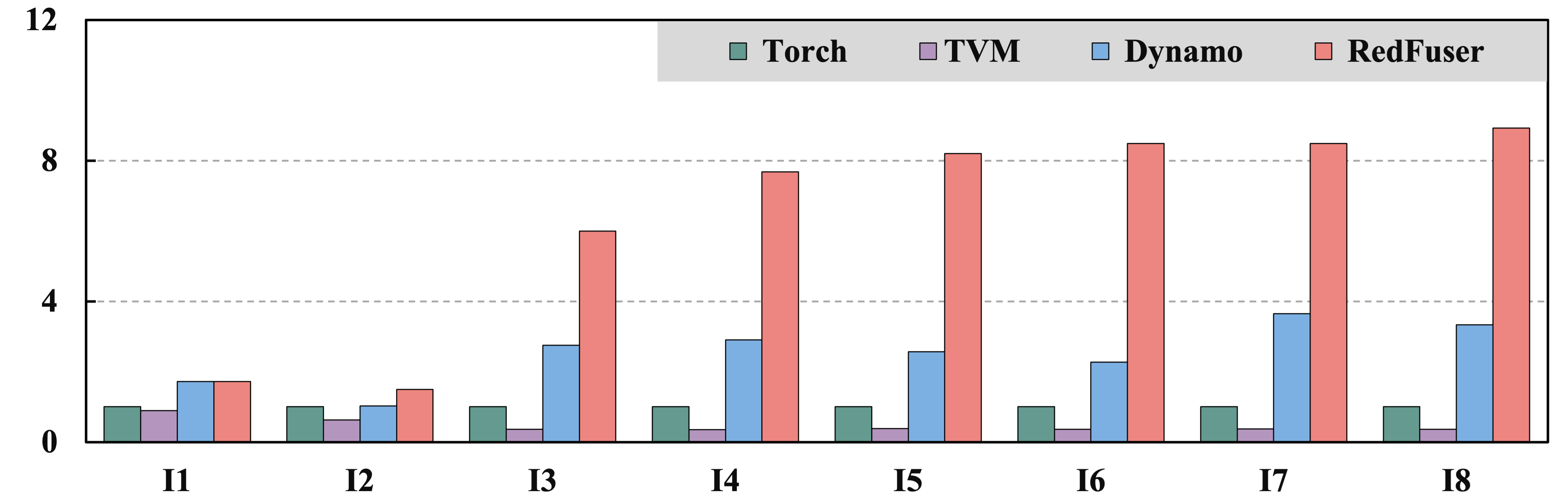}
        \caption{Moment of inertia on A10}
        \label{fig:exp_moi_A10}
    \end{subfigure}
    \begin{subfigure}[t]{0.49\textwidth}
        \includegraphics[width=\linewidth]{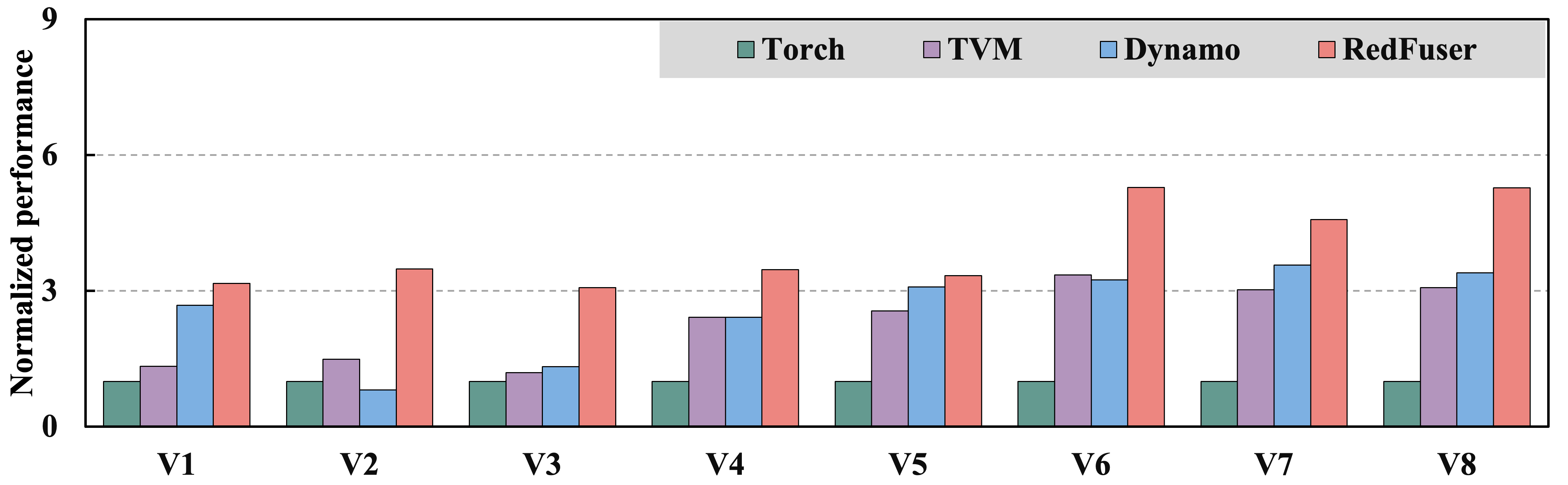}
        \caption{Variance on A100}
        \label{fig:exp_var_A100}
    \end{subfigure}
    \hfill
    \begin{subfigure}[t]{0.49\textwidth}
        \includegraphics[width=\linewidth]{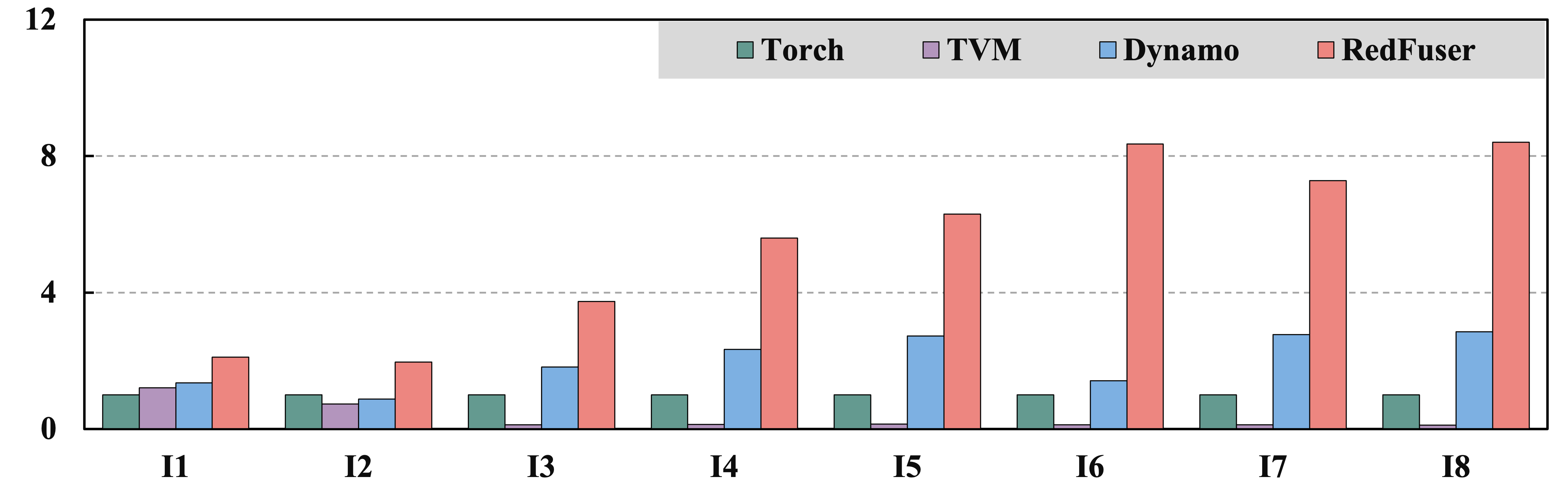}
        \caption{Moment of inertia on A100}
        \label{fig:exp_moi_A100}
    \end{subfigure}
    \begin{subfigure}[t]{0.49\textwidth}
        \includegraphics[width=\linewidth]{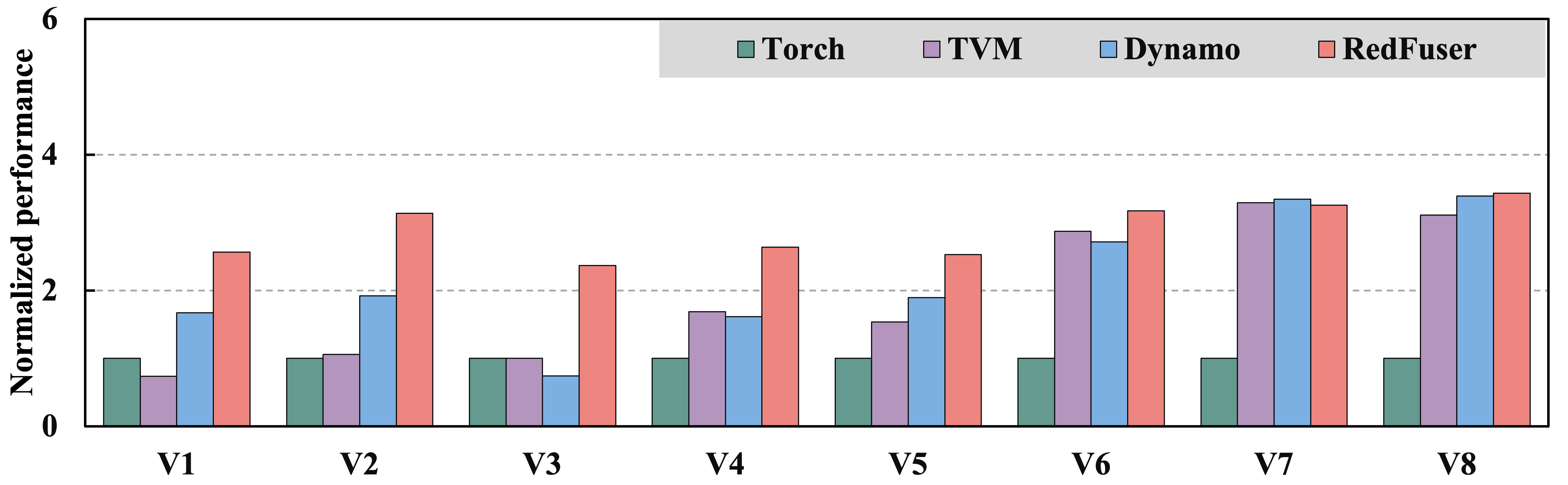}
        \caption{Variance on H800}
        \label{fig:exp_var_H800}
    \end{subfigure}
    \hfill
    \begin{subfigure}[t]{0.49\textwidth}
        \includegraphics[width=\linewidth]{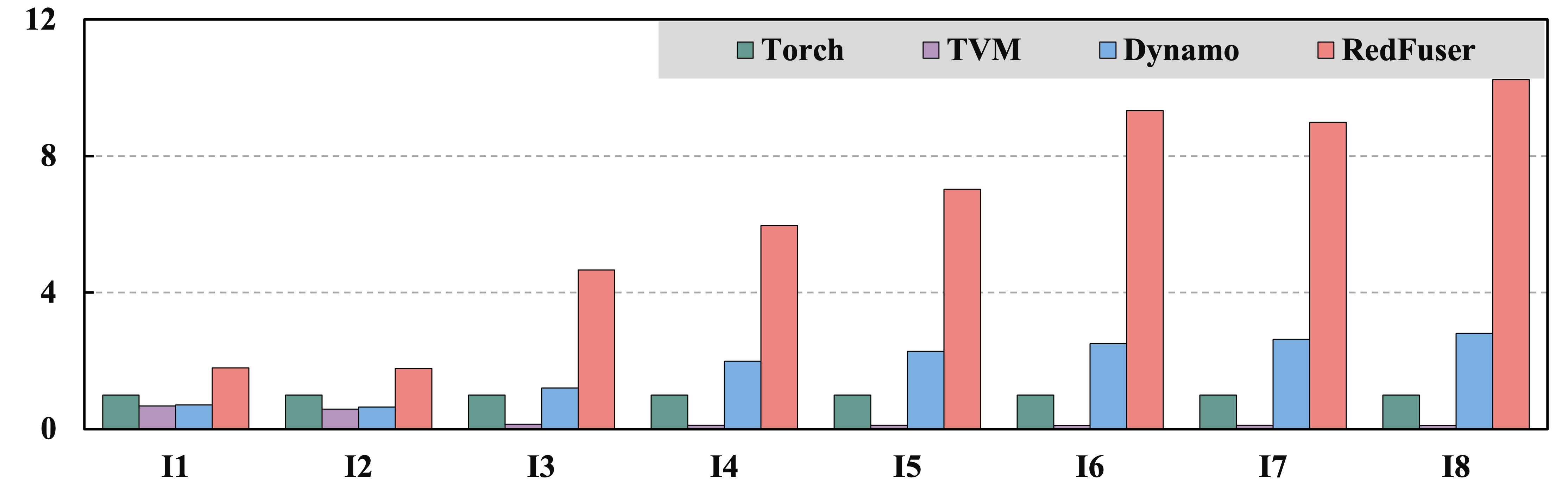}
        \caption{Moment of inertia on H800}
        \label{fig:exp_moi_H800}
    \end{subfigure}
    \begin{subfigure}[t]{0.49\textwidth}
        \includegraphics[width=\linewidth]{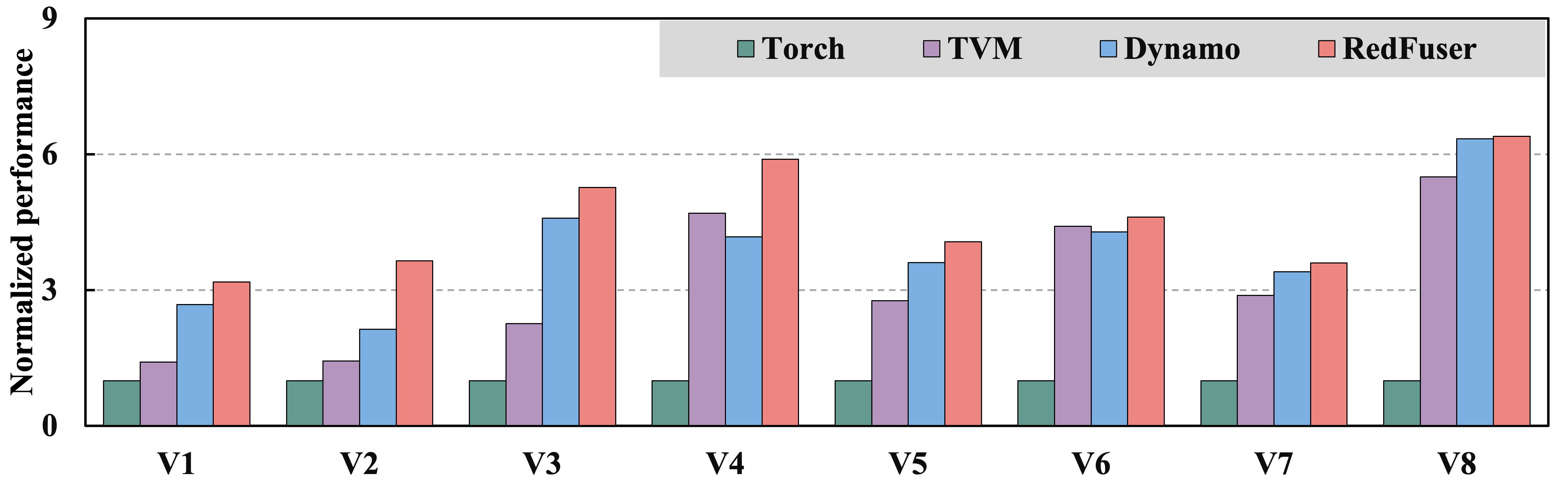}
        \caption{Variance on MI308X}
        \label{fig:exp_var_MI308X}
    \end{subfigure}
    \hfill
    \begin{subfigure}[t]{0.49\textwidth}
        \includegraphics[width=\linewidth]{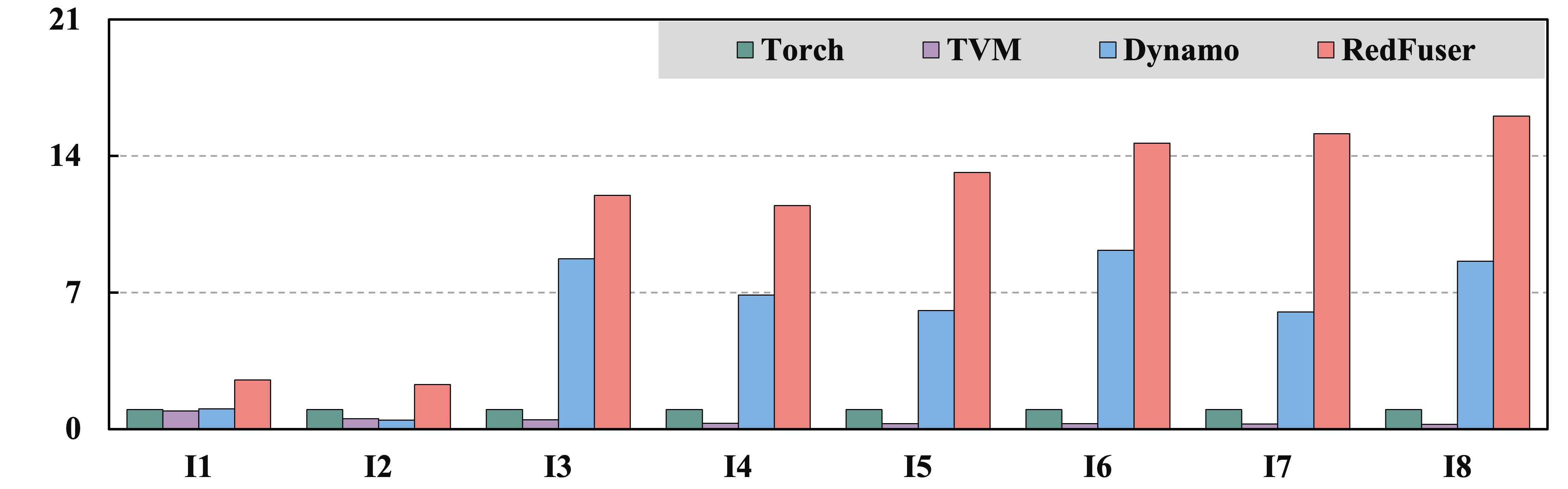}
        \caption{Moment of Inertia on MI308X}
        \label{fig:exp_moi_MI308X}
    \end{subfigure}
    \caption{The normalized performance of non-ML workloads across multiple platforms.}
    \Description{The normalized performance of non-ML workloads across multiple platforms.}
    \label{fig:exp_non_ml}
\end{figure*}

The results are shown in Figure~\ref{fig:exp_non_ml}. 
For variance computation, since variance frequently appears in ML workloads (e.g., in various normalization operators), 
PyTorch includes optimized implementations. Nevertheless, RedFuser achieves an average speedup of 4.8$\times$, 4.0$\times$, 2.9$\times$ and 4.6$\times$ over PyTorch Eager on A10, A100, H800 and MI308X, respectively. 
For moment of inertia computation, the presence of extensive element-wise operations between reductions prevents existing compilers from effectively fusing the reduction chains, resulting in suboptimal performance. 
In this case, RedFuser achieves an average speedup of 6.4$\times$, 5.5$\times$, 6.2$\times$ and 11.6$\times$ on A10, A100, H800 and MI308X, respectively.

\begin{table*}[tp]
    \centering
    \caption{The configurations of two non-ML workloads}
    \label{tab:non-ml-workload-configurations}
\begin{subtable}[t]{0.45\textwidth}
    \centering
    \begin{tabular}{lll}
        \toprule
        Name & bs & l \\
        \midrule
        V1 & 1 & 8192 \\
        V2 & 1 & 32768 \\
        V3 & 128 & 8192 \\
        V4 & 128 & 32768 \\
        V5 & 512 & 8192 \\
        V6 & 512 & 32768 \\
        V7 & 1024 & 8192 \\
        V8 & 1024 & 32768 \\
        \bottomrule
    \end{tabular}
    \caption{The configurations of Variance}
    \label{tab:var_configurations}
\end{subtable}
    \hfill
\begin{subtable}[t]{0.45\textwidth}
    \centering
    \begin{tabular}{llll}
        \toprule
        Name & bs & n & dim \\
        \midrule
        I1 & 1 & 8192 & 3 \\
        I2 & 1 & 32768 & 3 \\
        I3 & 128 & 8192 & 3 \\
        I4 & 128 & 32768 & 3 \\
        I5 & 512 & 8192 & 3 \\
        I6 & 512 & 32768 & 3 \\
        I7 & 1024 & 8192 & 3 \\
        I8 & 1024 & 32768 & 3 \\
        \bottomrule
    \end{tabular}
    \caption{The configurations of Moment of Inertia}
    \label{tab:moi_configurations}
\end{subtable}
\end{table*}

\subsection{Evaluation Across Multiple Platforms}
\label{Appendix:Additional_Evaluations}

We evaluate RedFuser on the four workloads from Section~\ref{chap:subgraph_perf} across NVIDIA A100, H800, and AMD MI308X GPUs. 
Compared to PyTorch Eager, RedFuser delivers average speedups of 6.7$\times$, 3.9$\times$, and 1.6$\times$ on MoE routing, 
and 6.9$\times$, 7.9$\times$, and 3.4$\times$ on MHA workloads, respectively. 
On FP8 per-token quantization with GEMM, it achieves a 2.0$\times$ speedup on MI308X.

\begin{figure*}[b]
    \centering
    \begin{subfigure}[t]{0.49\textwidth}
        \includegraphics[width=\linewidth]{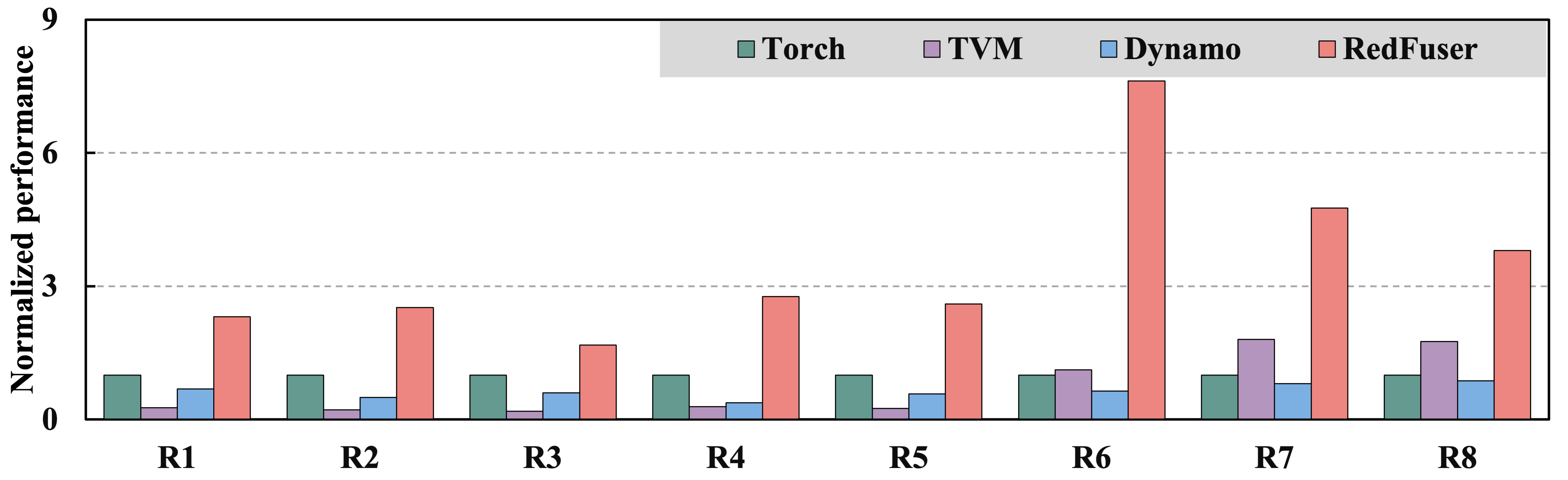}
        \caption{MoE routing on A100}
        \label{fig:moe_routing_a100}
    \end{subfigure}
    \hfill
    \begin{subfigure}[t]{0.49\textwidth}
        \includegraphics[width=\linewidth]{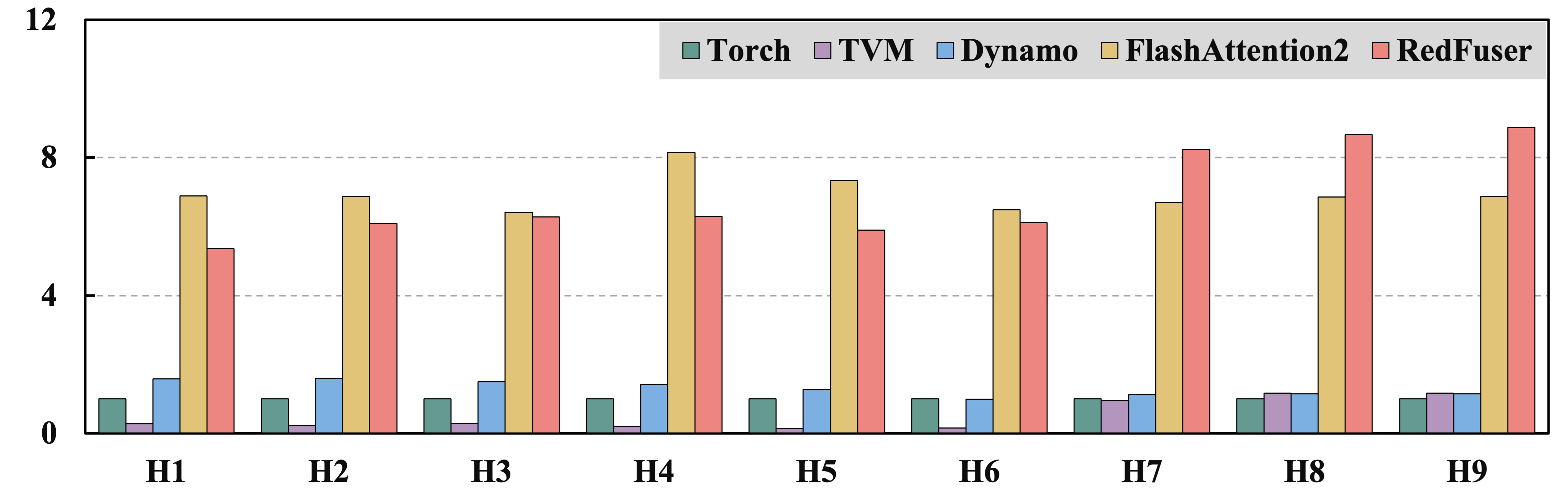}
        \caption{MHA on A100}
        \label{fig:mha_a100}
    \end{subfigure}

    \begin{subfigure}[t]{0.49\textwidth}
        \includegraphics[width=\linewidth]{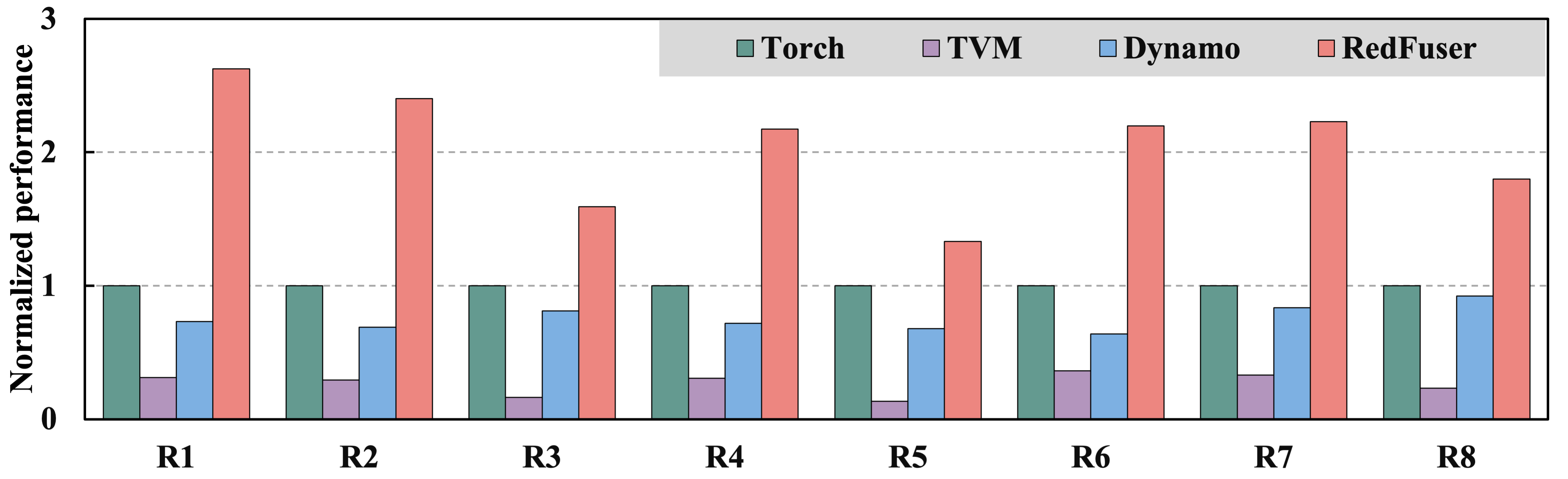}
        \caption{MoE routing on H800}
        \label{fig:moe_routing_h800}
    \end{subfigure}
    \hfill
    \begin{subfigure}[t]{0.49\textwidth}
        \includegraphics[width=\linewidth]{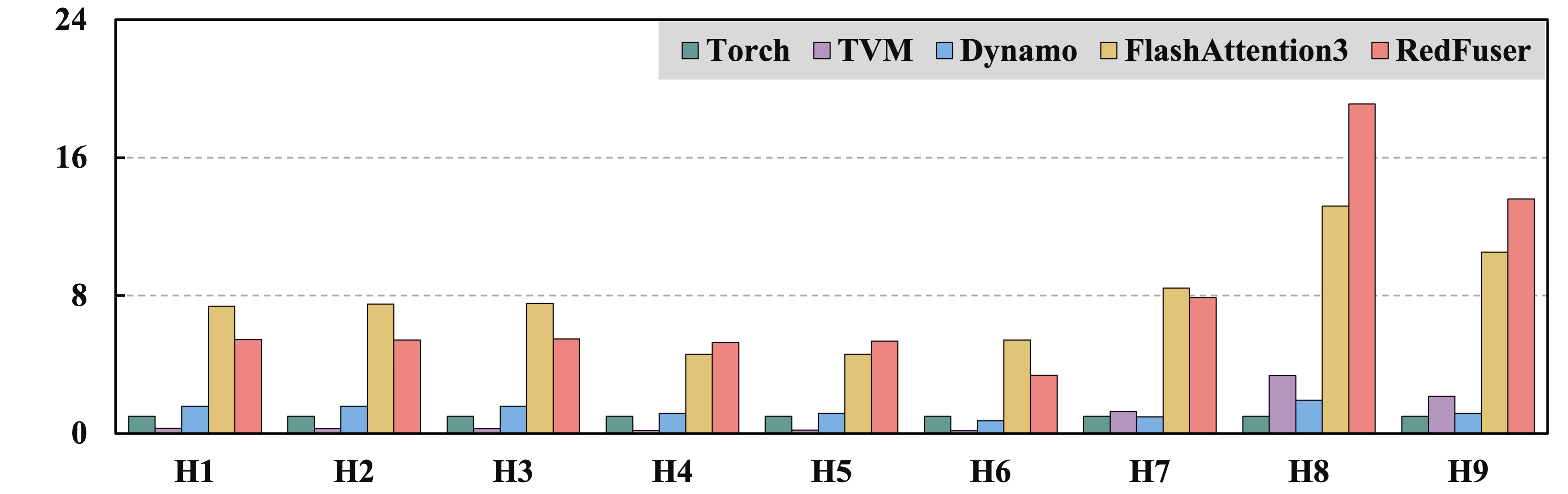}
        \caption{MHA on H800}
        \label{fig:mha_h800}
    \end{subfigure}

    \begin{subfigure}[t]{0.49\textwidth}
        \includegraphics[width=\linewidth]{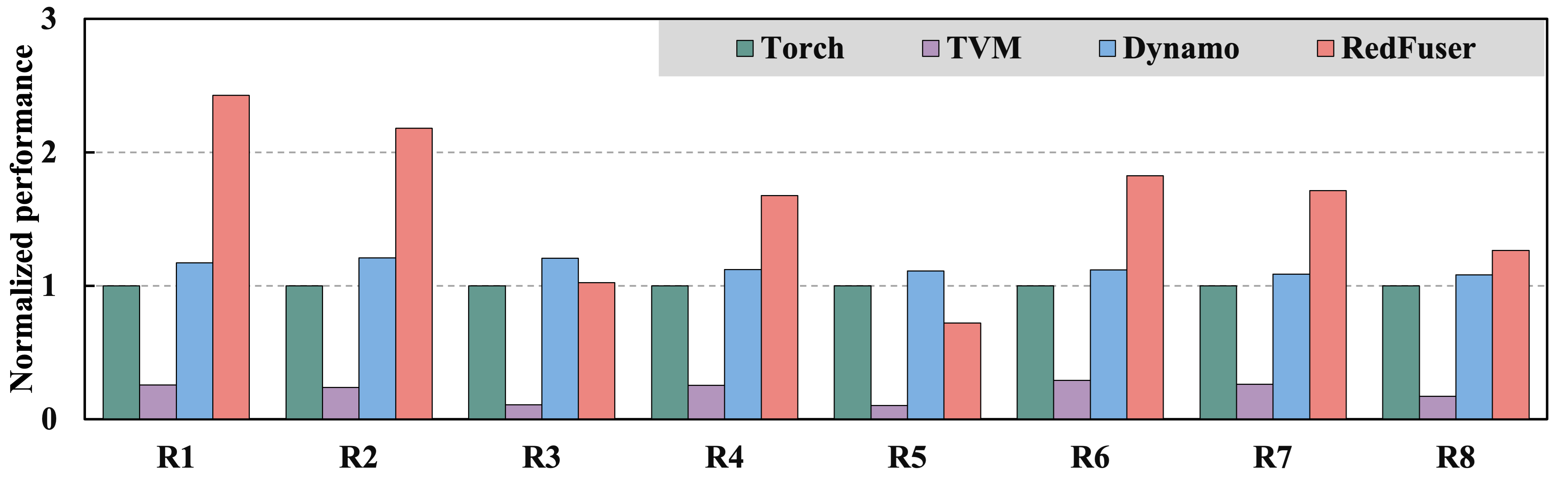}
        \caption{MoE routing on MI308X}
        \label{fig:moe_routing_mi308x}
    \end{subfigure}
    \hfill
    \begin{subfigure}[t]{0.49\textwidth}
        \includegraphics[width=\linewidth]{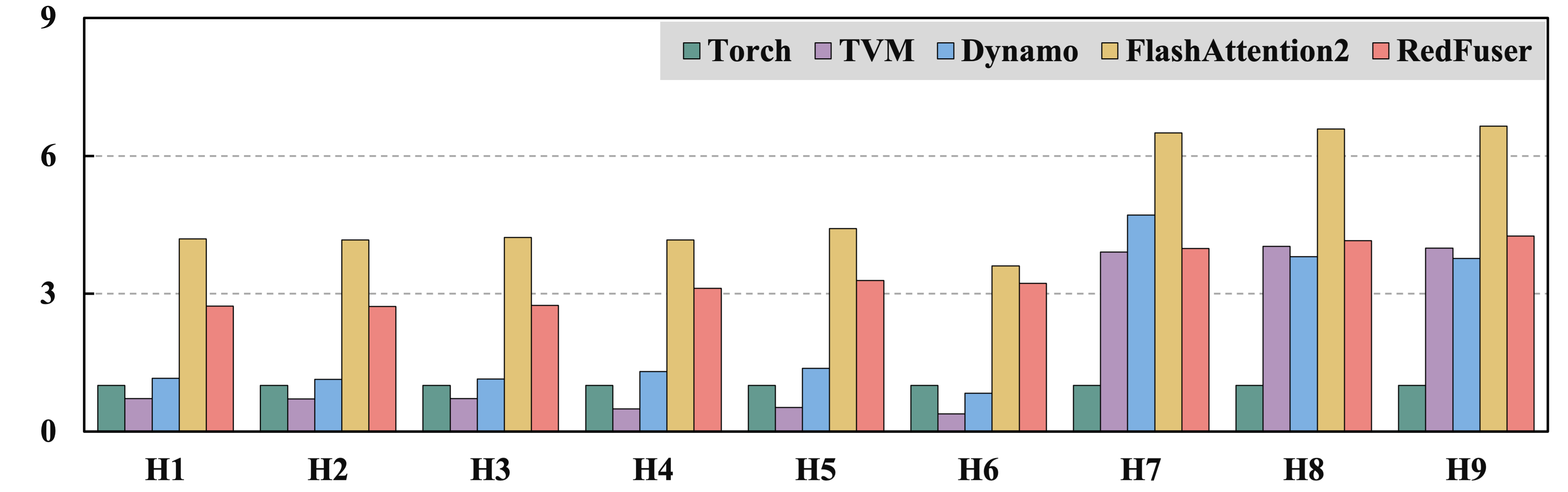}
        \caption{MHA on MI308X}
        \label{fig:mha_mi308x}
    \end{subfigure}

    \raggedright
    \begin{subfigure}[t]{0.49\textwidth}
        \includegraphics[width=\linewidth]{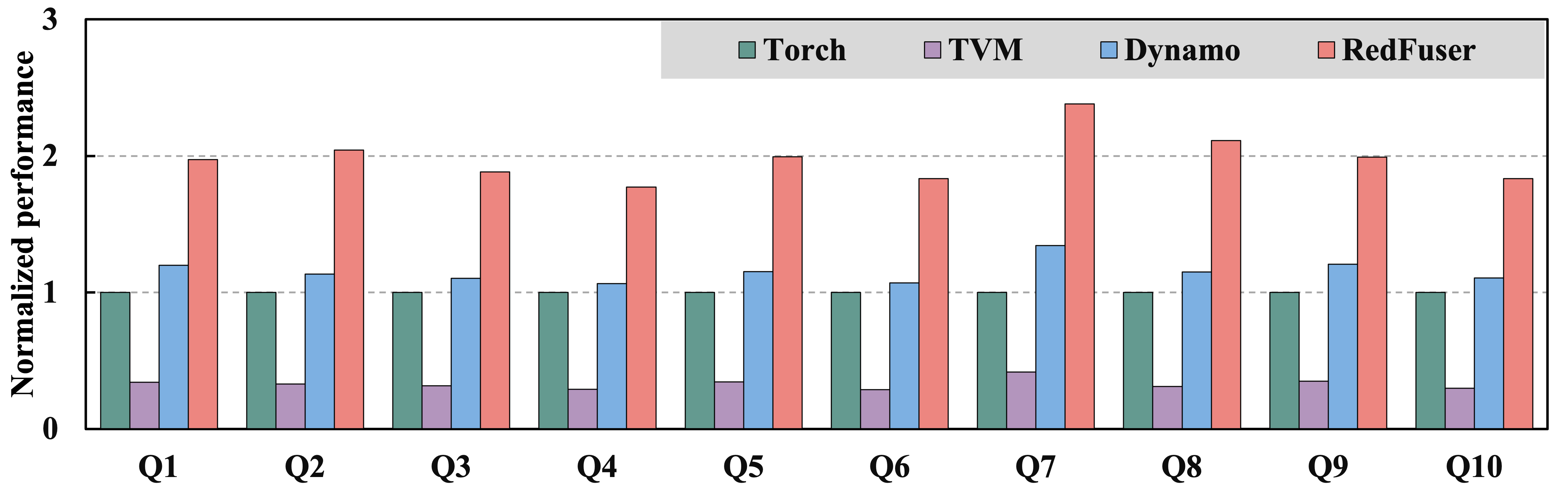}
        \caption{FP8 PerToken Quant $+$ GEMM on MI308X}
        \label{fig:quant_gemm_mi308x}
    \end{subfigure}
    \caption{The normalized performance of ML workloads across multiple platforms.}
    \Description{The normalized performance of ML workloads across multiple platforms.}
    \label{fig:ml_workloads_across_platforms}
\end{figure*}

\begin{figure*}[t]
    \begin{minipage}[t]{0.48\textwidth}
        \centering
        \begin{align*}
        \mathit{TileOp} &::= \texttt{copy}(\mathit{tile},\ \mathit{tile}) \\
        &\quad\mid \texttt{gemm}(\mathit{tile},\ \mathit{tile},\ \mathit{tile}) \\
        &\quad\mid \texttt{reduce}(\mathit{tile},\ \mathit{tile},\ \texttt{axis} = \mathit{lit},\ \mathit{op}) \\
        &\quad\mid \texttt{parallel}(\mathit{id}[\mathit{expr}+],\ \mathit{op}(\mathit{expr}*),\ \mathit{id}+,\ \mathit{range}+) \\
        &\quad\mid \texttt{fill}(\mathit{tile},\ \mathit{lit}) \\
        \mathit{tile} &::= \mathit{id}[\mathit{range}+] \\
        \mathit{range} &::= \mathit{expr} : \mathit{expr} \\
        \mathit{expr} &::= \mathit{lit} \mid \mathit{id} \mid \mathit{op}(\mathit{expr}*) \\
        \mathit{op} &::= \texttt{+} \mid \texttt{-} \mid \texttt{max} \mid \texttt{exp} \mid \dots
        \end{align*}
        \caption{TileOp Specification}
        \label{fig:tileop_specification}
    \end{minipage}
    \hfill
    \begin{minipage}[t]{0.48\textwidth}
    \centering
    \begin{lstlisting}
for qs in range(512):
    # reduction1: gemm(Q,K)
    for kvs in range(512):  
        for d in range(64): 
            P[qs, kvs] += Q[qs, d] * K[kvs, d] 
    for kvs in range(512):  # reduction2: max(P)
        pmax[qs] = max(pmax[qs], P[qs, kvs]) 
    # reduction3: sum(exp(P - pmax))
    for kvs in range(512):  
        psum[qs] += exp(P[qs, kvs] - pmax[qs])
    # reduction4: gemm(exp(P - pmax) / psum, V)
    for kvs in range(512):  
        for d in range(64):
            o[qs, d] += exp(P[qs, kvs] - pmax[qs]) / psum[qs] * V[kvs, d]
    \end{lstlisting}
    \caption{Unfused attention TIR}
    \Description{Unfused attention TIR}
    \label{fig:unfused_attention_TIR}
    \end{minipage}
\end{figure*}

\begin{figure*}[t]
    \begin{subfigure}[t]{0.99\textwidth}
    \begin{lstlisting}
for qs in range(512):
    for kvs in range(512):
        # reduction1: gemm(Q,K)
        for d in range(64):
            P[qs, kvs] += Q[qs, d] * K[kvs, d] 

        # reduction2: max(P)
        # step 1. store previous result
        pmax_prev[qs] = pmax[qs]
        # step 3. perform reduction
        pmax[qs] = max(pmax[qs], P[qs, kvs])
        
        # reduction3: sum(exp(P - pmax))
        # step 1. store previous result
        psum_prev[qs] = psum[qs]
        # step 2. apply correction
        psum[qs] *= exp(pmax_prev[qs] - pmax[qs])
        # step 3. perform reduction
        psum[qs] += exp(P[qs, kvs] - pmax[qs])

        # reduction4: gemm(exp(P - pmax) / psum, V)
        # step 2. apply correction
        for d in range(64):
            o[qs, d] *= exp(pmax_prev[qs] - pmax[qs]) * (psum_prev[qs] / psum[qs])
        # step 3. perform reduction
        for d in range(64):
            o[qs, d] += exp(P[qs, kvs] - pmax[qs]) / psum[qs] * V[kvs, d]          
    \end{lstlisting}
    \caption{FlashAttention scalar-level IR}
    \label{fig:flashattention_scalar}
    \end{subfigure}

    \begin{subfigure}[t]{0.99\textwidth}
    \begin{lstlisting}
bx = launch_thread("blockIdx.x", 4)
# we omit the alloc shared / fragment buffer for brevity
copy(Q[bx * 128:bx * 128 + 128, 0:64], Q_shared[0:128, 0:64])
for stage in range(4):
    copy(K[stage * 128:stage * 128 + 128, 0:64], K_shared[0:128, 0:64])
    copy(V[stage * 128:stage * 128 + 128, 0:64], V_shared[0:128, 0:64])
    # reduction1: gemm(Q,K)
    gemm(Q_shared, K_shared, P_frag)
    
    # reduction2: max(P)
    # step 1. store previous result
    copy(pmax, pmax_prev)
    # step 3. perform reduction                       
    reduce(P_frag, pmax, axis=1, op=max)                                           
    
    # reduction3: sum(exp(P - pmax))
    # step 1. store previous result
    copy(psum, psum_prev)
    # step 2. apply correction
    parallel(psum[i], psum[i] * exp(pmax_prev[i] - pmax[i]), i, 128)
    # step 3. perform reduction
    parallel(pexp[i, j], exp(P_frag[i, j] - pmax[i]), i, 128, j, 128)
    reduce(pexp, psum, axis=1, op=add)
    
    # reduction4: gemm(exp(P - pmax) / psum, V)
    # step 2. apply correction
    parallel(o_frag[i, j], o_frag[i, j] * exp(pmax_prev[i] - pmax[i]) * (psum_prev[i] / psum[i]), i, 128, j, 64)
    # step 3. perform reduction
    parallel(oexp[i, j], exp(P_frag[i, j] - pmax[i]) / psum[i], i, 128, j, 128)
    gemm(oexp, V_shared, o_frag)
copy(o_frag[0:128, 0:64], o[bx * 128:bx * 128 + 128, 0:64])
    \end{lstlisting}
    \caption{FlashAttention tile-level IR}
    \label{fig:flashattention_tile}
    \end{subfigure}
    \caption{FlashAttention IR}
    \Description{FlashAttention IR}
    \label{fig:flashattention_ir}
\end{figure*}

\begin{figure*}[t]
    \begin{subfigure}[t]{0.99\textwidth}
        \begin{lstlisting}
# partial segment
for qs, split in grid(512, 2):
    for kvs in range(256):
        offset = split * 256 + kvs
        for d in range(64):
            P[qs, offset] += Q[qs, d] * K[offset, d]
        pmax_part_prev[qs, split] = pmax_part[qs, split]
        pmax_part[qs, split] = max(pmax_part[qs, split], P[qs, offset])

        psum_part_prev[qs, split] = psum_part[qs, split]
        psum_part[qs, split] *= exp(pmax_part_prev[qs, split] - pmax_part[qs, split])
        psum_part[qs, split] += exp(P[qs, offset] - pmax_part[qs, split])

        for d in range(64):
            o_part[qs, d, split] *= exp(pmax_part_prev[qs, split] - pmax_part[qs, split]) * (psum_part_prev[qs, split] / psum_part[qs, split])
        for d in range(64):
            o_part[qs, d, split] += exp(P[qs, offset] - pmax_part[qs, split]) / psum_part[qs, split] * V[offset, d]
# combine
for qs in range(512):
    for split in range(2):
        pmax[qs] = max(pmax[qs], pmax_part[qs, split])
    for split in range(2):
        psum_part[qs, split] *= exp(pmax_part[qs, split] - pmax[qs])
        psum[qs] += psum_part[qs, split]
    for split in range(2):
        for d in range(64):
            o_part[qs, d, split] *= exp(pmax_part[qs, split] - pmax[qs]) * (psum_part[qs, split] / psum[qs])
        for d in range(64):
            o[qs, d] += o_part[qs, d, split]
        \end{lstlisting}
        \caption{FlashDecoding scalar-level IR}
        \label{fig:flashdecoding_scalar}
    \end{subfigure}

    \begin{subfigure}[t]{0.99\textwidth}
        \begin{lstlisting}
# partial segment
bx = launch_thread("blockIdx.x", 4)
by = launch_thread("blockIdx.y", 2)
copy(Q[bx * 128:bx * 128 + 128, 0:64], Q_shared[0:128, 0:64])
for stage in range(2):
    offset = by * 256 + stage * 128
    copy(K[offset:offset + 128, 0:64], K_shared[0:128, 0:64])
    copy(V[offset:offset + 128, 0:64], V_shared[0:128, 0:64])
    gemm(Q_shared, K_shared, P_frag)
    copy(pmax_frag, pmax_frag_prev)
    reduce(P_frag, pmax_frag, axis=1, op=max)
    copy(psum_frag, psum_frag_prev)
    parallel(psum_frag[i], psum_frag[i] * exp(pmax_frag_prev[i] - pmax_frag[i]), i, 128)
    parallel(pexp[i, j], exp(P_frag[i, j] - pmax_frag[i]), i, 128, j, 128)
    reduce(pexp, psum_frag, axis=1, op=add)
    parallel(o_frag[i, j], o_frag[i, j] * exp(pmax_frag_prev[i] - pmax_frag[i]) * (psum_frag_prev[i] / psum_frag[i]), i, 128, j, 64)
    parallel(oexp[i, j], exp(P_frag[i, j] - pmax_frag[i]) / psum_frag[i], i, 128, j, 64)
    gemm(oexp, V_shared, o_frag)
copy(pmax_frag[0:128], pmax_part[bx * 128:bx * 128 + 128, by:by+1])
copy(psum_frag[0:128], psum_part[bx * 128:bx * 128 + 128, by:by+1])
copy(o_frag[0:128, 0:64], o_part[bx * 128:bx * 128 + 128, 0:64, by:by+1])
# combine
bx = launch_thread("blockIdx.x", 4)
copy(pmax_part[bx * 128:bx * 128 + 128, 0:2], pmax_frag[0:128, 0:2])
copy(psum_part[bx * 128:bx * 128 + 128, 0:2], psum_frag[0:128, 0:2])
copy(o_part[bx * 128:bx * 128 + 128, 0:64, 0:2], o_frag[0:128, 0:64, 0:2])
reduce(pmax_frag, pmax, axis=1, op=max)
parallel(psum_frag[i, j], psum_frag[i, j] * exp(pmax_frag[i, j] - pmax[i]), i, 128, j, 2)
reduce(psum_frag, psum, axis=1, op=add)
parallel(o_frag[i, j, k], o_frag[i, j, k] * exp(pmax_frag[i, k] - pmax[i]) * (psum_frag[i, k] / psum[i]), i, 128, j, 64, k, 2)
reduce(o_frag, o_final, axis=2, op=add)
copy(o_final[0:128, 0:64], o[bx * 128:bx * 128 + 128, 0:64])
        \end{lstlisting}
        \caption{FlashDecoding tile-level IR}
        \label{fig:flashdecoding_tile}
    \end{subfigure}
    \caption{FlashDecoding IR}
    \Description{FlashDecoding IR}
    \label{fig:flashdecoding_ir}
\end{figure*}

\end{document}